\documentclass[12pt,twoside,openright,final]{report}
\usepackage{thesis}

\usepackage[ps2pdf,
            bookmarks=true,
            bookmarksnumbered=true,
            hypertexnames=false,
            linkbordercolor={0 0 1},
            pdfborder={0 0 10.0}]{hyperref}
\hypersetup{
pdfauthor = {Arnaud Koetsier},
pdftitle = {MSc Thesis},
pdfsubject = {Matrix Models of 2D String Theory},
pdfkeywords = {2D String Theory, Matrix Models, Tachyons, MQM, Random Matrices, Backgrounds},
pdfcreator = {LaTeX with hyperref package},
pdfproducer = {dvips + ps2pdf}}

\newlength{\myVSpace}
\newcommand{\xstrut}{
    \raisebox{-1 \myVSpace}
    {\rule{0pt}{\myVSpace}}%
    }
\newcommand{\sep}{\xstrut \\} 

\newcommand{\be}{\begin{equation}}
\newcommand{\ee}{\end{equation}}
\newcommand{\beq}{\begin{eqnarray}}
\newcommand{\eeq}{\end{eqnarray}}
\newcommand{\bea}[2]{\be\label{#2}\begin{array}{#1}}
\newcommand{\eea}{\end{array}\ee}
\newcommand{\eq}{@{\;=\;}}
\newcommand{\bde}{\begin{deqarr}}   
\newcommand{\ede}{\end{deqarr}}     
\newcommand{\nn}{\nonumber}         

\def\refeq#1{eq. (\ref{#1})}
\def\re#1{(\ref{eq:#1})}

\def\figlabel#1{
\refstepcounter{figure}
 \label{#1}
 \addtocounter{figure}{-1}
}
\def\tablabel#1{
\refstepcounter{table}
 \label{#1}
 \addtocounter{table}{-1}
}

\def\EndOfChapter{
\
\centerline{\rule{0.6\textwidth}{0.6pt}}
\
}

\DeclareMathOperator*{\sumL}{\mbox{\large $\sum$}}        
\DeclareMathOperator{\order}{\mathcal{O}}                 
\DeclareMathOperator{\Tr}{\,{\rm Tr}\,}
\DeclareMathOperator{\tr}{\Tr}
\DeclareMathOperator{\diag}{{\rm diag}}
\DeclareMathOperator{\p}{\partial}

\DeclareMathOperator{\dom}{{\rm Dom}}
\DeclareMathOperator{\vol}{\,{\rm Vol}\,}
\DeclareMathOperator{\pv}{\,{\rm P.\!V.}}

\def\({\left(}
\def\){\right)}
\def\[{\left[}
\def\]{\right]}
\def\lb{\lbrace}
\def\rb{\rbrace}
\def\llb{\left\lbrace}

\def\la{\langle}
\def\ra{\rangle}
\def\lla{\left\langle}
\def\rra{\right\rangle}

\newcommand{\nf}[2]{\nicefrac{#1}{#2}}                    
\newcommand{\fracL}[2]{\frac{\mbox{\raisebox{3pt}{\normalsize $#1$}}}
{\mbox{\raisebox{-4pt}{\normalsize $#2$}}}}               
\def\geq{\geqslant}                       
\def\leq{\leqslant}                       
\def\hbar{\hslash}
\def\oint{\ointctrclockwise}              
\newcommand\der[2]{\frac{\p #1}{\p #2}}   
\newcommand\derL[2]{\fracL{\p #1}{\p #2}} 

\newcommand{\snd}[1]{#1$^{\rm nd}$}

\newcommand{\sth}[1]{#1$^{\rm th}$}
\newcommand{\gr}[2]{\mathsf{#1}(#2)}
\def\hf{\frac{1}{2}}

\def\suml{\sum\limits}
\def\prodl{\prod\limits}
\def\intl{\int\limits}
\def\qqquad{\qquad\qquad}

    \def\a{\alpha}
    \def\b{\beta}       
    \def\g{\gamma}
    \def\d{\delta}
    \def\e{\varepsilon}
    \def\eps{\epsilon}
    \def\et{\eta}
    \def\i{\iota}    
    \def\k{\kappa}
    \def\l{\lambda}     

    \def\na{\aleph}
    
    \def\r{\rho}        
    \def\vr{\varrho}
    \def\s{\sigma}
    \def\t{\tau}
    \def\th{\theta}
    
    \def\G{\Gamma}
    \def\D{\Delta}

    \def\Sig{\Sigma}
    
    \def\O{\Omega}      
    \def\o{\omega }     
    \def\dd{\nabla}

   \def\bbR {\mathbb{R}}
   \def\bbC {\mathbb{C}}
   \def\bbI {\mathbbm{1}}       

    \def\bD{\bar \Delta }

    \def\bL{\bar L}
    \def\bH{\overline H}
    \def\bM{\overline M}
    \def\bW{ \overline W}
    \def\bCW{\overline \CW}

    \def\bw{\overline w}

    \def\bz{\bar z}
    \def\bu{\bar u}
    
    \def\bpsi{\overline \psi}
    \def\dl{\Phi} 

\def\ho{\hat\o}
\def\hh{\hat{h}}
\def\hCR{\widehat{\CR}}
\def\hp{\widehat{p}}
\def\hPi{\widehat{\Pi}}
\def\hHMQM{\widehat{H}_{\rm MQM}}
\def\htlPi{\widehat{\tPi}}
\def\hS{\widehat{S}}
\def\hE{\hat{E}}

\def\dM{\dot M}
\def\dx{\dot x}
\def\dO{\dot \Omega}
\def\dal{\dot \alpha}
\def\dbe{\dot \beta}

\def\dtbe{\dot{\tilde{\beta}}}

    \def\tlH{\widetilde H}

    \def\tlT{\widetilde T}        \def\tlV{\widetilde V}
        \def\tlX{\widetilde X}    
    
\def\tbe{\tilde{\beta}}
\def\tPi{\widetilde{\Pi}}

   \def\CD {{\cal D}}   \def\CE {{\cal E}}   \def\CF {{\cal F}}
         
         \def\CL {{\cal L}}
      \def\CN {{\cal N}}   \def\CO {{\cal O}}
   \def\CP {{\cal P}}      \def\CR {{\cal R}}
      \def\CT {{\cal T}}   
      \def\CW {{\cal W}}

\def\MA{M_{\rm A}}
\def\MB{M_{\rm B}}
\def\HMQM{H_{_{\rm MQM}}}
\def\Hcol{H_{_{\rm eff}}}
\def\Seff{S_{_{\rm eff}}}
\def\Phis{\Phi^{(\rm sing)}}
\def\Psis{\Psi^{(\rm sing)}}
\def\gstr{g_{\rm str}}
\def\Hss{\widehat{\mathscr{H}}}  
\def\fermiE{\varepsilon_{\rm F}}

\def\Xp{X_{\!_+}}
\def\Xm{X_{\!_-}}
\def\Xpm{X_{\!_\pm }}

\def\Phipm{\Phi_{\!_\pm}}

\def\ppl{p_{\!_+}}
\def\pmi{p_{\!_-}}
\def\ppm{p_{\!_\pm}}

\def\vphip{\varphi_{\!_+}}
\def\vphim{\varphi_{\!_-}}
\def\vphipm{\varphi_{\!_\pm}}
\def\vpp{\vphip}
\def\vpm{\vphim}
\def\vppm{\vphipm}

\def\xp{x_{\!_{+}}}
\def\xm{x_{\!_{-}}}
\def\xpm{x_{\!_{\pm}}}
\def\xmp{x_{\!_{\mp}}}

\def\hxp{\hat{x}_{\!_{+}}}
\def\hxm{\hat{x}_{\!_{-}}}
\def\hxpm{\hat{x}_{\!_{\pm}}}
\def\hxmp{\hat{x}_{\!_{\mp}}}

\def\delpm{\hat{\p}_{\!_{\pm}}}

\def\psip{\psi_{\!_+}}
\def\psim{\psi_{\!_-}}
\def\psipm{\psi_{\!_{\pm}}}
\def\psimp{\psi_{\!_{\mp}}}

\def\Psipm{\Psi_{\!_{\pm}}}

\def\tpsip{\widetilde{\psi}_{\!_{+}}}
\def\tpsim{\widetilde{\psi}_{\!_{-}}}
\def\tpsipm{\widetilde{\psi}_{\!_{\pm}}}

\def\tpmn#1{t_{\pm #1}}

\def\Ep{E_{\!_+}}
\def\Em{E_{\!_-}}
\def\Epm{E_{\!_\pm}}

\def\alp{\alpha_{\!_+}}
\def\alm{\alpha_{\!_-}}
\def\alpm{\alpha_{\!_\pm}}
\def\almp{\alpha_{_\mp}}

\def\CWpm{\CW_{\pm}}

\def\Lpm{L_{\pm}}

\def\wpm#1{w_{\pm#1}}

\def\colfield{\varphi}  
\newcommand{\phm}{_{n\phantom{-}}} 

\begin{document}



\pagenumbering{roman}
\setlength{\parskip}{0pt}
\topmargin=0in
\enlargethispage{3cm}
\thispagestyle{empty}

\newlength{\boddsidemargin}\setlength{\boddsidemargin}{\oddsidemargin} 
\addtolength{\evensidemargin}{-\bodyshift}
\setlength{\oddsidemargin}{\evensidemargin}

\newlength{\rulespace}\setlength{\rulespace}{5pt}       
\newlength{\ruleoffset}\setlength{\ruleoffset}{5pt}     
\newlength{\vrulelength}\setlength{\vrulelength}{165pt} 
\newlength{\hrulelength}\setlength{\hrulelength}{400pt} 

\vspace*{-20mm} \hspace*{0pt}
\hspace{-4mm}\rule[-27pt]{4pt}{73pt} 
\begin{flushleft}
  \raisebox{37pt}[0pt][0pt]{ 
    \hspace{-17pt}
   \includegraphics[scale=1]{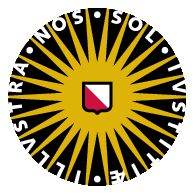}       
    \raisebox{32pt}{\hspace{-20pt}\BemboBold{19} University of Utrecht}
    \hspace{-320pt} 
    \raisebox{5pt}{\rule{15.5cm}{0.5pt}}       
   }
\end{flushleft}

\begin{center}
    \vskip 30pt
    {\fontsize{18pt}{0pt}\fontseries{bx}  \selectfont
    MSc Thesis \\[6pt] (Theoretical Physics)
    }
    \vskip 75pt
\end{center}
\newlength{\tempaa}
\setlength{\tempaa}{0pt}                        
\hspace{\tempaa} \rule[0ex]{\hrulelength}{0.5pt} \vskip -\baselineskip \vskip \rulespace  \addtolength{\tempaa}{-\ruleoffset}
\hspace{\tempaa} \rule[0ex]{\hrulelength}{0.5pt} \vskip -\baselineskip \vskip \rulespace  \addtolength{\tempaa}{-\ruleoffset}
\hspace{\tempaa} \rule[0ex]{\hrulelength}{0.5pt} \vskip -\baselineskip \vskip \rulespace  \addtolength{\tempaa}{-\ruleoffset}
\hspace{\tempaa} \rule[0ex]{\hrulelength}{0.5pt} \vskip -\baselineskip \vskip \rulespace  \addtolength{\tempaa}{-\ruleoffset}
\hspace{\tempaa} \rule[0ex]{\hrulelength}{0.5pt} \vskip -\baselineskip \vskip \rulespace  \setlength{\tempaa}{0mm}
\newlength{\tempab}\setlength{\tempab}{0pt}\addtolength{\tempab}{\vrulelength}\addtolength{\tempab}{-5\rulespace}
\addtolength{\rulespace}{-8.33pt}                
\hskip 0pt                                      
\raisebox{-\tempab}[0pt][0pt]{
\rule[\tempaa]{0.5pt}{\vrulelength} \hskip \rulespace \addtolength{\tempaa}{\ruleoffset}
\rule[\tempaa]{0.5pt}{\vrulelength} \hskip \rulespace \addtolength{\tempaa}{\ruleoffset}
\rule[\tempaa]{0.5pt}{\vrulelength} \hskip \rulespace \addtolength{\tempaa}{\ruleoffset}
\rule[\tempaa]{0.5pt}{\vrulelength} \hskip \rulespace \addtolength{\tempaa}{\ruleoffset}
\rule[\tempaa]{0.5pt}{\vrulelength} \addtolength{\tempaa}{\ruleoffset}
}

\begin{center}
\raisebox{-40pt}{                                   
 \colorbox[rgb]{1,1,1}{
  \parbox[c]{125mm}{                                 
   \GaramondBold{18}
   \begin{changemargins}{0pt}{0pt}{30pt}
     \centering Matrix Models of \\ 2D String Theory in Non{\char'226}trivial Backgrounds
   \end{changemargins}
  }
 }
}\\[150pt]                                        
{\fontsize{18pt}{0pt} \fontseries{bx} \selectfont
Arnaud Koetsier}

\vspace{\stretch{1}}
{\fontsize{14pt}{0pt} \fontseries{bx} \selectfont
Supervisor: Prof. G. 't Hooft} \vspace*{\stretch{1}}
\end{center}
\setlength{\topmargin}{\bodytopmargin}
\setlength{\parskip}{\bodyparskip}
\newpage\thispagestyle{empty}\normalcolor

\begin{center}
\vspace*{\stretch{1}}
Submitted in partial fulfilment of the requirements for the degree of
Master of Science (Theoretical Physics) at the University of
Utrecht in the Netherlands.
\end{center}

\vspace{\stretch{7}}
Version: $8^{\rm th}$ March, 2005.\\[30pt]
Typeset with \LaTeXe{} in Computer Modern Roman, 12pt. All figures created
using jfig 2.22, Paint Shop Pro 8 and Wolfram Mathematica 5. Feynman
diagrams constructed using \texttt{feynmf} package.
\addtolength{\evensidemargin}{\bodyshift}
\setlength{\oddsidemargin}{\boddsidemargin}

\newpage\thispagestyle{empty}

\vspace*{\stretch{1}}
{\fontsize{18pt}{0pt} \fontseries{bx} \selectfont
\centerline {Abstract}}

After a brief review of critical string theory in trivial
backgrounds we begin with introduction to strings in non--trivial
backgrounds and noncritical string theory. In particular, we relate
the latter to critical string theory in a linear dilaton background.

We then show how a black hole background arises from 2D string
theory and discuss some of its properties. A time--dependant tachyon
background is constructed by perturbing the CFT describing string
theory in a linear dilaton background. It is then explained that the
T--dual of this theory with one non--vanishing tachyon coupling,
which is a sine-Liouville CFT, is seemingly equivalent to the exact
CFT describing the Euclidean black hole background.

Subsequently, we launch into a review of some important facts
concerning random matrix models and matrix quantum mechanics (MQM),
culminating in an MQM model of 2D string theory in a dynamic tachyon
background. We then solve this theory explicitly in the tree level
approximation for the case of two non--vanishing tachyon couplings,
which generalises the case of sine-Liouville CFT previously
considered in the literature.

\vspace*{\stretch{1}}

\setlength{\topmargin}{\bodytopmargin}
\setlength{\parskip}{\bodyparskip}
\newpage\thispagestyle{empty}\normalcolor

\vspace*{\stretch{1}} {\fontsize{18pt}{0pt} \fontseries{bx}
\selectfont \centerline {Acknowledgements}}

First of all, I would like to extend my sincere gratitude to my
supervisor, Prof. Gerard 't Hooft for his guidance and the
enlightening discussions, and to my co--supervisor, Dr. Sergei
Alexandrov who suggested the topic, for his boundless tolerance to
my unending questions and for proofreading my thesis. I am also
grateful to Dr. Zoltan Kadar for introducing me to my supervisors,
and for lending a willing ear when I got stuck.

I would also like to thank my colleges Gabriel Grigorescu, Edmundo
Sanchez, Igor Shokarev, Aaron Swaving, Xuhui Wang and Sasha Zozulya
for the all the helpful hints, for the friendly working atmosphere
and mutual encouragement.


\vspace{\stretch{1}}



\newpage\thispagestyle{empty}
\setlength{\parskip}{\tocparskip} \tableofcontents
\newpage\thispagestyle{empty}
\cleardoublepage

\setlength{\parskip}{\bodyparskip}
\setlength{\topmargin}{\bodytopmargin}
\setcounter{prefpage}{\value{page}}\addtocounter{prefpage}{-1}
\pagenumbering{arabic}
%
%
%
%

\chapter{String theory}
\label{ch:STR}
String theory is almost 40 years old and was originally an
interesting interpretation of models for the nuclear strong force.
It had many drawbacks at its inception but, as time passed, new
ideas revolutionised string theory resolving many of its
inconveniences and today it is a very rich theory holding much
promise to describe all facets of fundamental interactions.

\section{Introduction}
\subsection{Dual Models -- The Early days of string theory}
\label{sec:hist}
In many ways, string theory began from a model that Gabriele
Veneziano, then at CERN, wrote down in 1968 to aid in describing the
veritable zoo of strongly interacting nuclear particles which were
being discovered in accelerators since the 1950's. These meson and
baryon resonances are characterised in part by their inherent
angular momentum  or {\em spin} $J=\a(s)$, where $s=M^2$ is the mass
squared in the centre of mass frame, and are organised according to
the empirical `Regge trajectory' of Regge-pole theory:
\be
\label{eq:regge}
\a(s) = \a(0) + \a's
\ee
$\a(0)$ is known as the `Regge intercept' and $\a'$ is the `Regge
slope'.

In quantum field theory, the leading non-trivial contributions to
the amplitude of an elastic scattering process come from the $s$--
and $t$--channel tree diagrams of fig. \ref{fig:stu}, depicting the
elastic scattering of two particles with incoming momenta
$p_1^{\text{in}}$, $p_2^{\text{in}}$ and outgoing momenta
$p_1^{\text{out}}$, $p_2^{\text{out}}$.

\setlength{\unitlength}{0.6mm} 
\begin{figure}[!hc]
\figlabel{fig:stu}
\begin{center}
\begin{fmffile}{stu}
\begin{eqnarray*}
\hspace{5mm}\parbox{40mm}{\fmfframe(0,1)(0,1){ 
\begin{fmfgraph*}(40,22) \fmfpen{thin}
    \fmfleft{i1,i2}
    \fmflabel{$p_1^{\text{in}}$}{i1}
    \fmflabel{$p_2^{\text{in}}$}{i2}
    \fmfright{o1,o2}
    \fmflabel{$p_1^{\text{out}}$}{o1}
    \fmflabel{$p_2^{\text{out}}$}{o2}
    \fmfblob{0.4w}{v}
    \fmf{fermion}{i1,v}
    \fmf{fermion}{i2,v}
    \fmf{fermion}{v,o1}
    \fmf{fermion}{v,o2}
   \end{fmfgraph*}
   }} \hspace{-9mm}=\hspace{3mm}
   \parbox{43mm}{\fmfframe(0,1)(0,1){
\begin{fmfgraph*}(53,22) \fmfpen{thin}
    \fmfleft{i1,i2}
    \fmflabel{$p_1^{\text{in}}$}{i1}
    \fmflabel{$p_2^{\text{in}}$}{i2}
    \fmfright{o1,o2}
    \fmflabel{$p_1^{\text{out}}$}{o1}
    \fmflabel{$p_2^{\text{out}}$}{o2}
    \fmf{fermion}{i1,v1}
    \fmf{fermion}{i2,v1}
    \fmf{fermion}{v2,o1}
    \fmf{fermion}{v2,o2}
    \fmf{photon,label=\underline{\small$s$-channel}}{v1,v2}
    \fmfdot{v1,v2}
   \end{fmfgraph*} }} \hspace{-10mm}+
\parbox{43mm}{\fmfframe(19,1)(0,1){
\begin{fmfgraph*}(30,27) \fmfpen{thin}
    \fmfleft{i1,i2}
    \fmflabel{$p_1^{\text{in}}$}{i1}
    \fmflabel{$p_2^{\text{in}}$}{i2}
    \fmfright{o1,o2}
    \fmflabel{$p_1^{\text{out}}$}{o1}
    \fmflabel{$p_2^{\text{out}}$}{o2}
    \fmf{fermion}{i1,v1,o1}
    \fmf{fermion}{i2,v2,o2}
    \fmf{photon,label=\underline{\small$t$-channel},label.side=left}{v1,v2}
    \fmfdot{v1,v2}
\end{fmfgraph*} }}
\hspace{-9mm}+ \cdots
\end{eqnarray*}
\end{fmffile}
\bf\caption{\rm Leading contributions to the elastic scattering of
two particles.}
\end{center}
\end{figure}
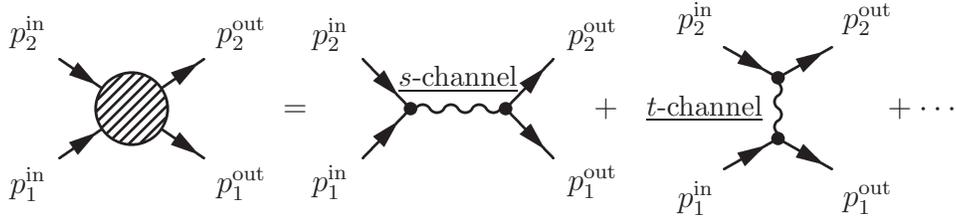

Veneziano postulated \cite{Venez} that the scattering amplitude of
the $s$- and $t$-channels could be made equal given a clever
choice for the masses and couplings of the exchange particles;
\beq \label{eq:Venez}
A_{s}(s,t)=A_{t}(t,s)=\frac{\G(-\a(s))\G(-\a(t))}
{\G(-\a(s)-\a(t))}
\eeq
where $\Gamma(x)$ is the gamma function, obeying $\Gamma(n)=(n-1)!$, $n \in
1,2,\ldots$, and $s$ and $t$ are the usual Mandelstam variables
\be
s = -(p_1^{\text{in}} + p_2^{\text{in}})^2,\quad
t = -(p_2^{\text{in}} + p_1^{\text{out}})^2,\quad
u = -(p_1^{\text{in}} + p_2^{\text{out}})^2, \nn
\ee
for particles labeled ``1'' and ``2'' as in fig. \ref{fig:stu} which
obey $s+t+u=\sum m_i^2$

Thus, the $s$- and $t$-channel scattering amplitudes would be
equivalent or {\em dual} interpretations, under the dual
transformation $s\leftrightarrow t$, of a single underlying physical
process and one need only sum over either $s$- or $t$-channel
amplitudes in the perturbation expansion\footnote{The $u$-channel
involves the crossing of the particles which amounts to a
four--quark intermediate state for meson scattering, for example.
This process is suppressed compared to the interactions of the $s$-
and $t$-channels which pass through a two--quark intermediary and
the contribution of the $u$-channel to the scattering amplitude is
therefore very small \cite{thooftpriv}.}. This is the ``duality
hypothesis''.

Two years later, Nambu \cite{Nambu}, Nielsen \cite{Nielsen} and
Susskind \cite{Susskind} independently recognised that
(\ref{eq:Venez}) corresponded to the scattering amplitude of tiny
relativistic strings which, if small enough, could be viewed as the
point particles which hadrons appeared to be. This was initially the
{\em raison d'\^etre} of string theory as a more fundamental
explanation. By using strings, the $s$- and $t$- channels can be
seen to be topologically equivalent to two open strings representing
the incoming particles merging to a single open string which then
splits again as the outgoing particles (fig. \ref{fig:stringstu}).

\begin{figure}[!hc]
\figlabel{fig:stringstu}
\begin{center}
\begin{eqnarray*}
\begin{fmffile}{stringstu1}
\hspace{20mm}\parbox{40mm}{\fmfframe(0,1)(0,1){ 
\begin{fmfgraph*}(40,22) 
    \fmfleft{i1,i2}
    \fmfright{o1,o2}
    \fmf{fermion}{i1,v1}
    \fmf{fermion}{i2,v1}
    \fmf{fermion}{v2,o1}
    \fmf{fermion}{v2,o2}
    \fmf{photon}{v1,v2}
    \fmfdot{v1,v2}
\end{fmfgraph*}   }}
\end{fmffile}
\hspace{-5mm} \leftrightarrow
\raisebox{-18mm}{\includegraphics[height=40mm]{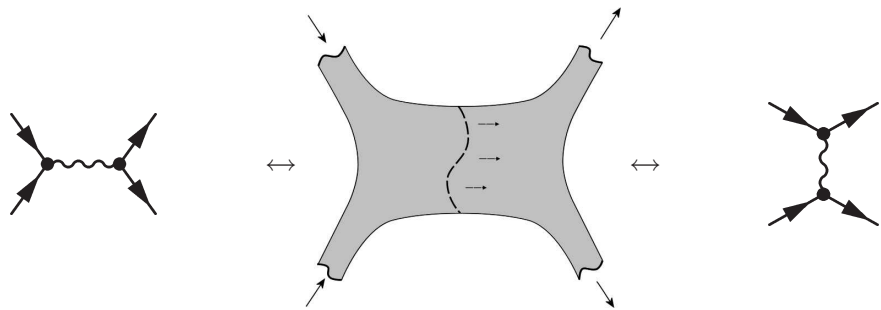}}
\leftrightarrow
\begin{fmffile}{stringstu2}
\parbox{43mm}{\fmfframe(19,1)(0,1){
\begin{fmfgraph*}(30,27) 
    \fmfleft{i1,i2}
    \fmfright{o1,o2}
    \fmf{fermion}{i1,v1,o1}
    \fmf{fermion}{i2,v2,o2}
    \fmf{photon}{v1,v2}
    \fmfdot{v1,v2}
\end{fmfgraph*} }}
\end{fmffile}
\end{eqnarray*}
\bf\caption{\rm Topological equivalence of open strings with the
$s$- and $t$- channels.}
\end{center}
\end{figure}

However appealing, there were several problems with Veneziano's
amplitude. Most notably, the Veneziano amplitude's high--energy
behaviour (with $\nf{s}{t}$ fixed) falls off exponentially, whereas
experiments conducted at the time in SLAC indicated that the strong
interaction scattering amplitudes fall off according to a power law.
This behaviour is better understood in terms of parton models such
as quantum choromodynamics (QCD) which describes hadrons in terms of
more fundamental constituents or ``partons'' --- quarks. String
theory (then comprising only ``bosonic strings'') also suffered from
some major drawbacks, such as its tachyonic ground state and its
reliance on the existence of 22 dimensions in excess of the four we
are accustomed to --- all so far completely unobserved. Despite much
initial excitement, the dual models and string theory were abandoned
as a description of the strong interaction several years later in
favour of QCD.

\subsection{Towards Gravity}
QCD greatly enriched the Standard Model which is today's benchmark
theory of particle physics. Whilst the Standard Model has proved
very successful --- indeed, it has now been tested and vindicated to
as far an accuracy as technology permits --- it views the elementary
constituents of the universe as point--like, with no internal
structure. One of the vices of quantum theory in its present form is
its complete failure to describe the gravitational force. Attempts
to blindly introduce gravity terms into a quantum field theory lead
to unrenormalizable divergences because of violent fluctuations in
the spacial fabric that appear at distances shorter than the Plank
length\footnote{These appear as ultra--violet divergences
manifesting themselves in Feynman graphs when loop momenta go to
infinity with the external momenta fixed.}. 
To examine the universe at such a scale, the assumption that
fundamental particles are point--like is apparently a flawed one.

Here, a relevant feature of dual models and hence early string
theory was their prediction of a variety of massless particles, as
poles in dual models or as excited modes in string theory, none of
which were present in the hadronic world. Initially considered a
nuisance, some of these particles were later given a physical
interpretation which gave rise to a flurry of research.

Out of dual models came the dilaton field which proved very similar
to the Brans--Dicke scalar; the $g_{44}$ component of the
compactified five dimensional $M^4 \times S^1$ space metric
$g_{MN}$, $M,N=0,\ldots,4$ of Kaluza--Klein theory
\cite{GSW,Kaluza,Klein}. Such particles arise in grand unified
theories which often draw upon Kaluza--Klein--type compactification
schemes, where they reside in the compactified dimension.

There was also a spin two massless particle and suggestions that
this particle may be a candidate for the graviton provided the
initial impetus behind research into string theory as a quantum
theory of gravity.

\subsection{Why 2D string theory?}
\label{sec:2Dgrav}
In an attempt to overcome the difficulties associated with
quantising 4D gravity, particles are replaced by one dimensional
extended objects --- strings, which trace out a two dimensional
manifold $\Sigma$, the ``world--sheet'', as they evolve in time, in
much the same way as a particle traces out a world-line in
space--time.

The world--sheet can be viewed as embedded in a $D$--dimensional
ambient ``target space''. The coordinates of the string in the
target space are parameterised by $X^\mu (\s,\t)$, $(\s,\t) \in
\Sigma$ being some coordinates on the world--sheet, $\mu =
0,1,\ldots,D-1$.

The string world--sheet is always two dimensional, but when the
target space is two dimensional, the resulting ``2D string theory''
affords us some insight unobtainable in string theories formulated
in higher dimensions \cite{FGZ94,Klebanov91}. Because it is
formulated in one space and one time dimension, the string can have
no transversal excitations and is classically completely described
by its centre--of--mass coordinates (although remnants of the
transversal excitations appear as ``discrete states''
\cite{Polyakov91}). Through the mathematical technique of matrix
models, many difficult problems in string theory turn out to be
exactly solvable in two dimensional string theory, such as the
calculation of the $S$--matrix (see \S\ref{sec:smatrix}).

If the world--sheet is itself treated as a two dimensional
space--time then such theories, as Polyakov showed, could be
interpreted as theories of {\em two dimensional} quantum gravity,
$X^\mu (\s,\t)$ being considered as matter fields on $\Sigma$ (see
\S\ref{sec:sta2dg}). Viewed as this dual picture, we are effectively
solving problems of $(1 + 1)$ quantum gravity. It serves also as an
alternative description of topological strings and supersymmetric
gauge theories \cite{Vafa,Dijk}.

Another interesting application of two dimensional strings was
discovered by \mbox{'t Hooft} \cite{tHooft74b}. In studying the
strong interaction, he considered mesons in one space and one
time dimension where the number of colours diverges, $N_{\rm
c}\rightarrow\infty$, as a simple toy model. The mass of the two
particle states was found to obey a straight Regge trajectory,
\be
M_k^2\sim k,\qquad k=1,2,\ldots \ \nn.
\ee
As in QCD, the potential between the two quarks in the bound meson
state was found to vary linearly with the separation distance and
these results were confirmed by numerical calculations. Furthermore,
in the four dimensional world, the infra--red instability of a the
bound two quark system squeezes the gauge fields together into a
structure resembling a string \cite{thooft74}. This all pointed to a
possible description of the strong force in terms of strings.

\newpage
\section{Bosonic Strings}
Bosonic strings were part of the `first era' of string theory,
before the advent of supersymmetry which resolved several pernicious
issues arising in the quantisation of classical strings. In the next
two sections give brief review of the essentials of string theory
which will be used in this thesis.

\subsection{Classical Strings}
\label{sec:action}
Let the coordinates in the $D$--dimensional target space--time of a
point $(\s,\t)$ on the string's world--sheet $\Sigma$ be $X^\mu
(\s,\t)$, $\mu = 0,1,\ldots,D-1$.

For a relativistic particle, the {\em geodesic action} is
proportional to the length of its world--line. Similarly, for a
string, the action is given by the area of its world--sheet.
We thus get the bilinear action, called the Nambu--Goto action
\be
S_{\text{NG}} =
-\frac{1}{2\pi \a'} \displaystyle{\int_\Sigma}d\s d\t\, \sqrt{-h},
\qquad
h=\det h_{ab},
\label{eq:SNG}
\ee
where $\a'$ is related to the Regge slope of dual models; it is the
{\em string scale}, of dimension of length squared. The matrix
$h_{ab}$ is the induced metric on the world sheet
\be
\label{eq:indmet}
h_{ab}=\et_{\mu\nu}\p_a X^{\mu}\p_b
X^{\nu},
\ee
and $\et_{\mu\nu}$ is the Minkowskian metric of the flat target
space.

Rewriting (\ref{eq:SNG}) to eliminate the square--root which makes
it non-linear, we hereby obtain the Polyakov action

\be S_{\rm P}=-\frac{1}{4\pi \a'} \int_\Sig d\tau d\sigma \, \sqrt{-h}\,
h^{ab}\et_{\mu\nu}\p_a X^{\mu}\p_b X^{\nu}. \label{eq:SPOL} \ee

The induced metric $h^{ab}$ is now to be thought of as a dynamical
variable, it is related to the (\ref{eq:indmet}) by
\be
h_{ab} = C(\s,\t)\p_a X^\mu \p_b X_\mu, \nn
\ee
where $C(\s,\t)$ is a conformal factor which cancels out in \refeq{eq:SPOL}.

In fact, \refeq{eq:SPOL} is invariant under the class of Weyl transformations
\be
h_{ab} \mapsto e^{\phi} h_{ab}. \nn
\ee
A {\em conformal gauge} is a choice\footnote{In noncritical string
theory, Weyl invariance may be broken by the addition of a
cosmological constant term in the action and the `Liouville field'
$\phi$ becomes a dynamical variable $\phi(\s,\t)$. This will be
explained later in \S\ref{sec:tdst}}
\be
\label{eq:confG}
h_{ab} \mapsto e^{\phi(\s,\t)} \hh_{ab},
\ee
where $\hh_{ab}$ is the background metric which gauge fixes
world--sheet diffeomorphisms. We may take advantage of Weyl
invariance to set $\phi = 0$. Then, locally, the conformal gauge can
be chosen as
\be
\label{eq:confGn}
h_{ab} = \et_{ab},
\ee
$\et_{ab}$ being the Minkowski or Euclidean metric, as the case may
be\footnote{Note that in light--cone coordinates
$\s^{\pm}=\frac{1}{\sqrt{2}}(\t\pm\s)$, the Minkowski metric
$\(\begin{array}{cc}-1&0\\0&1\end{array}\)$ becomes
$\(\begin{array}{cc}0&-1\\-1&0\end{array}\)$}. Thus, because of Weyl
symmetry and diffeomorphism invariance, the Nambu--Goto action
(\ref{eq:SNG}) is classically equivalent to the Polyakov action
(\ref{eq:SPOL}).

\subsubsection{Boundary Conditions}
So far, we have local, classical equations of motion for a string.
We should also impose some boundary conditions on the fields
$X(\s,\t)$. Without loss of generality, we may consider the
perimeter of the string to be of length $\pi$ as in
\cite{GSW,thooft}. If we consider closed strings, we impose periodic
boundary conditions; $X(\s,\t) = X(\s+\pi,\t)$. Open strings need
not be periodic but we have to impose momentum conservation which
implies Neumann boundary conditions; $\left.\p_\s X^\mu
\right|_{\s=0,\pi} = 0$ so no momentum can flow in or out of the
string at the edges.

\subsection{String Theory as 2D Gravity}
\label{sec:sta2dg}
Since we think of the world--sheet metric $h_{ab}$ as field itself,
we can think in terms of the dual picture alluded to in
\S\ref{sec:2Dgrav} where the world--sheet is treated as a
two--dimensional space--time and the $X^\mu$ are matter scalar
fields. In that case, we are forgetting about strings altogether and
dealing instead with two--dimensional gravity coupled to matter
fields. We may also add the Einstein--Hilbert action
term\footnote{\label{fn:norm}The conventional normalisation of the
Einstein--Hilbert action is $\frac{1}{2\pi}$, however, our
normalisation of $\frac{1}{4\pi}$ normalisation will avoid factors
of $\sqrt{2}$ when we later formulate the action for this theory. It
can be achieved by rescaling the metric appropriately.}
\be
\label{eq:e-h}
\chi=\frac{1}{4\pi}\displaystyle{\int_\Sig} d\tau d\sigma\,
\sqrt{-h}\, \CR,
\ee
describing the contribution to the action of a curved space--time,
where $\CR(h_{ab})$ is the world--sheet Ricci scalar (the total
curvature). Through the Gauss--Bonnet theorem, the Einstein term in
two dimensions is equal to the Euler--Poincar\'e characteristic, a
topological invariant which depends, in the Euclidean signature,
only on the genus $g$ (the number of handles) and the number of
boundaries $b$ of the surface\footnote{Strictly speaking, one would
need to add a boundary term to \re{e-h} if $b\neq0$ which integrates
over the extrinsic curvature of the boundary.}
\be
\label{eq:Eulerchar}
\chi = 2 - 2g - b.
\ee

Allowing for a more general target space, the action becomes
\be
\label{eq:SPOLtot}
\Rightarrow S_{\rm Ptot}=-\frac{1}{4\pi\a'}
\int_\Sig d\tau d\sigma \, \sqrt{-h}\, \left( h^{ab}G_{\mu\nu}\p_a
X^{\mu}\p_b X^{\nu} + \a' \nu \CR \right).
\ee
with $G_{\mu\nu}(X^\r)$ the target space metric and $\nu$ a coupling
constant (later to be identified with the scalar dilaton).

To avoid two--dimensional metrics of topologically non--trivial
surfaces which are always singular in the Minkowskian signature, one
transforms to the Euclidean action $S_{\text{P}} \mapsto
S_{\text{P}}^{\text{(E)}}$ by performing an analytic continuation to
Euclidean time on the world--sheet $\t \mapsto -i\t$. Having done
this, we may write an expression for the partition function in terms
of path integrals over non--singular metrics and matter fields,
\be
\label{eq:formalZ}
Z = \sum\limits_{\text{Topologies}} \int\CD\vr(h_{ab})\,\int\CD
X^\mu\, e^{-S_{\text{Ptot}}^{\text{(E)}}[X^\mu,h_{ab}]}
\ee
In two dimensions, the sum over all topologies can be regarded as a
sum surfaces of genus $g$ if the surfaces are orientable and closed.
The path integral over metrics is formal at this stage and shall
later be defined in terms of a sum over discretisations of space
which in turn is equivalent to a sum over planar diagrams, as
explained in Chapter \ref{ch:MAT}.

\subsection{Quantisation}
There are several schemes for quantising the classical bosonic
string theory developed so far. All of them lead to the same
conclusions regarding the dimension of space--time and the presence
of a tachyonic ground state. This will be demonstrated through one
of the approaches to quantisation: light--cone quantisation.

By varying the Polyakov action $S_{\text{P}}$ and applying boundary
conditions, we see that the $X^\mu$ are solutions of a free massless
wave equation on shell in a flat target space
\be\label{eq:fmwe}
(\p^2_\s + \p^2_\t)X^\mu = 0.
\ee
In light--cone quantisation, we exploit diffeomorphism invariance to
reparameterize the world--sheet coordinates $\s$ and $\t$. Here
boundary conditions mandate that the reparameterization map is
identical for both $\s$ and $\t$ and it follows that this map must
also be a solution of the same free massless wave equation.
Therefore, we can choose the reparameterization to be precisely one
of the $X^\mu$ coordinates for each of $\s$ and $\t$. In practice,
we choose the $X^\mu$ in question to be light--like, so that their
square in Minkowski space vanishes
\beq
\t = f(X^+),\quad\s=f(X^-) \nn\\
X^\pm = (X^0 \pm X^{D-1})/\sqrt{2}, \eeq where $f(X)$ is a linear
function. Thus, the metric of a flat target space with a Minkowskian
signature $G_{\mu\nu} = \et_{\mu\nu}$ looks like

\be \label{eq:mtLC}\et= \left( \begin{array}{ccccc}
0   &   -1  &   0   &   \cdots  &   0\\
-1  &   0   &   0   &           &    \\
0   &   0   &   1   &           &\vdots\\
\vdots &    &       &   \ddots  &    \\
0   &       & \cdots &     &   1   \\
\end{array}\right).
\ee

This gauge choice which fixes the reparameterization degree of
freedom is called the {\em light--cone gauge}. Quantisation now
proceeds through the promotion of Poisson brackets to commutator
relations via the familiar correspondence principle
\be\label{eq:copri}
\lbrace\ ,\ \rbrace_{\rm PB} \rightarrow \frac{1}{i\hbar}[\ ,\ ].
\ee

The classical constraints \re{fmwe} correspond to the ``Visaroso
conditions'' $T_{ab} = 0$ where the world--sheet energy--momentum
tensor of the theory
\be
\label{eq:EMtensor}
T_{ab} \equiv -\frac{2\pi}{\sqrt{h}}\frac{\d S_\text{P}}{\d h^{ab}}.
\ee

The advantage of the light--cone gauge choice is that the classical
constraints (the Virasoro conditions) can be expressed as
relationships between the transverse components
$(X^1,\ldots,X^{D-2})$ and the anomalous $X^+, X^-$ associated to
the off--diagonal components of the metric tensor \re{mtLC}.

The coefficients of the mode expansions for $X^\pm$ function become
raising and lowering operators of string excitations in the Fock
space. If the resulting quantum theory were physical then the string
states should form representations of the Poincar\'e group. When we
compare commutation relations of target--space energy--momentum
tensor of string theory with the Lie algebra of the Lorentz group,
we find an inconsistency which vanishes only if we impose the number
of dimensions $D$ to be 26. This is called the {\em critical
dimension} of string theory; $D_{\text{cr}}=26$.

The 22 dimensions superfluous to the four that we are familiar with
are thought of as compactified {\em \`{a} la Kaluza--Klein}
\cite{Kaluza,Klein}. However, there is a priori no preference for
how this is to be done. Further, the mass of the ground state for an
open bosonic string is $M^2=-1$ so this mode is a tachyon. The
higher excitation modes yield a plethora of particles including the
aforementioned spin--two graviton candidate.

But at this stage, quantised bosonic string theory has too many
vices. A resolution to some of those problems came with the
discovery of a special symmetry between bosons and fermions which we
now turn our attention to.

\newpage
\section{Superstrings}
Superstring theory is a generalisation of bosonic string theory
which includes fermionic degrees of freedom. In the $\CN = 1$
Ramond--Neveu--Schwarz (RNS) model described here, fermionic fields
$\psi^\mu (\s,\t)$ which are supersymmetrically related to the
space--time coordinates $X^\mu(\s,\t)$ are introduced {\em on the
world--sheet}. There is another way which introduces supersymmetry
directly in the target space, namely, the {\em Green--Schwarz}
formulation where, in addition to the fields $X^{\mu}$, one adds one
or two sets of world--sheet scalars $\th^{\rm A}$, corresponding to
$\CN=1$ or $\CN=2$ supersymmetry, which form a spinor in the target
space. As with the RNS formulation, the string spectrum is better
behaved and begins with massless modes instead a tachyon.
Unfortunately, the Green--Schwarz formulation is inextricably
complicated for real calculations and the RNS formulation is more
common in practice \cite{GSW}.

\subsection{The Supersymmetric String Action}
These fermionic fields we introduce are expressed in terms of a
two--dimensional Majorana spinor; a real, two component spinor
\cite{Polbook},
\beq
\psi^\mu &=&
\left(\begin{array}{c}
\psi^\mu_1 (\s,\t) \\
\psi^\mu_2 (\s,\t)
\end{array}\right).
\eeq
The $\psi^\mu$ are anticommuting with respect to all indices
\be
\lbrace \psi^\mu_i , \psi^\nu_j \rbrace =
\psi^\mu_i\psi^\nu_j+\psi^\nu_j\psi^\mu_i= 0,\qquad i,j\in\lb 1,2\rb
\ee
(Hereafter, the spin index will usually be suppressed).

In a flat target space, the supersymmetric string action can be
written in a reparameterization--invariant way as
\be
\label{SSUSY}
S_{\text{SUSY}} = -\frac{1}{2\pi} \int d^2 \s\, \sqrt{h}\,
h_{ab}\left(\p^a X^\mu \p^b X_\mu - i \bpsi^\mu
e^{A\a}\vr^A\p_\a\psi^\mu\right),
\ee
where the {\em vielbein fields}\footnote{``Zweibein'' would perhaps
be more appropriate in two dimensions.} $e^{A\a}, A=1,2$ span the
locally flat tangent space of the world--sheet manifold
\be
h_{ab} = \eta_{AB}e^A_a e^B_b,\qquad h^{ab} = \eta_{AB}e^{Aa}
e^{Bb},
\ee
and in two dimensions we have two gamma matrices obeying the
Clifford algebra
\be
\vr^0 =
\left(
\begin{array}{cc}
0 & -i \\
i & 0
\end{array}
\right),\quad
\vr^1 =
\left(
\begin{array}{cc}
0 & i \\
i & 0
\end{array}
\right),\quad\lbrace \vr^a,\vr^b \rbrace = -2\eta^{ab}.
\ee
So, in terms of the internal Lorentz indices \textsc{a,b}, the gamma
matrices in curved space appear as written within (\ref{SSUSY});
\be
\vr^a(x) = e^{Aa}(x)\vr^A,
\ee
and the symbol $\bpsi$ denotes $\psi^\dag \vr^0$ as usual.

Subject to certain boundary conditions, the action (\ref{SSUSY}) is
invariant under the infinitesimal global supersymmetry
transformations of the fields
\beq
\d X^\mu &=& \overline{\eps}\psi^\mu \\
\d \psi^\mu &=& -i\vr^a\p_a X^\mu \eps,
\eeq
where $\eps=\(\!\begin{array}{c}\eps_1\\\eps_2
\end{array}\!\)$ is a constant Majorana spinor.

\subsection{Boundary Conditions}
There are several sectors of string theory because there are
different ways to impose the condition that the action vanish under
infinitesimal variations of the fermionic fields $\psi^\mu$.
Consider the variation of the fermionic part of the action
(\ref{SSUSY}) in the conformal gauge ($\det h_{ab} = 1$)
\be
\frac{\d S}{\d \psi^\mu} = \frac{i}{2\pi}\int d^2\s\,
\left(2\d\bpsi^\mu (\vr^a\p_a)\psi^\mu + \p_a(\bpsi^\mu \vr^a
\d\psi^\mu) \right) = 0 \qquad \forall\quad \d \psi^\mu,\,\,
\d\bpsi^\mu.
\ee
This yields the mass--shell condition $i\vr^a\p_a\psi^\mu = 0$ (the
Dirac equation) as expected but for the open string there are also
boundary contributions from the ends at $\s = 0,\pi$. Hence we
should require that these surface terms vanish\footnote{ Notice that
we included only the $a=\s$ contribution from the $\p_a(\bpsi^\mu
\vr^a \d\psi^\mu)$ in \re{bob}. This is because there are no
boundary contributions from $\t\rightarrow\pm\infty$.}
\be\label{eq:bob}
\left. \bpsi^\mu\vr^1\d\psi^\mu \right|^{\s = \pi}_{\s = 0} = 0.
\ee
An acceptable condition at one end may be chosen arbitrarily as
$\psi_1^\mu (\t,0) = \psi_2^\mu (\t,0)$, leaving two possible
choices for the other end leading to two distinct open string
sectors:

\begin{description}
\item[Ramond sector:]
$\psi_1^\mu (\t,\pi) = \psi_2^\mu (\t,\pi)$ --- periodic,
\item[Neveu--Schwarz sector:]
$\psi_1^\mu (\t,\pi) = -\psi_2^\mu (\t,\pi)$ --- anti--periodic.
\end{description}

For closed strings, the surface terms vanish when the boundary
conditions are periodic or antiperiodic for each component of
$\psi^\mu$ separately. So we have the following possible boundary
conditions to choose

\begin{description}
\item[Ramond sector:]
$\psi_1^\mu (\t,\s) = \psi_1^\mu (\t,\s + \pi)$ --- periodic, {\bf
or}
\item[Neveu--Schwarz sector:]
$\psi_1^\mu (\t,\s) = -\psi_1^\mu (\t,\s + \pi)$ --- anti--periodic
\end{description}
and similarly for $\psi^\mu_2$. We need both $\psi^\mu_1$ and
$\psi^\mu_2$ so this leads to four distinct sectors formed by
combinations of the Ramond and Neveu--Schwarz boundary conditions;
the R--R, R--NS, NS--R and NS--NS sectors of closed string theory.

With the boundary conditions at hand, we may now write down
classical solutions in terms of (anti)periodic series expansions and
quantise the theory through the correspondence principle
(\ref{eq:copri}). Having done this, we may again construct the
generators of Lorentz transformations and compare them with the Lie
algebra of the Poincar\'e group. It then appears that, in order for
the anomaly to disappear, we require $D_{\text{cr}} = 10$.

\subsection{The GSO projection}
Computing the spectra of the various sectors, we find the following:

{\bf Ramond:}
    \begin{itemize}
        \item There are zero mass modes which are $(D-2)$--fold degenerate,
        \item The $\text{energy}/\text{mass}^2$ is raised by integer units,
        \item The ground state is a degenerate fermion,
        \item The spin modes are $\nf{1}{2}$--integer making this sector {\bf
        fermionic}.
    \end{itemize}
{\bf Neveu--Schwarz:}
    \begin{itemize}
        \item There no zero mass modes,
        \item The $\text{energy}/\text{mass}^2$ is raised by $\nf{1}{2}$--integer units,
        \item The ground state is a unique, non--degenerate spin--0 boson,
        \item The spin modes are integer making this sector {\bf
        bosonic}
    \end{itemize}

The current state of affairs is rather complicated, there are open
strings and closed strings, each with their various sectors.
Furthermore, computing the mass squared of the open string
NS--sector ground state reveals that it is a tachyon (the same is
true of the ground state of the closed string NS--NS sector). So the
RNS model is still an inconsistent theory unless further constraints
are imposed.

In 1977 Gliozzi, Scherk and Olive (GSO) \cite{GSO1,GSO2} found a way
to truncate the string spectrum such that supersymmetry was apparent
in the target space by constructing a projection operator which acts
on the states and projects out two chiralities of which only one is
kept. This is the {\em GSO condition}. By so doing, we get an equal
number of bosonic and fermionic states at each mass level --- a
necessary condition for supersymmetry, and the tachyon is also
projected out.

\subsection{Five String Theories and M Theory}
\label{sec:Mtheory}

Classically, there are many different string theories. One has great
latitude in choosing the ingredients used in formulating a string
theory: open and closed strings, oriented and non--oriented strings
(ie: symmetry under parity reversal or lack thereof), $\CN = 1$ and
$\CN = 2$ supersymmetric extensions (supersymmetry with one or two
independent fermionic degrees of freedom, respectively) and leeway
in choosing the chirality of left and right modes in closed
strings\footnote{Fields on the world--sheet are typically decomposed
into left-- and right--moving modes, $X^\mu(\s,\t) = X_{\rm
L}^\mu(\t + \s) + X_{\rm R}^\mu(\t - \s)$. String theories where
these modes are indiscernible by parity inversion are referred to as
`non--chiral'.}. One may also introduce $\gr{U}{N}\times \gr{U}{N}$
Yang--Mills gauge symmetry by adding charges to ends of strings in
the form of Chan--Paton factors.

In the quantum theory, however, anomaly cancelation in higher--order
loop expansions makes $\gr{SO}{32}$ the only admissible gauge group
(apart from `heterotic strings' where the right modes are fermionic
in 10 physical dimensions and left modes are bosonic on a self--dual
compactified 16--torus which admits the $\mathsf{E}_8 \times
\mathsf{E}_8$ gauge group as well). This leads to five consistent
string theories each with their respective Ramond and Neveu--Schwarz
sectors and associated spectrum of particles:

\begin{description}
\item[Type I:] $\CN = 1$ non--oriented and non--chiral open and closed strings,
\item[Type IIA:] $\CN = 2$ oriented non--chiral closed strings,
\item[Type IIB:] $\CN = 2$ oriented chiral closed strings,
\item[$\mathbf{\gr{SO}{32}}$ Heterotic:] $\CN = 1$ heterotic closed strings with the $\gr{SO}{32}$ gauge group,
\item[$\mathbf{\mathsf{E}_8\times \mathsf{E}_8}$ Heterotic:] $\CN = 1$ heterotic
closed strings with the $\mathsf{E}_8\times \mathsf{E}_8$ gauge
group.
\end{description}

These five string theories, whilst each consistent, dramatically
differ in their basic properties, such as world-sheet topology,
gauge groups and supersymmetries. Also, the string theories are
defined as perturbative expansions in the string coupling constant,
amounting to sum over Riemann surfaces of different genera for
closed strings, and of different boundaries for open strings. This
perturbation expansion is only asymptotic, so that the theory is at
best incomplete.

One promising way to resolve both problems came from the discovery
of a set of dualities connecting all the string theories and it was
postulated that the five string theories were thus dual to a
hitherto unknown 11--dimensional `M Theory', where, to quote Witten,
``The M is for Magic, Mystery, or Membrane according to taste''
(others would add `M(atrix)' to the list given its possible
realisation in terms of matrix models \cite{Banks} --- see
\S\ref{sec:matM}).

There are many conceptions of M theory at present. M theory is
regarded by some as a merely another point in the space of vacuum
configurations which contains singular points about which exist
perturbative expansions of the known string theories. Others regard
it as the overarching ``theory of everything'' (TOE), out of which
the various string theories and supergravity emerge as limiting
cases. Whilst M theory occasionally reveals tantalising hints of its
promise to becoming the {\em Tao} of physics, to date it remains a
largely intangible theory whose equations have yet to be written.

\newpage
\section{Strings in Background Fields}
So far, only a flat target space has been considered in the
formulation of string theory. By adding certain ``background terms''
in the action, we are able avert strings which would otherwise
venture into a region of the target space where the string coupling
diverges by means of a potential wall which precludes the
penetration into that region, making the theory well behaved.

The background fields which will be introduced into the string
theory Lagrangian actually emerge out of the dynamics of string
theory itself. In turn, the presence of background fields changes
the target space and determines the dynamics of the string. It
suffices to consider bosonic strings only, as the extension to
superstrings following the RNS formalism is relatively pedestrian
and will not affect our conclusions.

\subsection{The Nonlinear $\s$--Model}
\label{subsec:nonlinsigma}
A curved space--time should arise as a consequence of gravitons
present in the string spectrum. In the low energy limit, only
massless modes are seen. Closed bosonic strings have graviton zero
mass modes in the form of traceless symmetric tensor fields
$G_{\mu\nu}(X^\r)$. In addition, is an antisymmetric tensor field
$B_{\mu\nu}(X^\r)$ (the {\em axion}) and a scalar field $\dl(X^\r)$
(the {\em dilaton})\footnote{In Type IIA superstring theory, for
example, these fields arise from the NS--NS sector. The other
sectors contain associated chiral superpartners, the ``gravitino'',
and the ``dilatino'', and various other chiral tensor fields.}. As
it is the dynamics of the string which should determine the possible
background fields, to write down an action in a general target
space--time with metric tensor $G_{\mu\nu}(X^\r)$, one must include
the other zero modes present, namely, $B_{\mu\nu}(X^\r)$ and
$\dl(X^\r)$, as part of the background, leading to the interacting
action of the {\em non--linear sigma model}\footnote{It is no
coincidence that action is renormalizable; one must ensure that the
fields are included in a renormalizable and
reparameterization--invariant way. In particular, including
$G_{\mu\nu}(X^\r)$ involves generalising the Einstein--Hilbert term
(\ref{eq:e-h}) promoting the coupling $\nu$ to a field $\dl(X^\r)$
\cite{GSW}. Now, being in affect a running coupling constant,
$\dl(X^\r)$, as the other zero mode fields here, will have
$\b$--functions associated to it.}

\be
 S_\sigma =
\frac{1}{4\pi\a'}\int d^2 \sigma\,\sqrt{h}
\left[\left(h^{ab}G_{\mu\nu}(X)+i\eps^{ab}B_{\mu\nu}(X)\right) \p_a
X^{\mu} \p_b X^{\nu} +\a'\CR \dl(X)\right],
\label{eq:SigmaModel}
\ee
here, a Euclidean world--sheet metric is implied and $\eps^{ab}$ is
the Levi--Civita
tensor\footnote{$\eps^{ab}=\left(\begin{array}{cc}0&1\\-1&0\end{array}\right)$.}.

One may view this action as describing a quantum field theory on the
world--sheet which leads to a perturbative expansion in $\a'$, or as
a topological expansion in the target space most accurate for
long--range interactions or in the long--wavelength domain. As the
field theory is an expansion in the string scale $\a'$, the internal
structure of the string has no bearing on large--distance physics
where we obtain an effective theory which is simply an effective
field theory of massless string modes \cite{CFMP}. The action for
this effective theory is best understood after considering the
constraints of the theory.

\subsection{Constraints}
At this stage, we have introduced three new fields, $G_{\mu\nu}$,
$B_{\mu\nu}$ and $\dl$, all so far unconstrained. In order for the
action (\ref{eq:SigmaModel}) to have a sensible physical
interpretation, one needs to ensure that it is Weyl invariant. To
each field (or coupling in the world--sheet picture), we can
associate a renormalization group $\b$--function. It can be shown
through dimensional regularisation of \refeq{eq:SigmaModel} that
demanding Weyl invariance necessarily implies the vanishing of the
$\b$--functions \cite{Fried}, and this forms the constraint of the
theory\footnote{The vanishing of the $\b$--functions makes the
world--sheet energy--momentum tensor vanish. Its trace is
$$
4\pi T^a_a= \b^{\dl}\sqrt{h} \CR +
\b_{\mu\nu}^G \sqrt{h} h^{ab}\p_a X^{\mu} \p_b X^{\nu} +
\b_{\mu\nu}^{B} \e^{ab}\p_a X^{\mu} \p_b X^{\nu}
$$
}. For this effective field theory, we can consider the long range
interactions and need only consider expressions for the
$\b$--functions to zeroth order in $\a'$. They are \cite{FGZ94}:
\beq
\b_{\mu\nu}^G &=& \CR_{\mu\nu}+2 D_{\mu} D_{\nu} \dl-
\frac{1}{4}H_{\mu\s\r}{H_{\nu}}^{\s\r}+\CO(\a')=0,
\nn \\
\b_{\mu\nu}^{B}&=&
D_{\r}{H_{\mu\nu}}^{\r}-2{H_{\mu\nu}}^{\r}D_{\r}\dl
 + \CO(\a')=0,
\label{eq:betafunctions} \\
\b^{\dl}&=& \frac{1}{\a'}\frac{D-26}{48\pi^2} +
\frac{1}{16\pi^2}\left[ 4(D_\mu\dl)^2 - 4 D_\mu D^\mu \dl -\CR
+\frac{1}{12}H_{\mu\nu\r}H^{\mu\nu\r}\right]+\CO(\a') = 0, \nn
\eeq
where $D_\mu$ is the covariant derivative in $D$
dimensions\footnote{For example, for contravariant vector $A^\r$ in
$D=4$, the covariant derivative of $A^\r$ is \mbox{$D_\mu A^\r
\equiv \frac{\p A^\r}{\p X^\mu} + \G^\r_{\phantom{\r}\nu\mu}A^\nu$},
with the affine connection $\G^\r_{\phantom{\r}\nu\mu}$ being the
usual Christoffel symbols.}, $\CR_{\mu\nu}$ is the Ricci tensor,
$\CR$ is the scalar curvature and
\be
H_{\mu\nu\r}=\p_{\mu} B_{\nu\r}+\p_{\r} B_{\mu\nu} +\p_{\nu}
B_{\r\mu}
\ee
is the field strength for the antisymmetric tensor field $B_{\mu\nu}$.

These $\b$--functions are precisely the equations of motion for
Einstein gravity coupled to an antisymmetric tensor gauge field
$H_{\mu\nu\r}$ and to the dilaton. They are to zeroth order the
Euler--Lagrange equations of the $D$--dimensional effective {\em
target space} action describing the interactions of massless modes
of the closed bosonic string in the long--wavelength limit,
\be
S_{\rm eff}=\hf \int d^D X \, \sqrt{-G}\, e^{-2\dl } \left[
-\frac{2(D-26)}{3\a'}+ \CR + D_\mu \dl D^\mu \dl -\frac{1}{12}
H_{\mu\nu\lambda}H^{\mu\nu\lambda}\right].
\label{eq:Seff}
\ee
Any solution of (\ref{eq:betafunctions}) describes a consistent,
Weyl invariant string theory.

We can reproduce the Einstein--Hilbert action term (\ref{eq:e-h}) by
absorbing the factor $e^{-2\dl }$ into the definition of the
target--space metric which shows that the $\s$--model in the
low--energy domain is a theory of gravity.

\subsection{The Tachyon}
\label{sec:other}
It is also possible to introduce a nonderivative interaction terms
in the sigma model \re{SigmaModel}. Any such term is equivalent to
including the tachyonic ground state field of the bosonic string to
the sigma model action and leads to a nonzero expectation value for
the tachyon. (As the tachyon vertex operator $e^{ik_\mu X^\mu}$
which will be derived below is nonderivative, nonderivative terms
present in the $\s$--model Lagrangian would lead to a non--zero
expectation value for the Tachyon). This is not usually considered
part of the general $\s$--model since tachyons are in most cases
nonphysical particles that travel backwards and time and have a
complex mass. However, it will be shown below that in two dimensions
the tachyon is massless and is therefore a physical particle.
Further, it will serve the practical purpose of preventing strings
from venturing into the strong--coupling region.

Just such a nonderivative contribution arises out of term that could
have been added to \refeq{eq:SPOLtot} --- the cosmological constant
term. Working with a Euclidean world--sheet metric, this looks like
\be
\label{eq:Ssig-cosmo}
S_\s^{\rm cosmo} = \frac{1}{4\pi\a'}\int d^2\s \, \sqrt{h} T(X^\mu).
\ee

If (\ref{eq:Ssig-cosmo}) is included in the $\s$--model, the
$\b$--functions will have to be modified. For the case when the
$B_{\mu \nu}$ tensor field is zero\footnote{We will not need this
field when we consider 2D string theory later.}, they become
\cite{GinMoore}
\beq
\b_{\mu\nu}^G &=& \CR_{\mu\nu}+2 D_{\mu} D_{\nu} \dl
- D_\mu T D_\nu T
+\CO(\a') = 0,
\nn \\
\b^{T}&=& -2 D_\mu D^\mu T + 4 D_\mu \dl D^\mu T
- 4T
+ \CO(\a') = 0,
\label{eq:betafunctionsT} \\
\b^{\dl}&=& \frac{D-26}{3\a'} - \CR + 4(D_\mu\dl)^2
- 4 D_\mu D^\mu \dl
+ (D_\mu T)^2 - 2T^2
+\CO(\a') = 0,
\nn
\eeq
and the effective action (\ref{eq:Seff}) out of which they ought to
arise as Euler--Lagrange equations needs to be extended by adding
\be
\label{eq:Seff-tach}
S_{\rm eff}^{\rm tach} = -\frac{1}{2}\int d^D X\,\sqrt{-G} e^{-2\dl}
\left( (D_\mu T)^2 - \frac{4}{\a'} T^2\right).
\ee

\subsection{Background Solutions}
By setting the background metric to $G_{\mu\nu}=\eta_{\mu\nu}$ in
the critical dimension $D=D_{\text{cr}}$\footnote{Critical string
theory is Weyl invariant string theory. For bosonic strings, we
usually have $D_{\text{cr}}=26$.}, we are working in flat space
where the Ricci tensor $\CR_{\mu\nu}$ and curvature $\CR$ vanish.
The trivial solution of \refeq{eq:betafunctions} is then
\be
\label{eq:NoDilBG}
 G_{\mu\nu}   = \eta_{\mu\nu},\qquad
 B_{\mu\nu}   = 0, \qquad
 \dl   = 0,
\ee
describing the absence on any background. Our sigma model
(\ref{eq:SigmaModel}) then reduces to the Polyakov action
(\ref{eq:SPOL}).

By inspection, we immediately see a nontrivial solution of
\refeq{eq:betafunctions}, namely the constant dilaton background
\be
\label{eq:ConstDilBG}
 G_{\mu\nu}   = \eta_{\mu\nu},\qquad
 B_{\mu\nu}   = 0, \qquad
 \dl   = \nu.
\ee
This reproduces the Polyakov action with the Einstein--Hilbert action
term, \refeq{eq:SPOLtot}.

A more interesting solution is the linear dilaton background (which
we will relate to Liouville gravity and non--critical strings in
\S\ref{sec:LiouvilleTh})
\be
\label{eq:LinDilBG}
 G_{\mu\nu}   = \eta_{\mu\nu},\qquad
 B_{\mu\nu}   = 0, \qquad
 \dl   = l_\mu X^\mu.
\ee
Substituting this solution into the tachyonic effective action, \refeq{eq:Seff-tach},
we obtain the following equation of motion on shell
\be
\p^2 T - 2l^\mu \p_\mu T + \frac{4}{\a'}T = 0,
\ee
which has the general solution
\be
\label{eq:tachBG}
T=\mu\exp(k_{\mu}X^{\mu}), \qquad (k_\mu -
l_\mu)^2=\frac{2-D}{6\a'}.
\ee
This is the asymptotic form of the tachyon vertex operator. Together
with \refeq{eq:LinDilBG}, it defines a ``generalised linear dilaton
background''. In this background, the $\s$--model action
\ref{eq:SigmaModel} takes the form (again, with a Euclidean
world--sheet metric)
\be
\label{eq:Slindil}
S^{\rm LD}_{\s}= \frac{1}{4\pi\a'}\int d^2 \s\,\sqrt{h}
\left[h^{ab}\p_a X^{\mu} \p_b X_{\mu} +\a'\CR l_{\mu}X^{\mu}+\mu
e^{k_{\mu}X^{\mu}} \right], \quad (k_\mu -
l_\mu)^2=\frac{2-D}{6\a'}.
\ee
It is important to note that (\ref{eq:tachBG}) and
(\ref{eq:LinDilBG}) {\em do not} constitute a solution to the
equations of motion derived from $S_{\rm eff} + S_{\rm eff}^{\rm
tach}$, in other words, it does not make the $\b$--functions
(\ref{eq:betafunctionsT}) vanish. This can be blamed on the fact
that the effective action reproduces only the zeroth order terms in
$\a'$ and that the generalised linear dilaton background ---
(\ref{eq:LinDilBG}) with (\ref{eq:tachBG}) --- in fact make vanish
the exact $\b$--functions written to all orders of $\a'$
\cite{GinMoore}. In any case, \refeq{eq:Slindil} defines an exact
conformal field theory and is therefore physical.

\newpage
\section{The String Coupling}
\label{sec:stringcoupling}
In order to see how the tachyon aids in subduing the strong coupling
behaviour of the theory, we must first examine the string coupling.
Moving away from the world--sheet dynamics of the background for a
moment and concentrating on string dynamics in the target space, let
us examine a scattering of $N$ gravitons via 3--line vertices as
depicted in fig. \ref{fig:tubes}. Each vertex contributes a factor
of the coupling constant $\k_{\rm cl}$ for closed strings.

\begin{figure}[!hc]
\figlabel{fig:tubes}
\begin{center}
\includegraphics[scale=0.3]{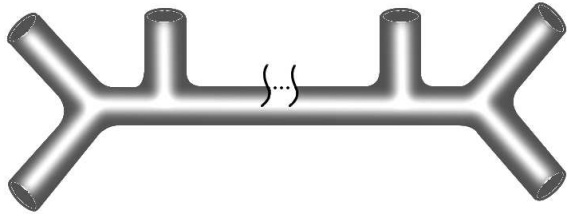}
\bf \caption{ \rm Scattering of $N$ closed strings at tree--level.
Each 3--vertex contributes a factor $\k_{\rm cl}$}
\end{center}
\end{figure}
Note that this could be viewed as scattering though a single
$N$--vertex, but in the stringy world, this is topologically
equivalent to scattering through $N-2$ elementary 3--vertices. Thus,
the diagram is proportional to $\k_{\rm cl}^{N-2}$.

We could also add some loops as in fig. \ref{fig:tubes-loop}.

\begin{figure}[!hc]
\figlabel{fig:tubes-loop}
\begin{center}
\includegraphics[scale=0.3]{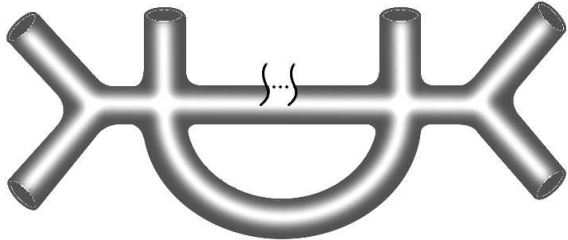}
\bf \caption{ \rm Each loop contributes a factor of $\k_{\rm
cl}^2$.}
\end{center}
\end{figure}
Each loop needs two 3--vertices and thus contributes $\k_{\rm
cl}^2$. For a general scattering of $N$ closed strings with $g$
loops, the diagram will be proportional to
\be
\k_{\rm cl}^{N-2 + 2g} = \k_{\rm cl}^N \cdot \k_{\rm cl}^{-2(1 -
g)}. \nn
\ee
The factor $\k_{\rm cl}^N$ can be absorbed in the definition of the
$N$ external vertices. Hence, for any process involving closed
strings generating a Riemann surface of genus $g$, the partition
function \re{formalZ} corresponding to this process will be weighted
by an overall factor of
\be
\label{eq:Zweight}
 \k_{\rm cl}^{-2(1-g)}.
\ee
By examining \refeq{eq:SPOLtot} and \refeq{eq:formalZ} we see that
the partition function is weighted by a purely topological factor
$e^{\nu \chi}$. For the case of closed strings, where there are no
boundaries ($b=0$) we could write the weighting (\ref{eq:Zweight})
compactly in terms of the Euler characteristic $\chi = 2-2g$ and
therefore make the association
\be
\k_{\rm cl}^{-2(1 - g)} \sim e^{-\nu \chi} \Rightarrow \k_{\rm cl}
\sim e^\nu.
\ee
Following a similar reasoning with open strings, we may conclude
that
\be
\k_{\rm op}^{-2(1 - b)} \sim e^{-\nu \chi} \Rightarrow \k_{\rm op}
\sim e^{\nu/2},
\ee
where $\k_{\rm op}$ is the open string coupling constant.

\subsection{Strong Coupling Region}
\begin{figure}[!b]
\figlabel{fig:lwall}
\begin{center}
\includegraphics[scale=0.5]{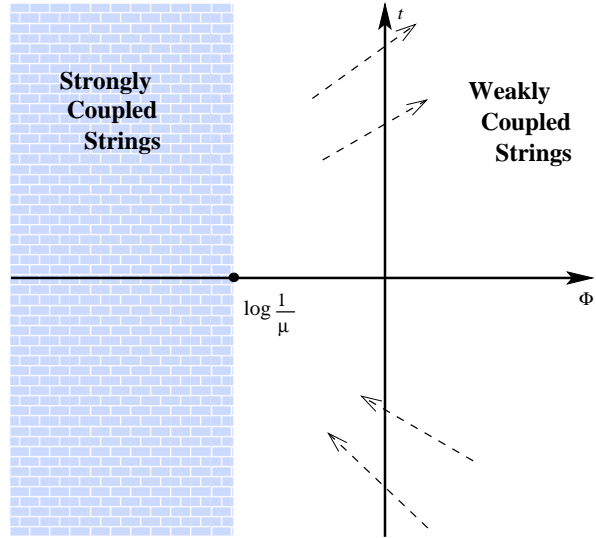}
\bf \caption{ \rm The target space of 2D string theory in a linear
dilaton background with a cosmological constant term. The tachyon
mode due to this term produces a potential wall prohibiting strings
from venturing to the strong coupling region (see text). The dotted
lines represent the scattering off the wall. (The choice of
coordinates here is made in anticipation of the next section, where
we will see make the associations $X^\mu=(it,\phi)$, $l_\mu=(0,-Q)$,
where $Q>0$ is a constant so the coupling will be $\propto
e^{-Q\phi}$.)}
\end{center}
\end{figure}
In \S\ref{subsec:nonlinsigma} we considered closed strings in the
context of the sigma--model. Inspecting \refeq{eq:SigmaModel}, we
see that the dilation field, which is a generalisation of the
coupling constant $\nu$ in the Einstein--Hilbert term
(\ref{eq:SPOLtot}), thus {\em defines} the coupling constant
$\k_{\rm cl}$, i.e. for closed strings, with $\nu = \dl(X^\rho)$;

\be\label{eq:strdilatoncoupling}
\k_{\rm cl} \sim e^\dl.
\ee

But then, in the linear dilaton background $\dl   = l_\mu X^\mu$,
there is a region in the target space where the $X^\mu$ is such that
the coupling diverges and string perturbation theory fails.

Now the usefulness of including the cosmological constant term
$$
S_\s^{\rm cosmo} = \frac{1}{4\pi\a'}\int d^2\s \, \sqrt{h}\
\mu\exp(k_{\mu}X^{\mu})
$$
in the sigma--model action $S_{\s}^{\rm LD}$ becomes apparent. Since
the interaction turns on exponentially, there is effectively a wall
--- the ``Liouville wall''
(see fig. \ref{fig:lwall}) placed in the target space at
\be
\label{eq:LW}
X^\mu \sim \frac{1}{k_\mu}\log
\frac{1}{\mu}=-\frac{k^\mu}{k^2}\log\mu
\ee
precluding strings from venturing into the strong coupling region
(we will always assume that the cosmological constant satisfies
$\mu>0$).

We did not restrict ourselves to the critical dimension $D=D_{\rm
cr} = 26$ here and, just as the linear dilaton background exits in
any dimension, the potential wall exists in dimensions $D$ other
than the critical dimension \cite{GinMoore}.

\newpage
\section{Two Dimensional String Theory}
\label{sec:tdst}
In \S\ref{sec:2Dgrav} one possible interpretation of noncritical
string theory was mentioned, that is, string theory formulated in a
space--time of dimension other than $D_{\rm cr} = 26$, where we
considered 2D string theory which admits an interpretation as a
theory of 2D gravity. Noncritical dimensions lead to a Weyl anomaly,
but some interpretations do not necessitate Weyl invariance from the
outset. For example, there is not reason to demand Weyl invariance
in a system where we interpret the string world--sheets as
statistical system of two--dimensional surfaces embedded in a
$d$--dimensional target space. This viewpoint leads to a conformal
field theory known as Liouville gravity.

Before considering the ramifications of surrendering Weyl
invariance, we start by showing how noncritical string theory in
$d$--dimensions relates to $(d+1)$--dimensional critical string
theory in a linear--dilaton background.

\subsection{Liouville Gravity and Conformal Field Theory}
\label{sec:LiouvilleTh}
The most convenient way to quantise a two dimensional system
described by the generally covariant Polyakov action (\ref{eq:SPOL})
in the absence of Weyl symmetry is to fix a conformal gauge
\cite{FGZ94}
\be
\label{eq:confgaugeL}
h_{ab} = e^{\g\phi(\s,\t)}\hh_{ab},
\ee
where $\hh_{ab}$ is a stationary background metric. The {\em
conformal mode}, $\phi$, has now become a dynamical field, called
the {\em Liouville field}, and a parameter $\g$ was inserted for
later convenience. The resulting system is described by the
conformal mode $\phi$ and target space coordinates $X^\mu$ (or {\em
matter fields} on the world--sheet, as the case may be) existing in
the presence of the background metric $\hh_{ab}$.

In order to write an effective action for this theory having now
fixed the gauge (\ref{eq:confgaugeL}), a measure of integration over
the space of all possible metrics must be carefully constructed
taking into account Faddeev--Popov ghosts introduced by the gauge
fixing, and a cosmological constant term, absent on the classical
level, must be added for renormalizability. This was implemented by
Polyakov in \cite{Polyakov81}. Working in a $d$--dimensional target
space with a Euclidean signature, we write the result
as\footnote{Other references \cite{FGZ94,GinMoore,DistlerKawai}
sometimes use a different overall normalisation which means that the
fields may be rescaled and there may be factors in front of some of
the terms in the Lagrangian of (\ref{eq:SCFT}). See footnote
\ref{fn:norm} on page \pageref{fn:norm}.}
\be
\label{eq:SCFT}
 S_{\rm CFT} = \frac{1}{4\pi\a'} \int \!\!d\s^2\, \sqrt{\hh}
\left( \hh^{ab}\p_a X^\mu \p_b X_\mu + \hh^{ab}\p_a \phi \p_b \phi -
\a' Q\hCR \phi + \mu e^{\g \phi} +\!\!\!
\begin{tabular}{c}\footnotesize ghost \\[-4pt] \footnotesize terms \end{tabular}
\!\!\!\right)
\ee
where $\hCR(\hh^{ab})$ is the total curvature of the background
metric. Also, $\phi$ has been scaled to get the conventionally
normalised kinetic term \cite{GinMoore} (prior to this, there is a
$d$--dependant factor in front of the kinetic term of $\phi$ which
is responsible for its disappearance in the critical dimension).

This action describes a {\em conformal field theory}, it is
invariant under conformal transformations of the fields
\footnote{Actually, \re{conftr} is only true for primary (highest
weight) fields, but we are only interested in those fields
here.} \cite{Gab}
\be\label{eq:conftr}
\xi_j(z,\bz) \mapsto \xi_j(w,\bw) = \left(\frac{\p z}{\p
w}\right)^{\D_j} \left(\frac{\p \bz}{\p \bw}\right)^{\bD_j}
\xi_j(z,\bz),
\ee
where the complex coordinates $z,\bz,w,\bw$ are related to the
world--sheet coordinates in a similar fashion as \re{cmplxcoo}, and
$\xi_j$ stands in place of the various fields of (\ref{eq:SCFT}).
The $\D_j$ and $\bD_j$ are called the holomorphic and
antiholomorphic weights of the fields $\xi_j$ respectively.

To determine $Q$ we note that the original theory (\ref{eq:SPOL}) is
dependent only on $h_{ab}$, so our action (\ref{eq:SCFT}) should be
invariant under the simultaneous shift
\be
\label{eq:shifth}
\hh_{ab} \mapsto e^{\rho(\s,\t)} \hh_{ab},\qquad \phi(\s,\t) \mapsto
\phi(\s,\t) -\rho(\s,\t)/\g.
\ee
Given that the $\phi$--integration measure is invariant under
(\ref{eq:shifth}), the invariance thus requires simply that $S_{\rm
CFT}[\phi-\rho(\s,\t)/\g,e^{\rho}\hh_{ab}]=S_{\rm
CFT}[\phi,\hh_{ab}]$, in other words, that the total conformal
anomaly vanishes. Inspection of the ghost and matter integration
measures under conformal transformations as we fix the gauge reveals
that the central charge of the matter field is $c_{\rm matter} = d$,
and of that of the ghost field is $c_{\rm ghost} = -26$.

The central charge of the Liouville field part, $c_{\rm Liouville}$,
is found through an operator product expansion. To see how this
works for open strings, for example, we first Wick--rotate the
world--sheet $(\s,\t)$ coordinates to $(\s,i\t)$. Then, in the case
of a simply connected open string world--sheet, we can then perform
a conformal transformation which maps the world--sheet into a disk
in Euclidean space. This amounts to a transition to the variables
$(z,\overline{z})$ where
\be
\label{eq:cmplxcoo}
z = e^{i\s^+},\qquad \overline{z} = e^{i\s^-}, \qquad \s^\pm =
i\t\pm\s.
\ee
Given that the energy--momentum tensor $T$ of the theory is
holomorphic\footnote{More specifically, only the $T_{++}$ (or
$T_{zz}$) component is holomorphic, the $T_{--}$ ($T_{\bar{z}\bar
{z}}$) component, being antiholomorphic, will have a Laurent
expansion in $\overline{z}$. But for the purposes of calculating the
central charge, we may use either component. Recall also that
$T_{+-}=T_{-+}=0$.}, it will have a Laurent expansion
\be
T_{zz}=T(z) = \sum\limits_{n \in \mathbb{Z}} L_n z^{-n-2}, \qquad
L_n^\dag = L_{-n},
\ee
where the {\em Virasoro generators}, $L_n$, may be calculated from
Cauchy's formula
\be \label{eq:cauchy}
L_n = \oint_C\frac{dz}{2\pi i}\, z^{n+1}T(z),
\ee
where $C$ is a contour encircling the origin, oriented
counterclockwise. The Lie algebra of conformal transformations is
\be
\label{eq:virasoro}
[L_m ,L_n] = (m-n)L_{m+n} + \frac{c}{12}(m^3 - m)\delta_{m+n,0}\;.
\ee
The anomalous term in (\ref{eq:virasoro}) is known as the {\em
Schwinger term}. This is called the Virasoro algebra and it closes
under the Lie bracket so long as we include the commuting
$\delta_{m+n,0}$ central extension in our set of generators. This
algebra is embodied in the operator product expansion \cite{FMS86}
\be
T(z)T(w) \sim \frac{c}{2(z-w)^4} + \frac{2}{(z-w)^2}T(w) +
\frac{1}{z-w}\p_w T(w),
\ee
where ``$\sim$'' indicates equality up to non--singular terms. One
may think of the operator product expansion as generating functions
for the commutators in \re{virasoro}. The two are linked through
\re{cauchy}.

Taking the energy--momentum tensor of the Liouville part of
\re{SCFT} and computing its operator--product expansion, the
Schwinger term of the Virasoro algebra is found to have $c_{\rm
Liouville} = 1 + 6\a'Q^2$. This conformal anomaly describes the lack
of invariance of the theory to deforming the background metric
$\hh_{ab}$.

As explained above, in order for \re{SCFT} to define a conformal
field theory, the sum total of central charges must vanish
\be
c_{\rm total}=c_{\rm matter}+c_{\rm ghost}+c_{\rm Liouville}=d-25 +
6\a' Q^2 = 0,
\ee
which leads to
\be
\label{eq:Q}
Q = \sqrt{\frac{25 - d}{6 \a'}},
\ee
where the choice of sign of the square root can be absorbed in a
redefinition of the sign of $\phi$.

Conformal symmetry also fixes the conformal dimensions of the
perturbing term $\mu e^{\g\phi}$; in order for the CFT action
\re{SCFT} to be covariant, the factor $e^{\g\phi}$ in the
transformation of the metric $\hh_{ab}$ in \refeq{eq:confgaugeL}
must have conformal holomorphic and antiholomorphic weights $\Delta
= \overline{\Delta} = 1$. Considering again the energy--momentum
tensor of the Liouville part of $S_{\rm CFT}$, its operator product
expansion with the deformation of the metric, obtained from the Ward
identity \cite{Polbook}, is
\beq
T(z)e^{\g\phi(w)} &\sim& \frac{\Delta e^{\g\phi(w)}}{(z-w)^2} +
\frac{\p_w e^{\g\phi(w)}}{z-w}+\cdots \\
\overline{T}(z)e^{\g\phi(w)} &\sim& \frac{\overline{\Delta}
e^{\g\phi(w)}}{(z-w)^2} +\frac{\p_w e^{\g\phi(w)}}{z-w}+ \cdots \; .
\eeq
Calculating this expansion yields conformal weights for $e^{\g\phi}$
of
\be
\label{eq:confweie} \Delta = \overline{\Delta} =
-\frac{1}{2}\g(\frac{1}{2\a'}\g + Q).
\ee
Therefore, solving (\ref{eq:confweie}) by requiring that $\Delta =
\overline{\Delta} = 1$ and using \refeq{eq:Q}, we get
\be
\g_{\mp} = -Q \mp \sqrt{Q^2 - \frac{4}{\a'}} =
-\frac{1}{\sqrt{6\a'}}\left(\sqrt{25 - d} \mp \sqrt{1-d}\right)
\ee
were the branch $\g = \g_{-}$ of the quadratic solution is to be
chosen for agreement with the semiclassical limit \mbox{$(d
\rightarrow -\infty)$} \cite{DistlerKawai}. This imposes the
``Seiberg bound'' removing the set of $\phi$ wave functions which
reside in the strong coupling region (see
\S\ref{ch:STR}.\ref{sec:Sieberg}).

Now, we make the following associations:
\be
\label{eq:associations}
D=d+1; \qquad
l_\mu =\left\lbrace
\begin{array}{l}
0,\quad \mu \neq D\\
-Q,\quad \mu = D
\end{array}\right. ;\qquad
k_\mu = \left\lbrace
\begin{array}{l}
0,\quad \mu \neq D\\
\g,\quad \mu = D
\end{array}\right. ;\qquad
X^D = \phi.
\ee
We then immediately see the correspondence between the CFT action
\re{SCFT} and the \mbox{$\s$--model} in the linear dilaton
background (\ref{eq:Slindil}).

Thus \re{SCFT} which describes Liouville theory coupled with $c=d$
matter, or noncritical string theory in $d$ dimensions, is none
other than $(d+1)$--dimensional critical string theory in a linear
dilaton background (\ref{eq:LinDilBG}) with $d$ matter fields and
the conformal mode field $\phi$, called the Liouville field.

\subsection{2D Strings in the Linear Dilaton Background}
\label{sec:Sieberg}
Let us now restrict ourselves to two dimensional bosonic strings in
the linear dilaton background (\ref{eq:LinDilBG}). Owing to the
associations made in the previous section, the dilaton becomes
\be
\dl = l_\mu X_\mu = -Q \phi,
\ee
with $Q$ from \refeq{eq:Q}.

If we redefine the Tachyon field as
\be\label{eq:redefT}
T = \eta e^{-Q\phi},
\ee
then the tachyon action (\ref{eq:Seff-tach}) in Euclidean
space--time\footnote{Note that the covariant derivatives in
\refeq{eq:Seff-tach} now become ordinary partial derivatives since
the metric is flat.} becomes the action of a scalar field $\eta$
\beq\label{eq:redefTs}
S_{\rm eff}^{\rm tach} &=& -\frac{1}{2}\int d^D X\, e^{2Q\phi}
\left[ (\p T)^2 - \frac{4}{\a'} T^2\right]\nn \sep
&=&-\frac{1}{2}\int d^D X\, \left[ (\p\eta)^2 + \left(Q^2 -
\frac{4}{\a'}\right) \eta^2\right].
\eeq
and we read off the mass of the tachyon to be
\be
m^2_{\rm tach} = \left(Q^2 - \frac{4}{\a'}\right) = \frac{2-D}{6\a'}
= 0,
\ee
where we used $Q$ from \re{Q}. So in two dimensions
($d=1\Leftrightarrow D=2$), the tachyon is {\em massless}. For
historical reasons, this massless scalar is still called a tachyon,
but in two dimensions it a physical particle describing the stable
vacuum of two dimensional bosonic string theory. This demonstrates
that the $D=2$ case is indeed special.

We can now summarise our conformal field theory for the
two--dimensional bosonic string in Euclidean space--time. For $\a' =
1$,
\be
\label{eq:CFT}
S_{\rm CFT} = \frac{1}{4\pi}\int d^2\s \sqrt{\hh}\left( \hh^{ab}
\p_a X \p_b X + \hh^{ab} \p_a\phi\p_b\phi - 2\hCR\phi + \mu
e^{-2\phi} +
\begin{tabular}{c}\footnotesize ghost \\[-4pt] \footnotesize terms
\end{tabular}\!\!\right)
\ee
\subsubsection{Seiberg Bound}
In the region $\phi\rightarrow\infty$, the Liouville interaction
term $\mu e^{-2\phi} $ becomes negligible and (\ref{eq:CFT})
represents a free theory. However, the interaction imposes a
potential wall precluding states from venturing to the
$\phi\rightarrow -\infty$ region. Therefore, to discuss the spectrum
of states, we may in fact simply consider a free theory with $\mu =
0$ and impose the {\em Seiberg bound} which removes states from the
$\phi\rightarrow -\infty$ region \cite{Seibbound}.

Finally, note that the above results were obtained working in
Euclidean space-time. The continuation to Minkowski space--time is
easily made by the analytic continuation
\be
X \mapsto it.
\ee
\subsection{2D Strings in a Black Hole Background}
Two dimensional string theory embodies a microscopic description of
black holes, as can be seen already in the low--energy limit, where
we have an effective target--space action composed of
(\ref{eq:Seff}) and (\ref{eq:Seff-tach}); a two dimensional
effective string action in the presence of general backgrounds,
\be
\label{eq:Seff2D}
S_{\rm eff} = \frac{1}{2} \int d^2 X\, \sqrt{-G}e^{-2\dl}\left[
\frac{16}{\a'} + \CR + (D_\mu \dl)^2 - (D_\mu T)^2 +
\frac{4}{\a'}T^2\right].
\ee
(Since there can be no antisymmetric 3--tensor in two dimensions, we
omit the $B_{\mu\nu}$--field and its field strength 3--tensor
$H_{\mu\nu\r}$).

For $T=0$, this theory is classically exactly solvable. In this
case, writing the vector $X^\mu$ as
$\(\!\!\begin{array}{c}t\\r\end{array}\!\!\)$ and taking into
account \re{associations}, we get the classical solutions to the
metric equations of motion \cite{Mandal91}
\be
\label{eq:SCsol}
ds^2= -\left(1-e^{-2Qr}\right)dt^2+\frac{1}{1-e^{-2Qr}}dr^2, \qquad
\dl=\vr_0 - Q r.
\ee
with $ Q = \sqrt{4/\a'}$ in two dimensions and an integration
constant $\vr_0$.

The resemblance with the radial part of the Schwarzschild metric --
the metric of a spherically symmetric black hole in $(3+1)$
space--time dimensions
\be\label{eq:SCm}
ds^2 = -(1-r_s/r)c^2 dt^2 + \frac{1}{1-r_s/r}dr^2 + r^2(d\theta^2 +
\sin^2\theta d\phi^2),\qquad r_s \equiv \frac{2GM}{c^2}
\ee
is apparent (with $G$ the gravitational constant, $M$ the black hole
mass, $\theta$ and $\phi$ spherical angles here).

In fact, (\ref{eq:SCsol}) describes a $(1+1)$--dimensional black
hole, with a singular space--time curvature for\footnote{The
coordinate $r$ here should not be confused with the positive
semidefinite radial coordinate which, by analogy with the
Schwarzschild metric \re{SCm}, would be $\sim e^{r}$ here.}
$r\rightarrow -\infty$ and a coordinate singularity at\footnote{It
must be purely a coordinate singularity since the scalar curvature
computed from (\ref{eq:SCsol}) is $\CR_G=4Q^2 e^{-2Qr}$ which is
regular at $r=0$.} $r=0$ defining the black hole horizon. As there
are no essential singularities, we may analytically continue past
the coordinate singularity by imitating the procedure for the
Schwarzschild solution where one migrates to Kruskal coordinates
\cite{W91}.

The black hole has a mass related to the integration constant by
\be
M_{_{\bullet}} = 2Qe^{-2\vr_0}
\ee
It emits Hawking radiation at a temperature of $T_{\rm H} = Q/2\pi$
and has a non--zero entropy \cite{NP92,GP92,CGHS91,Kumar95}
\be\label{eq:entbh}
S_{_{\bullet}}=2\pi\jmath,\qquad \jmath=2e^{-2\vr_0}.
\ee
It is interesting to note that the entropy of a 4D black hole is
proportional to its area, (the Bekenstein--Hawking entropy,
$S=A/4l_{\rm P}^2$ where $l_{\rm P}=\sqrt{\hbar G/c^3}$ is the Plank
length). In 2D the ``surface'' of a black hole is a
null--dimensional point and the notion of area is nonsense.
Nonetheless, by comparing the metrics \re{SCsol} and \re{SCm} we see
that the dilaton field at the horizon may be thought of as the
logarithm of an effective area $A\sim\jmath$ of the black hole in 2D
black hole physics \cite{Frolov92,CadoniMignemi99}. The entropy
\re{entbh} then becomes a 2D generalisation of the 4D
Bekenstein--Hawking entropy. This can also be confirmed through
dimensional reduction of heterotic string theory \cite{Cadoni99}.

The solution (\ref{eq:SCsol}) provides a background value for the
metric and the dilaton field in a world--sheet $\s$--model. Compared
with the backgrounds (\ref{eq:NoDilBG}--\ref{eq:LinDilBG}),
\refeq{eq:SCsol} is an example of a ``non--trivial background''.

As this theory is a well defined on curved space--times, one can
thus hope to achieve a microscopic description of black hole entropy
through string theory which is not possible though typically
ill--defined quantum field theoretic gravity extensions. In order to
do this, however, we require more than just an effective action, but
rather an exact CFT action that is valid in all energy domains.
Precisely such a CFT was found by Witten \cite{W91} which reduces to
leading order in $\a'$ to the world--sheet string action in the
presence of the black hole background (\ref{eq:SCsol}). This model
is called the $[\gr{SL}{2,\bbR}]_k/\gr{U}{1}$ coset $\s$--model, $k$
being the level of the representation for the algebra of the
generators associated to the $\gr{SL}{2,\bbR}$ symmetry current.

For interest, the background produced by this exact CFT \cite{DVV92}
in the notation of \cite{KT01} is\footnote{The normalisation of
units in the gauged Wess--Zumino--Witten model originally presented
in \cite{W91} would require the shift $r_{\rm WZW} = Qr,\quad
\th_{\rm WZW} = Q\th,\quad \a' Q^2 = \nf{1}{(k-2)}$}
\beq
\label{eq:DVV}
ds^2 = \frac{ \tanh^2 Qr }{ 1-p\tanh^2 Qr } d\th^2
+ dr^2 \label{eq:bh1}\\
\dl = \vr_0 - \log(\cosh Qr) - \frac{1}{4} \log(1-p\tanh^2 Qr)\label{eq:bh2}\\
p = \frac{2\a' Q^2}{1+2\a' Q^2},\qquad k = \frac{2}{p} = 2 +
\frac{1}{\a' Q^2}.\label{eq:bh3}
\eeq
In our two dimensional case, $p = \nf{8}{9}$ and $k=\nf{9}{4}$.
Since $\tanh r \rightarrow 1$ for large $r$, we get a cigar--shaped
manifold depicted in fig. \ref{fig:cigar} representing a Euclidean
black hole.
\begin{figure}[!t]
\figlabel{fig:cigar}
\begin{center}
\includegraphics[scale=.48,clip=true]{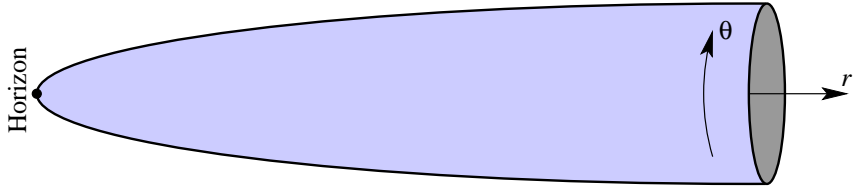}
\bf \caption{ \rm Semi--infinite cigar manifold --- the Euclidean
black hole.}
\end{center}
\end{figure}

In the limit $k \rightarrow \infty
\stackrel{(\footnotesize\ref{eq:bh3})}{\Leftrightarrow} p\rightarrow
0$ we get back the leading order behaviour of \cite{DVV92}, namely,
\be
ds^2 = dr^2 + \tanh^2Qr \,d\th^2.
\ee

To see how this reproduces the metric equations of motion
\re{SCsol}, we make the associations
\beq
r &\mapsto& \frac{1}{Q}\tanh^{-1}\sqrt{1-e^{-2Qr}} =
\frac{1}{2Q}\ln\left(\frac{1+\sqrt{1-e^{-2Qr}}}{1-\sqrt{1-e^{-2Qr}}} \right) \\
\theta &\mapsto& it
\eeq
then
\be
dr \mapsto \frac{dr}{\sqrt{1-e^{-2Qr}}}
\ee
and we thus retrieve the black hole background \re{SCsol}.

The flat space (i.e. the space--time in the absence of the black
hole) has the topology of a cylinder, $S^1 \times \bbR^1$, where the
Euclidean time is periodic with an arbitrary period,
\be
X = X + \b,\qquad \b = 2\pi R,\nn
\ee
and $R$ is the radius of $S^1$. But for the cigar topology of the
Euclidean black hole, the period for the Euclidean time $\th$ is
fixed at
\be
\b = \frac{2\pi}{Q\sqrt{1-p}}
\ee
in order to obtain a proper disc metric at the tip of the cigar,
otherwise the metric has a conical singularity where $r=0$
\cite{GP92}.

Unfortunately, attempts to calculate thermodynamic properties in
this background are thwarted by the lack of a exact target--space
action, similar to (\ref{eq:Seff2D}), for which
(\ref{eq:bh1}--\ref{eq:bh3}) gives a solution. But in the limiting
case $p\rightarrow 0$ ($k\rightarrow \infty$) it is clear that we
obtain the Hawking temperature $T_{\rm H} = \nf{Q}{2\pi}$ as the
local temperature given by the inverse periodicity of the Euclidean
time, $\b^{-1}$.

The Minkowskian version of the sigma model producing (\ref{eq:DVV})
was obtained by performing the analytic continuation\footnote{In
some treatments, it is the Liouville coordinate $r$ which is
analytically continued. But the physics of the Euclidean black hole
corresponds to the interpretation of $r$ as the spacial coordinate
\cite{W91}.} $\th = it$ and corresponds to the unitary coset
$\gr{SL}{2,\bbR}/\gr{U}{1}$. The Euclidean version corresponds to
the unitary coset $\mathsf{H}_3^+ /\gr{U}{1}$ where
\mbox{$\mathsf{H}_3 = \gr{SL}{2,\bbC}/\gr{SU}{2} = \mathsf{AdS}_3$}
\cite{KKK}, $\mathsf{AdS}_3$ being an three dimensional Anti-de
Sitter group manifold, in other words, a three dimensional space of
constant negative curvature.

We will see that, through the ``FZZ conjecture'', introducing
winding perturbations to the CFT action in the linear dilaton
background (\ref{eq:CFT}) creates a theory which appears to be
equivalent to the $\mathsf{H}_3^+ /\gr{U}{1}$ \mbox{$\s$--model} of
the Euclidean black hole but is solvable using matrix methods. The
perturbations in question are affected by relevant operators arising
out of the CFT action. Amongst those are the vertex and vortex
operators.

\subsection{Vertex and Vortex Operators}
\label{sec:vervor}
Working with the free theory by neglecting the Liouville interaction
momentarily, the asymptotic form of the tachyon momentum is found
from (\ref{eq:tachBG}). Writing $k_\mu = (k_X,k_\phi)$, in two
dimensions, where $l_{\mu} = (0,-Q)$,
\be
k_X^2 + (k_\phi + Q)^2 = 0,
\ee
and we should solve this such that the momentum conjugate to the
Liouville field is real. With $k \in \bbR$, we thus chose $k_X = i
k$, and obtain
\be
\label{eq:mom}
k_X = i k,\qquad k_\phi = \pm |k| - 2.
\ee
In the noncompactified theory, we use these momenta to form the
Euclidean vertex operators associated to the centre of mass motion
of the string
\be\label{eq:shmeahfoo}
V_k = \int d^2\s\, e^{ik X} e^{(|k| - 2)\phi}
\ee
where we imposed the Seiberg bound (see \S\ref{sec:Sieberg})
prohibiting operators growing as $\phi \rightarrow -\infty$ and thus
choose the positive solution in (\ref{eq:mom}).

To obtain the Minkowskian version, we analytically continue the
coordinates and momenta
\be\label{eq:Minkversion}
X\mapsto it,\qquad k \mapsto -ik.
\ee
For $k>0$, this implies
\be
k_{\rm X}\mapsto \pm k,\qquad k_{\phi}\mapsto -i k-2
\ee
leading to
\be\label{eq:vertexops}\begin{array}{rcl}
V_k^+ =\int d^2\s\,e^{-ik(t+\phi)}e^{-2\phi}\\
V_k^- =\int d^2\s\,e^{ik(t-\phi)}e^{-2\phi}.
\end{array}\ee
Inspecting the signs of $t$ and $k$ we see that $V_k^-$ describes
outgoing tachyons moving to the right (i.e. the in direction of
positive $\phi$) and $V_k^+$ describes incoming, left--moving
tachyons\footnote{In CFT nomenclature, $V_k^-$ is antiholomorphic
and $V_k^+$ is holomorphic.}.

\subsubsection{Winding Modes and Compactification}
There exist also ``vortex'' operators associated with winding
excitations that arise when the time coordinate $X$ is compactified
(there is no periodicity in the Liouville coordinate so it can not
be compactified). They correspond to configurations where the string
wrapped around the compactified dimension. Although it is perfectly
legitimate to introduce vortex perturbations to the bare CFT action
(\ref{eq:CFT}), their interpretation in terms of fields in the
target space is not per se clear, although they can be related to
vertex operators through T--duality. This opens the possibility of
solving a complicated problem involving vertex operators by
transiting to the dual picture where things may be simpler, then
transforming back. But, beyond this mention, winding modes will not
feature in this thesis. A discussion of winding modes and vortex
operators may be found in \cite{SergePhD}.

\subsection{Perturbed CFT model}
\label{sec:pertCFT}
In the theory compactified at radius $R$, the Euclidean time
coordinate $X$ is periodic,
\be
X = X + \b,\qquad \b = 2\pi R.
\ee
The tachyonic momentum spectrum, now obeying periodic boundary
conditions, is rendered discrete, $k_n = n/R$ and, consequently, the
vertex operators $V_{k_n}^\pm$ are also discretised. These operators
can be used to perturb the CFT action \re{CFT} which now takes the
form
\be\label{eq:pertCFT}
S=S_{\rm CFT} + \sum\limits_{n\neq 0} t_n V_n,\qquad  V_n =
\left\lbrace
\begin{array}{l}
V_{k_n}^+,\quad n >0\\
V_{k_n}^-,\quad n <0
\end{array}\right.
\ee
where the discrete vertex operators are labeled $V_n$ for
simplicity. The perturbations introduce a nonvanishing
time--dependent vacuum expectation value for the tachyon thereby
changing the background value of $T$, making it dynamic. These
tachyon ``condensates'' are time--dependent, unlike the cosmological
constant term $\mu e^{-2\phi}$ in the CFT action (\ref{eq:CFT})
which by itself produces the $T$ background (\ref{eq:tachBG}) and is
responsible for the Liouville wall \re{LW}.

The simplest case where $t_{\pm1}$ is called the ``sine--Liouville
CFT'' model. The T--dual theory was conjectured by V. Fateev, A.
Zamolodchikov and Al. Zamolodchikov (the ``FZZ conjecture'') to be
equivalent to the $\mathsf{H}_3^+ /\gr{U}{1}$ unitary coset CFT.
Regrettably, this
result was never published 
although it is described in several papers (eg: \cite{KKK}).

\subsubsection{The FZZ conjecture}
First, we decompose the matter field into left and right components
\be
X(\s,\t) = X_{\rm L}(\t + i\s) + X_{\rm R}(\t - i\s)
\ee
where the field $X$ is compactified with a compactification radius
$R$, and subsequently formulate a {\em dual field}
\be
\tlX(\s,\t) = X_{\rm L}(\t + i\s) - X_{\rm R}(\t - i\s).
\ee

The FZZ conjecture asserts that the $\mathsf{H}_3^+ /\gr{U}{1}$
unitary coset CFT, with the level of representation $k$ for the
underlying $\gr{SL}{2,\bbC}$ symmetry current algebra, is equivalent
to a CFT given by the sine--Liouville action\footnote{The classical
equation of motion derived from the \re{SLaction} contains a sinus
term $\sim\sin(R\tlX)$ like the sine--Gordon equation which is why
it is referred to as the sine--Liouville action.}
\be
\label{eq:SLaction}
S_{\rm SL}=\frac{1}{4\pi\a'}\int d^2\sigma\, \left[ (\p X)^2+(\p
\phi)^2 -\a' Q \hat{\CR} \phi +\l e^{\r\phi}\cos(R\tlX) \right].
\ee

If correct, the central charges of (\ref{eq:SLaction}) and the
$\mathsf{H}_3^+ /\gr{U}{1}$ coset CFT must coincide. From
\cite{W91,KKK},
\be
c_{\rm coset} = \frac{3k}{k-2} - 1,\qquad c_{\rm SL} = 2 + 6\a' Q^2
\ee
which yields
\be
\label{eq:FZZq}
Q=\frac{1}{\sqrt{\a' (k-2)}}.
\ee
in accordance with \re{bh3}.

In the asymptotic region ($\phi\rightarrow\infty$,
$r\rightarrow\infty$) the target space has a cylindrical topology in
both theories (from fig. \ref{fig:cigar} this is evident in the
coset CFT case). Setting
\be
\label{eq:FZZr}
R=\sqrt{\a' k}
\ee
ensures that both theories agree in this asymptotic region since the
level of representation $k$ scales the radius of the cigar of the
coset CFT \cite{KKK}. Finally, requiring that the holomorphic and
antiholomorphic dimension of the sine--Liouville interaction term be
equal to one and appealing to equations (\ref{eq:FZZq}) and
(\ref{eq:FZZr}), the final parameter is fixed,
\be
\rho=-\sqrt{\frac{k-2}{\a'}}.
\ee

So in our two dimensional case,
\be
Q=2/\sqrt{\a'}, \qquad R=3\sqrt{\a'}/2, \qquad k=9/4.
\ee

It was checked that the spectra of the two theories coincide, the
scaling dimensions of observables agree and two point correlators
agree \cite{KKK}. Although not a proof, this is very strong evidence
in favour of the FZZ conjecture. In fact, the conjecture has now
been explicitly proven in the case of supersymmetric strings
\cite{kapust}.

\EndOfChapter

In this chapter, we saw that by introducing the zero modes of the
bosonic string subject to certain constraints imposed by Weyl
invariance to the Polyakov action, we obtained a sigma model in the
low energy domain of $D$--dimensional critical string theory in a
linear dilaton background in \re{CFT}. The cosmological constant
term $\mu e^{-2\phi}$ was responsible for a potential wall in the
target space which guarded against a divergent string coupling.

It was then shown that Liouville gravity coupled to $c=(D-1)$ matter
fields or, equivalently, $(D-1)$--dimensional noncritical (non-Weyl
invariant) string theory was equivalent to $D$--dimensional critical
string theory in a linear dilaton background, the target space
coordinates being described by the matter fields $X^\mu,\
\mu=1,\ldots,(D-1)$ and a Liouville field $\phi$.

From the effective action of the CFT with background fields, it was
shown that, in two dimensions, the target space is curved with black
hole background described by the target space metric (in the absence
of the tachyon)
$$
G_{\mu\nu}=\left(\begin{array}{cc}-(1-e^{-2Qr})&0\\0&(1-e^{-2Qr})^{-1}\end{array}\right).
$$

In two dimensions, the only physical target space field is the
massless bosonic tachyon field which describes the centre of mass of
the string. With this in mind, we introduced time--dependant tachyon
perturbations to the CFT action \re{CFT}. The T--dual of this model
with linear order perturbations is seemingly equivalent to the exact
CFT (the $[\gr{SL}{2,\bbR}]_k/\gr{U}{1}$ coset $\s$--model, valid
not only in the low energy limit) which, to leading order in $\a'$,
produces the black hole background of the aforementioned effective
theory.

As we shall see, the perturbed CFT model has a realisation in terms
of matrix models and the theory under this vise is exactly solvable.

We now turn out attention to the theory of random matrices in the
next chapter. Subsequently, we will generalise random matrix theory
to the case of where the number of distinct matrices becomes
infinite. It is the latter model, called matrix quantum mechanics
(MQM) which we will use to examine the tachyon perturbations in the
final chapter.


\chapter{Random Matrices}
\label{ch:MAT}
In the 1950s, experiments probing heavy nuclei with neutron beams
revealed that the density of their energy levels was too complicated
to be described microscopically. But in 1951 Wigner suggested a
model to describe their distribution statistically using matrices of
large size \cite{Wigner}. This was the first example of a matrix
model in physics. In a matrix model, one typically writes an action
of some kind containing matrix variables. Often the matrices
themselves may be simply the Hamiltonian matrix of the system, or
they may be more abstract, as occurs in CFT and string theory
applications.

The purpose of this chapter is to introduce the mathematical
techniques of random matrices and to show how these techniques can
then be applied to physical problems. For a dedicated treatment, see
\cite{Eynard,Mehta}.

\section{General Applications of Random Matrices}
\label{sec:genapp}
The field of random matrices by far outgrew the original application
in the statistical analysis of heavy nuclei. Matrix models appear in
many areas of physics and mathematics, and some of those
applications are mentioned below before launching into a overview of
matrix model methods in the subsequent section which will be of use
in analysing the perturbed CFT model (\ref{eq:CFT}).

\subsection{Mathematics}
The study of random matrices proves particularly insightful in the
fields of combinatorics, knot theory and non--commutative
probabilities. Their application to discretised surfaces and
integrable hierarchies of partial differential equations is of
considerable use in analysing string theory in non--trivial
backgrounds as we shall soon see.

Another interesting application of random matrices lies in Riemann's
conjecture that all non--trivial zeros of the Riemann zeta function
\be
\zeta(s) = \sum\limits_{n=1}^{\infty}\frac{1}{n^s},\nn
\ee
when extended to the complex plane, lie on the line $\Re(s) =
\nf{1}{2}$. The distribution of those zeros seemingly obey Gaussian
unitary ensemble statistics, in other words, the distribution of
zeros appears to coincide with the distribution of eigenvalues of an
ensemble of hermitian matrices with an action which is left
invariant under unitary transformations of the matrices.

\subsection{Classical and Quantum Chaos}
A classically chaotic system, such as Sinai's billiard or the
stadium of fig. \ref{fig:sinai} is described in terms of a random
matrix Hamiltonian.
\begin{figure}[!hc]
\figlabel{fig:sinai}
\begin{center}
\begin{tabular}{cc}
\includegraphics[scale=0.2]{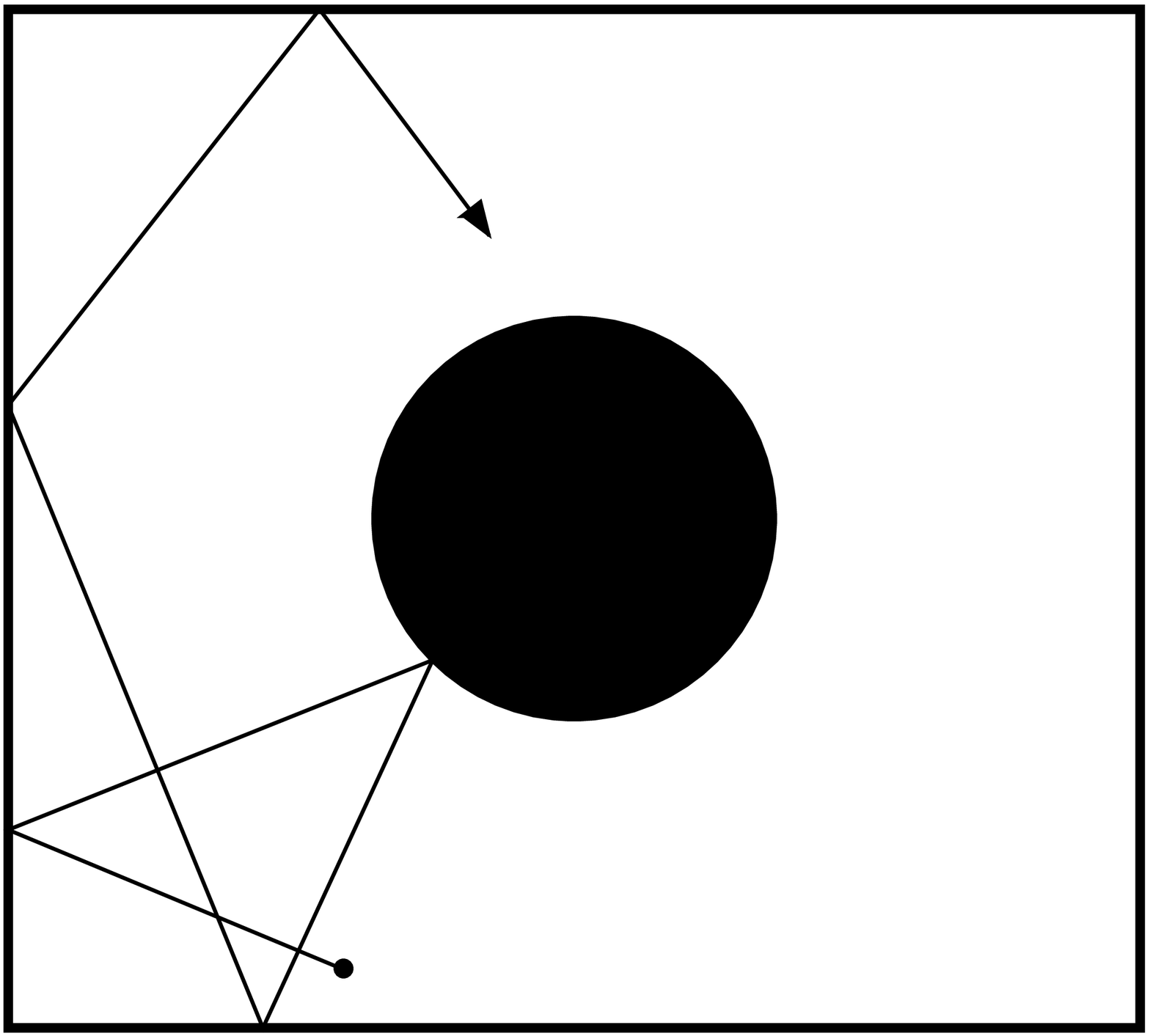}&
\includegraphics[scale=0.7,angle=180,origin=c,bb=0 0 260 120]{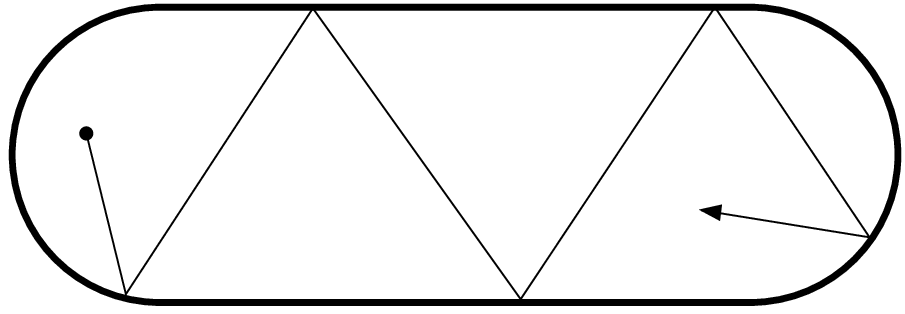}\\
{\bf (a)} & {\bf (b)}
\end{tabular}
\bf \caption{ \rm Classically chaotic systems: {\bf (a)} the Sinai
billiard, {\bf (b)} the stadium.}
\end{center}
\end{figure}
It is now possible to construct such systems on a quantum scale,
consisting for example of an electron in a microscopic Sinai cavity.
It has been conjectured and indeed observed but not rigorously
understood that the Hamiltonian matrix of a quantum system
corresponding to a chaotic classical system has the same spectral
distribution as that of a random matrix ensemble. Thus, although the
Hamiltonian matrix of the quantum system is itself not a random
matrix, one does in fact consider random Hamiltonian matrices in
studying such quantum chaotic systems \cite{Bohigas}. In studying
such systems we see certain remarkable features or matrix models
which are independent on the exact form of the physical potential.
For example, the energy levels of an electron in the Sinai cavity
vary chaotically as the radius of the central disk is altered, but
they seem to always repel each other and will never cross. Such
behaviour is termed {\em universal}. In \S\ref{sec:1MM} we will see
explicitly how a Coulomb repulsion between the matrix eigenvalues
arises in the one--matrix model.

\subsection{Quantum Choromodynamics}
\label{sec:QCD}
An application of random matrices to quantum choromodynamics (QCD)
was first explored by 't Hooft in 1974 \cite{thooft74} (see also
\S\ref{sec:2Dgrav}). In QCD, interacting gluons can be treated as
$N_{\rm c} \times N_{\rm c}$ chiral matrix fields, where $N_{\rm c}$
is the number of colours. These matrix fields are represented as
``fat Feynman diagrams'' (see \S\ref{sec:ffd}) consisting of double
lines where the oriented edges carry the index structure of the
matrices. By exploring the large $N_{\rm c}$ expansion of QCD, 't
Hooft discovered that the Feynman diagrams are weighted by the
topology of a surface dual to the Feynman diagrams. The power of
$N_{\rm c}$ associated to each graph is the Euler--Poincar\'e
characteristic (\ref{eq:Eulerchar}) of its associated dual surface.
In the limit $N_{\rm c}\rightarrow\infty$, only diagrams of maximal
Euler--Poincar\'e characteristic contribute which translates in the
dual picture to only planar diagrams surviving --- diagrams which
are nonintersecting when drawn on a 2--sphere (see
fig.\ref{fig:QCDdiag}). The seed was thus sown for an entire new
field of research in random matrices as a description of discretised
surfaces.

\begin{figure}[!hc]
\figlabel{fig:QCDdiag}
\begin{center}
\begin{tabular}{cc}
\includegraphics[scale=0.56]{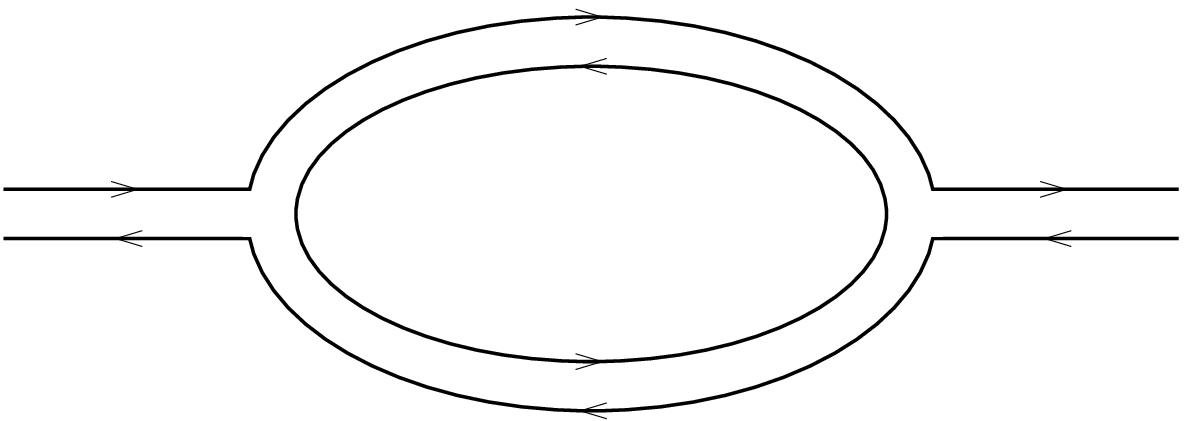}
&
\includegraphics[scale=0.56]{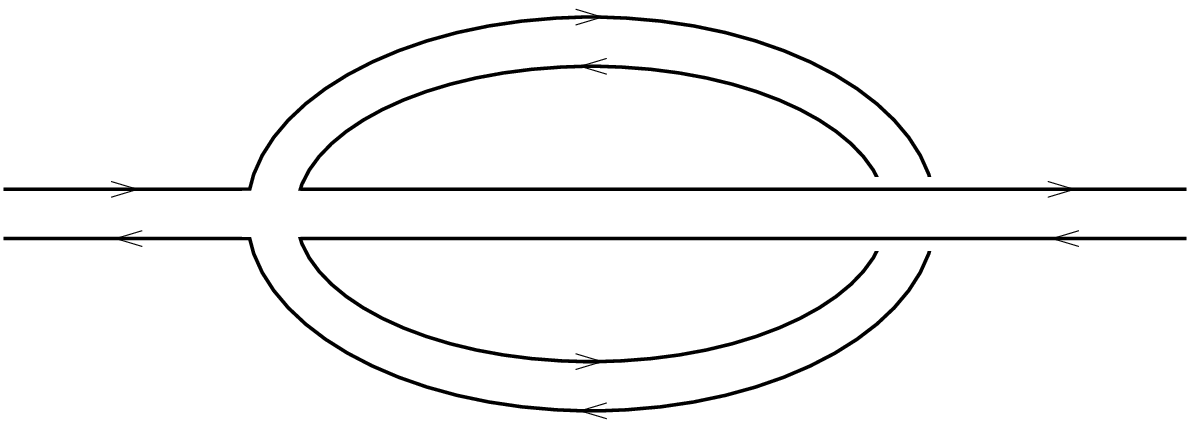}\\
{\bf (a)} & {\bf (b)}
\end{tabular}
\bf \caption{ \rm Fat Feynman graphs of different genera. {\bf (a)}
is planar with $\chi = 2$, {\bf (b)} is non--planar with $\chi=0$
and is suppressed by a factor of $\nf{1}{N_{\rm c}^2}$ relative to
{\bf (a)}. The arrows on the edges indicate the matrix index
structure carried by the graphs.}
\end{center}
\end{figure}

\subsection{String Theory}
Given their capacity to describe discretised surfaces, random
matrices lend themselves well to string theory by describing a
discretised world--sheet manifold. We will take the simultaneous
limit, called the ``double scaling limit'', sending the size of the
matrices to infinity and tuning the coupling constants to certain
critical values. The latter constitutes the continuum limit, where
average number of elementary discrete pieces goes to infinity. The
former limit is necessary in order for the free energy to remain
finite in the continuum limit and for the matrix path integral to be
well defined through its perturbative expansion for certain
potentials. This application, being the subject of this thesis, will
be elaborated in a later section.

\subsection{M Theory}
\label{sec:matM}
It has been argued \cite{Banks,BFSS} that a super Yang--Mills theory
formulated in terms of matrix models (where the matrices in this
case are D0--brane fields) underlies the various string perturbation
expansions which are currently known, and that Super--Poincar\'e
invariant physics arises as a limiting case which reduces to eleven
dimensional supergravity in the low energy domain. This in line with
the purported features of M theory (see \S\ref{sec:Mtheory}).
Further, the matrix model approach furnishes the theory with a
non--perturbative formulation of string theory endowed with much of
the string duality apparent in an explicit Lagrangian formulation.
It may well be that M theory is none other than a theory of random
matrices.


\newpage
\section{Matrix Models}
\label{sec:MM}
A matrix model is defined by
\begin{itemize}
\item A symmetry group $\mathsf{G}$,
\item An ensemble of matrices $M$ equipped with a measure $dM$ invariant under
$\mathsf{G}$,
\item A probability law $\CP(M)$ of the form of a Boltzmann weight
$e^{-N \tr V(M)}$ where $N$ is the size of the matrices and $V(M)$
is some potential.
\end{itemize}
The simplest example of a matrix model is the {\em one matrix model}
which is discussed below and in \S\ref{sec:1MM}.

\subsection{The Partition Function}
\label{sec:partfunction}
Physical properties of a system can be derived from the partition
function. For a canonical ensemble of $N\times N$ matrices $M$, the
partition function is taken to be
\be
\label{eq:1MM}
 Z=\int dM\, \exp\[ -N \tr V(M)\].
\ee
The potential is a polynomial in $M$ with coefficients (couplings)
$g_k$ and the measure $dM$ is given in terms of a product over all
matrix element infinitesimals
\beq
\label{eq:potential}
V(M)&=&\sum\limits_{k>0}\frac{g_k}{k}M^k \\
\label{eq:measure}
dM&=&\prod\limits_{i,j} dM_{ij}.
\eeq
For the theory to be unitary, the coupling constants $g_k$ must be
real.

The ensemble described by \refeq{eq:1MM} can be divided into three
broad classes depending on the symmetry group $\mathsf{G}$ of the
model .
\begin{description}
\item[Gaussian (flat) ensembles:] The eigenvalues of $M$ are in
$\bbR$, the real axis. The symmetry groups in
this class are:\\
\begin{center}
\begin{tabular}{|c|c|c|}
\multicolumn{3}{c}{\textbf{Gaussian Ensemble Symmetry Groups}}\\
\hline\hline $M$ & $\mathsf{G}=$ & Name:\\ \hline
real symmetric & $\gr{O}{N}$ & Gaussian Orthogonal Ensemble (GOE)\\
hermitian & $\gr{U}{N}$ & Gaussian Unitary Ensemble (GUE)\\
quaternionic & $\gr{Sp}{N}$ & Gaussian Symplectic Ensemble (GSE)\\
\hline
\end{tabular}\\
\end{center}
\item[Circular ensembles:] $M$ is a unitary matrix and its
eigenvalues lie on the unit circle in the complex plane. The
unitarity of $M$ preserves some positive norm.
\item[Hyperbolic ensembles:] $M\in SU(N,N)$ preserves a negative
norm. In fact, it is impossible to define a normalizable probability
for this ensemble and their dynamics consist of some stochastic
process described by a diffusion equation, for example.
\end{description}
We will consider only GUEs so our symmetry group elements will be
$\O\in\gr{U}{N}$;
\be
M\rightarrow \O^\dag M\O, \qquad \O^\dag\O = \bbI_{N\times N}
\ee

\subsubsection{The Multimatrix Model and MQM}
The one--matrix model can of course be generalised to any arbitrary
number of matrices. In that case, we could write
\be
Z=\int \prod\limits_k dM_k\,\exp[ -N W(M_1,\ldots,M_n)].
\ee
The potential $W(M_1,\ldots,M_n)$ can be chosen arbitrarily, however
not all choices are integrable\footnote{For a definition of
integrability, see \S\ref{sec:Toda}.}. For example the {\em linear
matrix chain} model given by
\be
\label{eq:linmch}
W(M_1,\dots,M_n)=\sum\limits_{k=1}^{n-1}c_k\tr (M_kM_{k+1})-
\sum\limits_{k=1}^{n}\tr V_k(M_k)
\ee
looses integrability if the chain is closed by the inclusion of the
term $\tr(M_kM_1)$ to the potential (\ref{eq:linmch}).

When the number of matrices is infinite, $n\rightarrow\infty$, the
label $k$ can be replaced by a continuous argument $t$. Matrix
quantum mechanics (MQM) such a model where $t$ is interpreted as
{\em time}. This will be the subject of chapter \ref{ch:MQM}.

\subsection{Fat Feynman Diagrams}
\label{sec:ffd}
Since the partition function \re{1MM} can not normally be evaluated
analytically, we are relegated to calculating it through a
perturbation expansion in the coupling constants. Following the
usual quantum field theory prescription, we write down the
propagator from the quadratic term in the potential
(\ref{eq:potential}):
\beq\label{eq:fattyP}
\raisebox{-10pt}{
\includegraphics[scale=0.5]{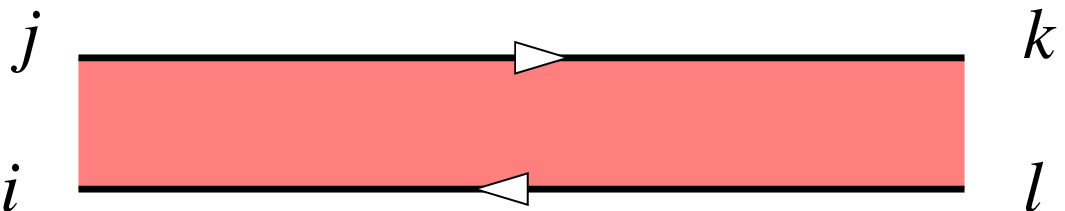}}
&=&
\frac{1}{N g_2}\d_{il}\d_{jk}
\eeq

And from the higher order terms, we get the $k$--vertices:
\beq\label{eq:fattyV}
\raisebox{-32mm}{
\includegraphics[scale=0.4]{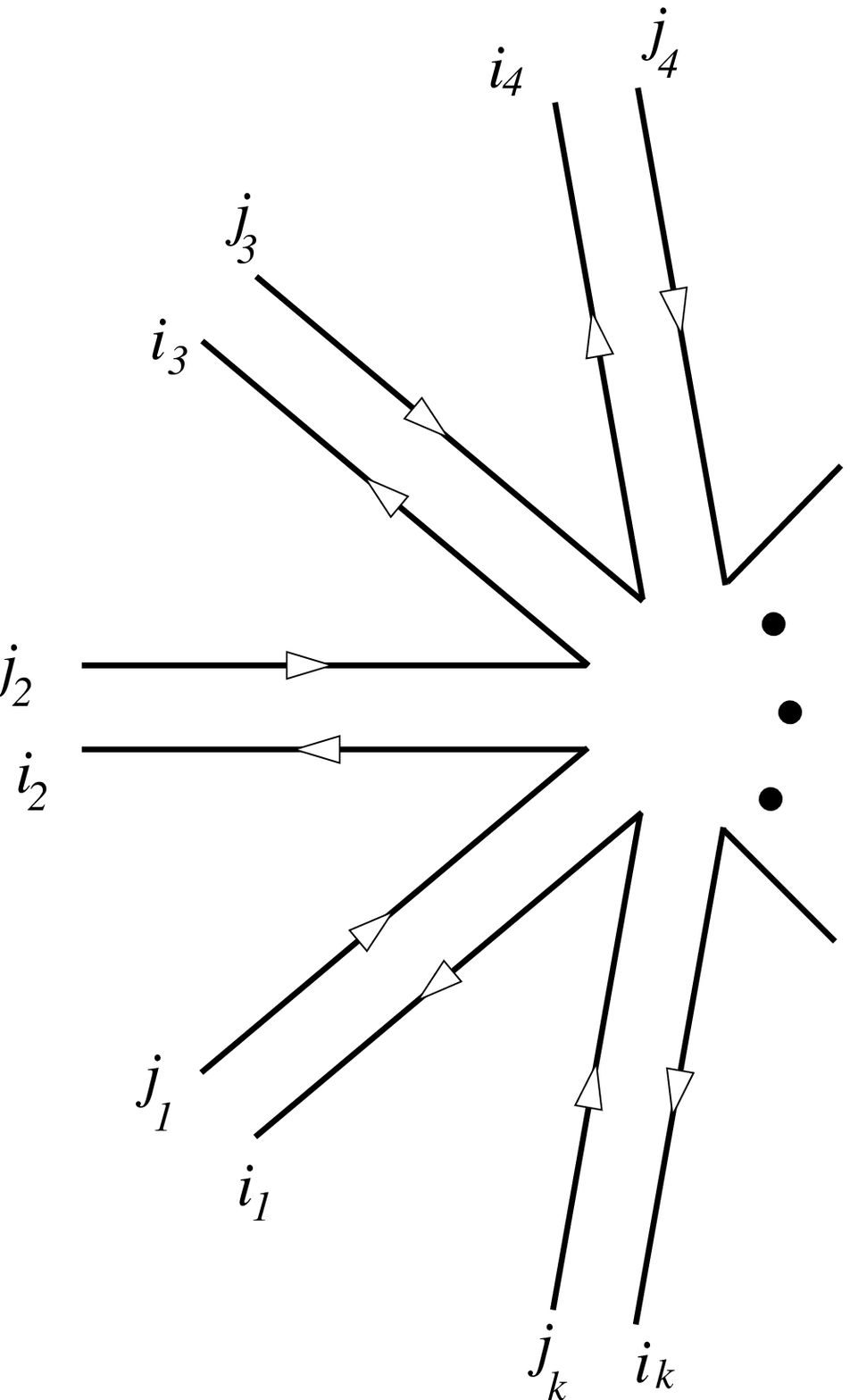}
}
&=&\frac{1}{N g_k} \d_{j_1 i_2} \d_{j_2 i_3} \cdots \d_{j_k i_1}\rule{0pt}{32mm} 
\eeq

Now, the partition function (\ref{eq:1MM}) takes a simpler form when
represented as a sum over diagrams which would most generally be
composed of $n_k$ $k$--vertices, $L$ loops, $P =
\frac{1}{2}\sum\limits_k kn_k$ propagators and $V=\sum\limits_k n_k$
vertices. Given that each of the $L$ loops contributes a factor of
$N$ from the contraction of indices in the Kronecker--$\d$ function,
from (\ref{eq:fattyP}) and (\ref{eq:fattyV}), the contribution from
each diagram to the partition function will be
\be
\label{eq:contribZ}
\frac{1}{s}N^{L+V-P}g_2^{-P}\prod\limits_{k>2} (-g_k)^{n_k}
\ee
where $s$ is a symmetry factor similar to the factor calculated
through combinatorics in quantum field theory to avoid
over--counting of identical diagrams (it will not be of physical
relevance in subsequent calculations).

\subsection{Discretised Surfaces}
\label{sec:descsurf}
There is an orientable discretised Riemann surface dual to each
(fat) Feynman graph as mentioned in \S\ref{sec:QCD}\footnote{If we
use instead real symmetric matrices rather than the hermitian
complex matrices M, the two matrix indices would be
indistinguishable and there would be no arrows in the propagators
and vertices of \re{fattyP} and \re{fattyV}. These non--oriented
vertices and propagators generate an ensemble of both orientable and
non--orientable surfaces such as the Klein bottle \cite{FGZ94}.}.
The correspondence between them is summarised in table
\ref{tab:dual}.

\begin{table}[!hb]\tablabel{tab:dual}
\begin{center}
\begin{tabular}{|l c l|}
\hline
\bf Feynman diagram &$\longleftrightarrow$& \bf Polygonal discretised surface \\
\hline
Propagator &$\longleftrightarrow$& Polygon edge \\
$k$--vertex &$\longleftrightarrow$& $k$--polygon \\
Loop &$\longleftrightarrow$& Surface vertex \\
Symmetry factor $s$ &$\longleftrightarrow$& Order of automorphism groups of surface \\
Number of propagators, $P$ &$\longleftrightarrow$& Number of edges, $P$ \\
Number of $k$--vertices, $n_k$ &$\longleftrightarrow$& Number of $k$--polygons, $n_k$\\
Total number of vertices, $V$ &$\longleftrightarrow$& Total number of faces, $V$\\
Number of loops, $L$ &$\longleftrightarrow$& Number of surface vertices, $L$\\
\hline
\end{tabular}
\bf\caption{\rm Duality between Feynman diagram and discretised surface.}
\end{center}
\end{table}
Thus, to construct this discretised surface we associate to
$k$--vertex a $k$--polygon and to each polygon edge a propagator.
For the one matrix model (\ref{eq:1MM}), this is illustrated in
fig.\ref{fig:surface}.
\begin{figure}[!hc] \figlabel{fig:surface}
\begin{center}
\begin{tabular}{cc}
\includegraphics[scale=0.25]{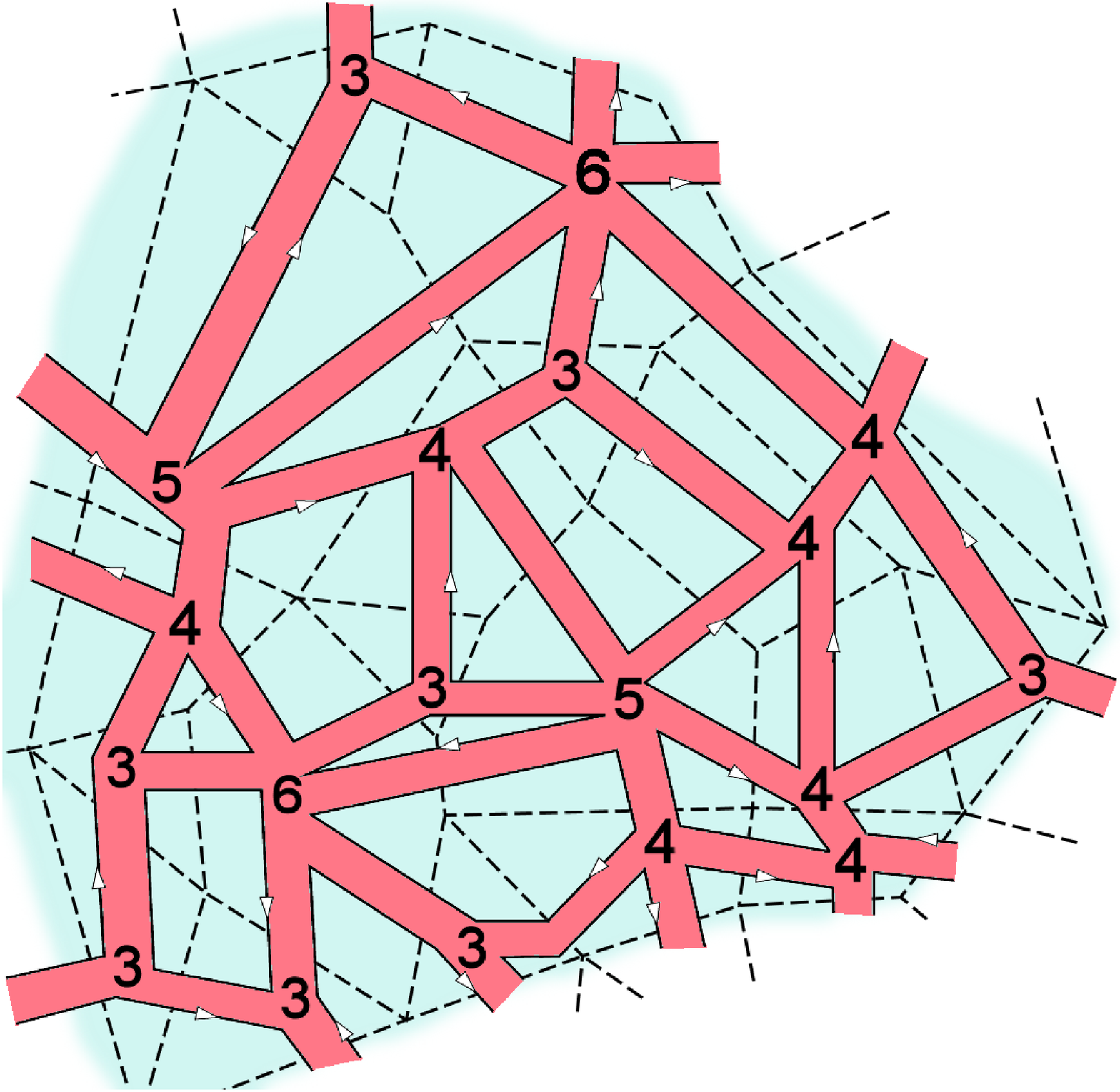} &
\raisebox{120pt}{
\begin{tabular}{|r@{ }l|}
\hline
\raisebox{-9pt}{\includegraphics[scale=0.3]{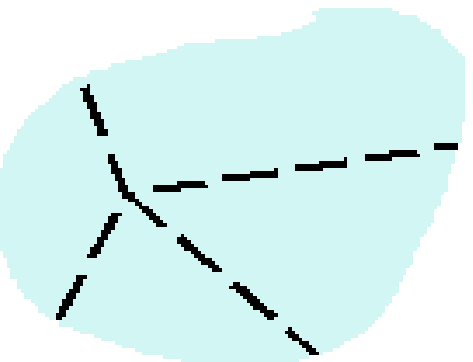}} & Discretised surface\rule{0pt}{8ex}\\
\raisebox{-7pt}[30pt][21pt]{\includegraphics[scale=0.3]{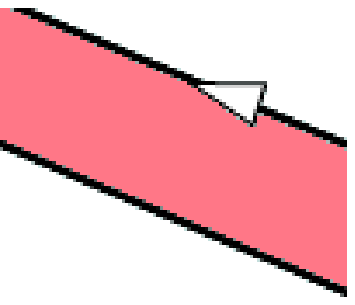}} & Dual Propogator\\
\hline
\end{tabular}
}
\end{tabular}
\bf \caption{ \rm Duality between discretised
surfaces and fat Feynman graphs as summarised in table
\ref{tab:dual}. The numbers indicate the order of the vertex (and
hence the number of sides which the underlying polygon has).}
\end{center}
\end{figure}

For an $\aleph$-matrix model where $\aleph >1$, to each matrix
variable is associated a set of propagators and vertices independent
from those of other matrix variables. In this case, the discretised
surface will carry some additional structure; it is not changed by
the presence of multiple distinct matrices, but it will be instead
dual to several distinct Feynman graph networks all lying atop each
other. Therefore, at each vertex (viz: to each polygon) one may
associate an additional variable which may take on $\aleph$ values.
For example, in the two matrix model, this variable can take two
values, say $\pm1$, yielding an Ising model on a lattice.

Clearly, the type of polygons which are available for use in
discretisation of the surface is controlled by the couplings $g_k$.
The simplest discretisation consists of $3$--polygons (triangles)
corresponding to $3$--vertices and hence a potential
(\ref{eq:potential}) of cubic order in the corresponding matrix
model. A curvature may be present on the surface wherever the edges
of more than two polygons meet. It can be calculated geometrically
from the deficit angle incurred at such a vertex were the polygons
to be laid on a flat surface, following a similar reasoning as the
derivation of the Gauss--Bonnet theorem. For a surface vertex where
the edges of $n_k$
$k$--polygons meet, the curvature is 
\be\label{eq:scalarcurvature}
\CR = 2\pi\frac{2-
\sum\limits_{k>2}\frac{k-2}{k}n_k}{\sum\limits_{k>2}\frac{1}{k}n_k}.
\ee

In the limit where the number of $k$--polygons used for the
discretisation diverges, called the {\em continuum limit}, the
infinitesimal volume element at the $i^{\rm th}$ vertex formed by
the meeting of $n_k^{(i)}$ $k$--polygons is
\be\label{eq:volvertex}
\sqrt{h_i} = \sum\limits_{k>2} \frac{1}{k} n_k^{(i)}
\ee
and the total curvature of the surface becomes a sum over the
vertices
\beq\label{eq:totalcurvature}
\chi &=& \frac{1}{4\pi}\iint\limits_{\rm Surface} \CR dA
=
\frac{1}{4\pi}\sum\limits_i\sqrt{h_i}\CR_i
\nn\\ &=&
\frac{1}{2}\sum\limits_i\left(2 - \sum\limits_{k>2}\frac{k-2}{k}n_k^{(i)}\right) \\
&=&
L-\frac{1}{2}\sum\limits_{k>2} (k-2)n_k
=
L - P + V \nn
\eeq
where $\CR_i$ is the scalar curvature at the $i^{\rm th}$ vertex,
defined by \refeq{eq:scalarcurvature} and $\chi$ is a topological
invariant, namely, the Euler--Poincar\'e characteristic
(\ref{eq:Eulerchar}) of the surface.

So we now see that because of the dualities in table \ref{tab:dual},
the power of $N$ in \refeq{eq:contribZ} gives the Euler--Poincar\'e
characteristic. With $L=$ the number of vertices, $V=$ the number of
faces, and $P=$ the number of edges,
\be
\label{eq:MMp}
L+V-P = \chi = 2-2g,
\ee
where $g$ is the genus of the discretised surface. In the dual
Feynman graph picture, $g$ is the minimal number of intersections
(places where propagators pass under each other) needed to draw the
graph on a 2--sphere.

\subsection{The Topological Expansion}
\label{sec:topexpan}
The partition function is a sum over topologically connected and
disconnected surfaces (diagrams). In quantum gravity, one typically
desires to calculate the free energy $F=\log Z$ since taking the
logarithm of the partition function automatically cancels the
contribution from disconnected surfaces in a similar fashion that
disconnected diagrams are canceled from the free energy by taking
the log of the partition function in quantum field theory. From
\refeq{eq:contribZ} and duality (table \ref{tab:dual}),
\be\label{eq:freeE}
F = \sum\limits_{\substack{\rm Connected\\ \rm diagams}}
\frac{1}{s}N^{L+V-P}g_2^{-P}\prod\limits_{k>2} (-g_k)^{n_k} =
\sum\limits_{\rm Surfaces}
\frac{1}{s}N^{L+V-P}g_2^{-P}\prod\limits_{k>2} (-g_k)^{n_k}
\ee

The relation (\ref{eq:MMp}) provides the option of viewing the
expansion of $F$ as a topological expansion in $N$
\beq\label{eq:topexpan}
F&=& \sum\limits_{g=0}^\infty N^{2-2g}\,F_g \nn\\ &=& N^2
\hspace{0pt}\raisebox{-10pt}{\includegraphics[width=32pt]{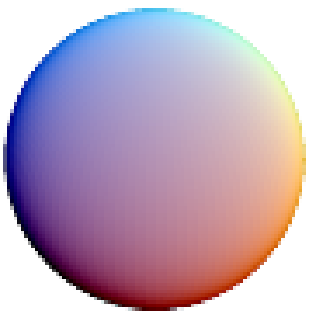}}
\ +\ N^0
\hspace{-0pt}\raisebox{-4mm}{\includegraphics[width=50pt]{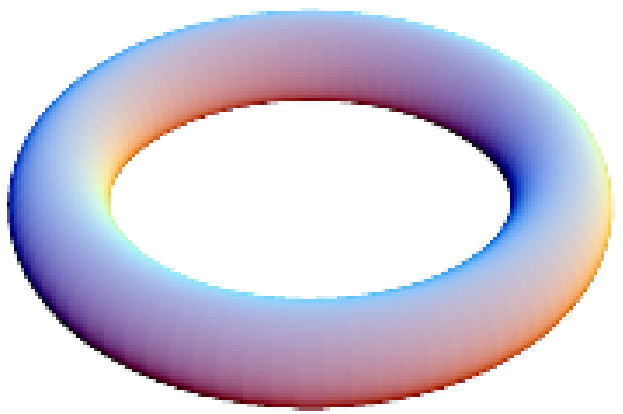}}
\ +\ N^{-2}
\hspace{-0pt}\raisebox{-4mm}{\includegraphics[width=80pt]{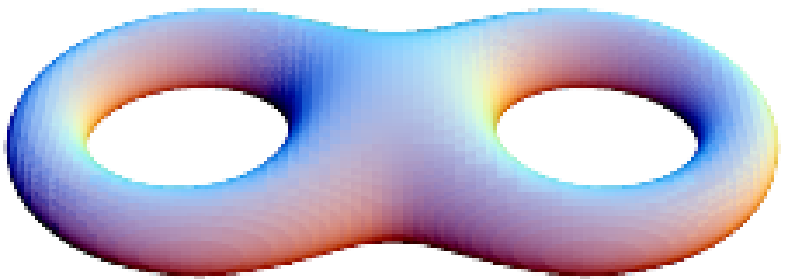}}
\nn\\&&\vspace{-15pt}
\ +\ N^{-4}
\hspace{-0pt}\raisebox{-6mm}{\includegraphics[width=80pt]{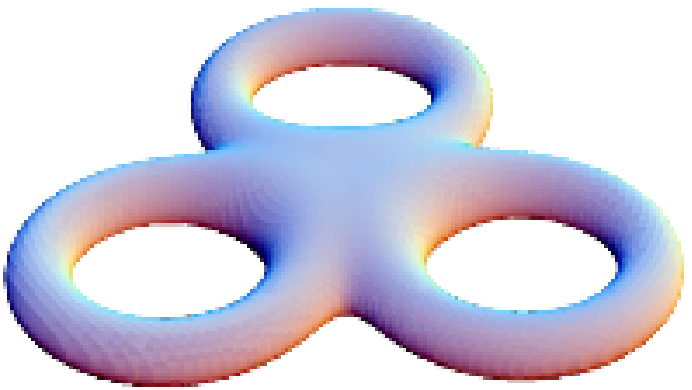}}
+ \cdots
\eeq
where $F_g$ is the free energy of discretised surfaces of genus $g$,
\be\label{eq:freeEg}
F_g = \sum\limits_{\substack{{\rm Discretisations }\\{\rm of\ genus
}\; g}} \frac{1}{s}g_2^{-P}\prod\limits_{k>2} (-g_k)^{n_k}.
\ee

\subsubsection{The Spherical Limit}
For real general couplings or for a potential of odd order which is
unbounded from below, the partition function (\ref{eq:1MM}) is not
well defined. It is nonetheless possible for the integral to be
defined through its perturbation expansion in $\frac{1}{N}$ if we
take the limit $N\rightarrow\infty$. From the expansion
(\ref{eq:topexpan}) we see that, in this limit, the genus zero
surfaces dominate. Hence this is termed the {\em spherical limit}
\cite{FGZ94,GinMoore,Brezin78}. It corresponds to keeping only
planar diagrams as explained in \S\ref{sec:QCD} and
fig.\ref{fig:QCDdiag} and is also occasionally referred to the {\em
planar} limit.

\subsection{The Continuum Limit}
\label{sec:contL}
The continuum limit describes the case where the number polygons
used in the discretisation diverges. This is achieved by fine tuning
the coupling constants $g_k$ in the potential (\ref{eq:potential}).
Indeed, from (\ref{eq:freeEg}), we see that the total number $n_k$
of $k$--polygons used in the discretisation of a genus $g$ surface
is
\be
\langle n_k \rangle = g_k \frac{\p}{\p g_k} \log F_g
\ee
Now, $F_g$ may itself be expanded in a perturbation series in the
coupling $g_k$. This expansion is decidedly nontrivial to calculate
but has been found \cite{Bessis}. For example, in the case of a
cubic potential where $g_2 = 1, g_3 =\l\neq 0$ and all other $g_k$
zero in (\ref{eq:potential}), viz.
\be
V(M) = \frac{1}{2} M^2 + \frac{\l}{3}M^3,
\ee
asymptotically this perturbation expansion looks like
\be\label{eq:Fg}
F_g \sim F_g^{\rm(reg)} + (g_c - g_3)^{\frac{\chi}{2}(2-\g_{\rm
str})},
\ee
where the {\em string susceptibility}, $\g_{\rm str}$, is a critical
exponent defining critical behaviour of the corresponding continuum
theory \cite{KazakovMigdal}, $g_c$ is a critical value for the
coupling constant and $F_g^{\rm(reg)}$ is a non--universal (i.e.
regular) contribution which dominates for the $g=0$ spherical
topology. With $F=\sum\limits_{g=0}^{\infty} N^\chi F_g$ from
(\ref{eq:topexpan}), the number of triangles (3--polygons), to the
$j^{\rm th}$ power is
\be
\langle n_3^j \rangle = \left(g_3 \frac{\p}{\p g_3}\right)^j\log F.
\ee
For a given $\g_{\rm str}$, we must choose a $j>1$ such that the
finite non--universal contribution $F_g^{\rm(reg)}$ does not
dominate $F_g$ and we then obtain
\be
\frac{\langle n^j_3 \rangle}{\langle n^{j-1}_3 \rangle} \sim
\frac{1}{g_{\rm c} - g_3}.
\ee
We take this quantity as a measure of the number of triangles in the
discretisation rather then simply $\langle n_3 \rangle$ since it is
dimensionally the same as $\langle n_3 \rangle$ and reveals the
universal behaviour associated to taking the continuum limit: it
shows that in the limit $g_3 \rightarrow g_{\rm c}$, the sum
(\ref{eq:topexpan}) is dominated by discretisations with divergent
number of polygons --- the discretisation becomes continuous.

\subsection{Double Scaling Limit}
\label{sec:DSL}
We have seen so far that in the spherical limit where
$N\rightarrow\infty$, we get merely genus zero surfaces in the
perturbative expansion of the free energy. Thus, if the discretised
surface under consideration is a string world--sheet, interactions
of strings could not be dealt with as they involve a change of
genus. Worse, the universal part of the free energy vanishes or
diverges for some genera $g$ when taking the continuum limit
$g_k\rightarrow g_{\rm c}$ depending on whether $\g_{\rm str}$ is
less than or greater than 2, respectively. Fortunately, it is
possible to take the spherical ($N\rightarrow\infty$) and continuum
($g_k\rightarrow g_{\rm c}$) limits {\em simultaneously} in such a
way that topological information encoded in the string world--sheet
is not lost.

Analogously to \refeq{eq:topexpan}, there is a perturbative
topological expansion in string theory where each term is weighted
by the string coupling constant $\k$ (see
\S\ref{sec:stringcoupling}). For example, the diagrams contributing
to the self--energy for closed strings are
\be\label{eq:strselfenergy}
\raisebox{-5pt}{\includegraphics[scale=0.1]{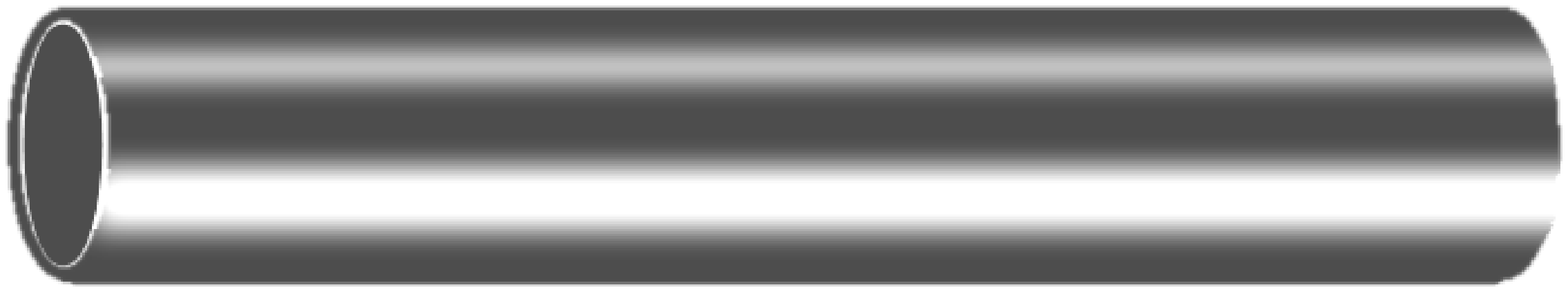}}\hspace{5pt}+
\k_{\rm cl}^2\;
\raisebox{-20pt}{\includegraphics[height=45pt,width=80pt]{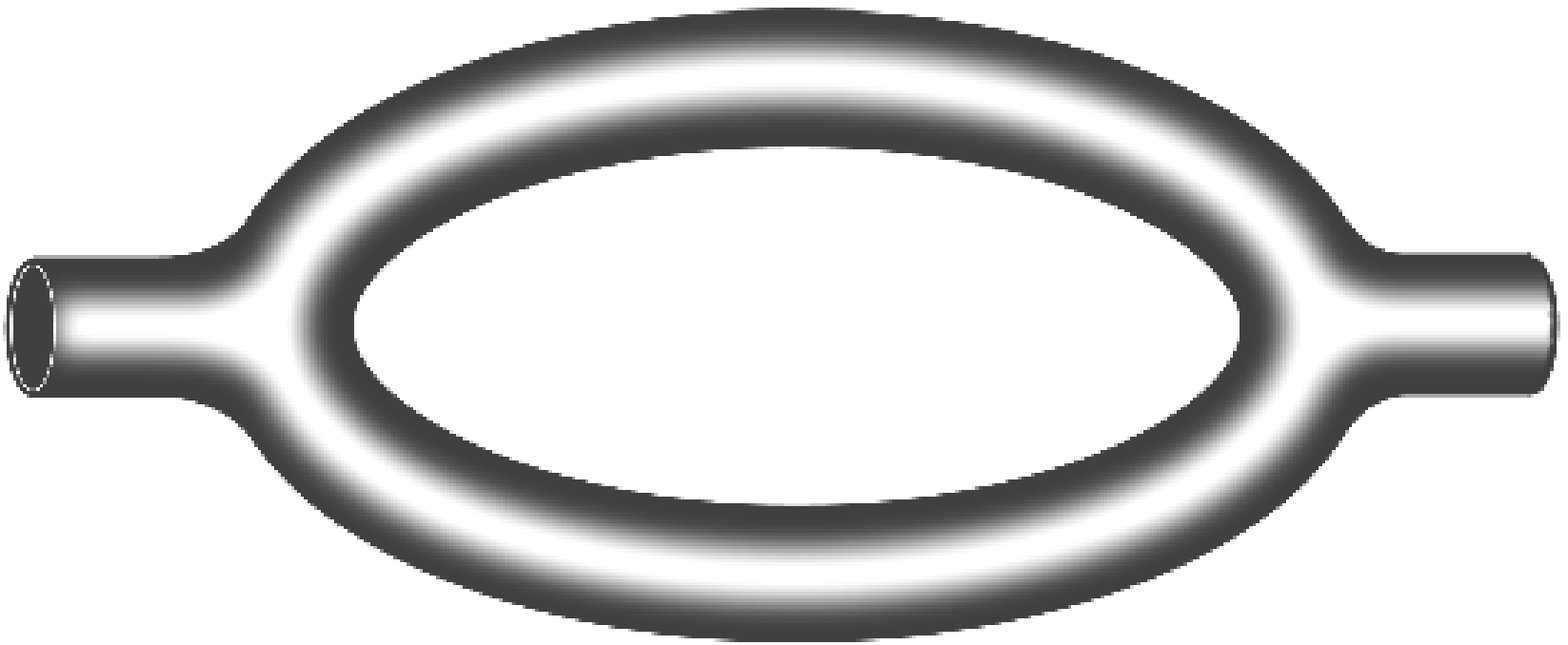}}\hspace{5pt}
+ \k_{\rm cl}^4
\raisebox{-19pt}{\includegraphics[height=45pt]{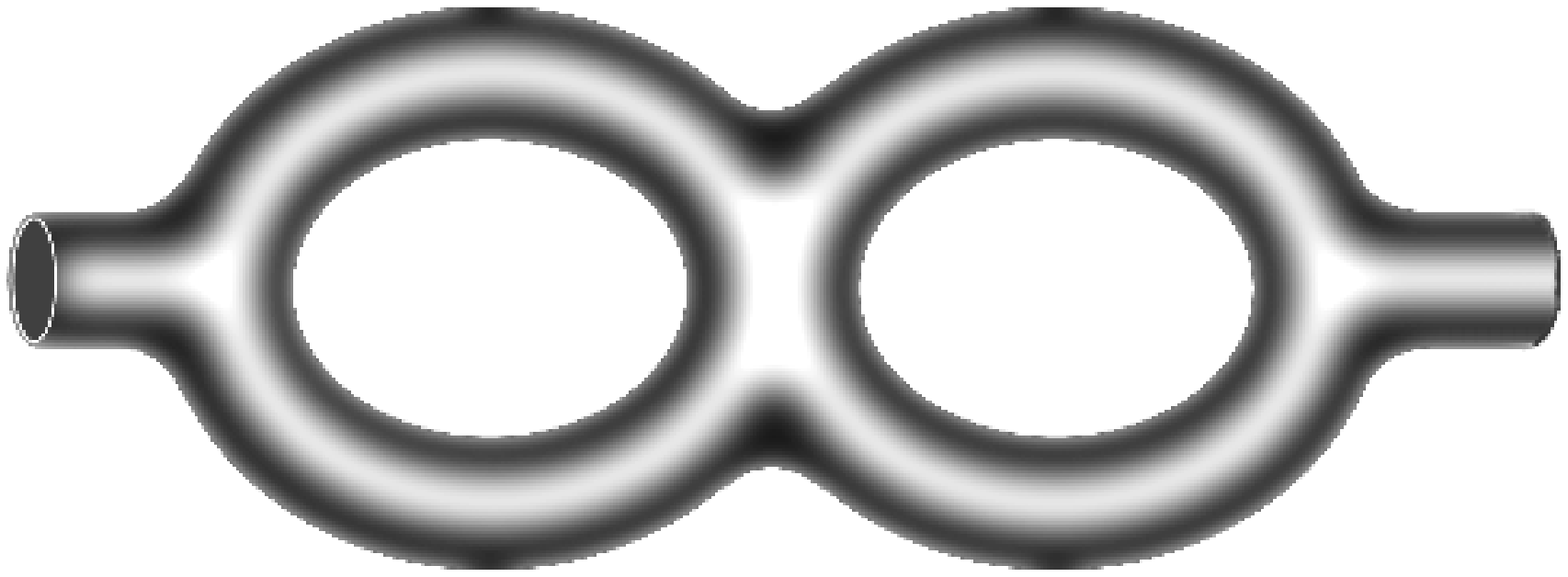}}+\hspace{5pt}
\cdots .\nn
\ee
In a random matrix model, where each diagram is weighted by a factor
$N^\chi$, it is tempting to make the association $N^\chi =
\k^{-\chi}$. But then in the spherical limit, we would have
$\k\rightarrow 0$. Instead, we fix
\be
\k^{-1} = N(g_{\rm c} - g_k)^{(2-\g_{\rm str})/2}
\ee
and take the limit $N\rightarrow\infty$ and $g\rightarrow g_{\rm c}$
while keeping $\k$ constant. Then we can write perturbation
expansions in string theory with contributions from manifolds of all
genera
\be\label{eq:genfunc}
F=\sum\limits_{\substack{{\rm Genus}\;g\\{\rm world-sheets}}}
\k^{-\chi} F_g,
\ee
$F_g$ being the generating function or free energy for a
diagrammatic contribution of genus $g$.

The generating function (\ref{eq:genfunc}) is the partition function
of 2D quantum gravity since, in two dimensions, the sum over all
topologies in \re{formalZ} can be regarded as a genus summation over
Riemann surfaces. We will see that, where the number of matrices in
a multimatrix model tends to infinity and we obtain MQM (see
\S\ref{sec:partfunction}), the double scaling limit yields 2D string
theory.


\subsubsection{Universality}
Note that generating function (\ref{eq:genfunc}) is independent of
the matrix coupling constants $g_k$ and hence independent on the
form of the potential (\ref{eq:potential}) or the type of polygons
used in the discretisation.

This remarkable fact is a central feature of matrix models related
to the large $N$ limit, namely, {\em universality}: in this limit
the distribution of eigenvalues tends towards some universal
distribution dependant solely on the symmetry properties of the
system rather than the details of the initial probability law. This
conjecture had been borne out in many matrix models, as in our case,
yet it remains to be proven in the general case.

Matrix models can be stratified into universality classes, each
class corresponding to a distinct physical system, which in our
context would correspond to different CFTs (see \S\ref{sec:u}).

\newpage
\section{One--Matrix Model}
\label{sec:1MM}
There are several methods of solving for the spectral distribution
and various other observables in matrix models. We present the
saddle point method here as it will serve to illustrate several
important features of matrix models such as the necessity of the
large $N$ limit and universality.

\subsection{Reduction to Eigenvalues}
In the one matrix model, the hermitian matrix is diagonalised
\be
M = \O^\dag x \O,\qquad x=\diag(x_1,\ldots,x_N),\quad \O^\dag\O =
\bbI,\quad x_i \in \bbR.
\ee
This produces a Jacobian in the matrix integral measure $dM$
\re{measure}. Since $M$ is Hermitian, its diagonal elements are real
and the only independent elements are $\lb M_{ij}\rb_{i\leq j}$.
Thus, it can be decomposed as
\be\label{eq:1MMmeasure}
dM = \prod\limits_{i,j}dM_{ij}= \prod\limits_i dM_{ii}
\prod\limits_{i<j} \Re(dM_{ij})\Im(dM_{ij}).
\ee
Now, consider the infinitesimal unitary transformation and small
variation in $x$
\be
\O=\bbI+d\o\Rightarrow\O^\dag=(\bbI+d\o)^{-1}=\bbI-d\o,\qquad dx =
\diag\lb dx_1,\ldots,dx_N\rb
\ee
The impact of this on $M$ is
\bea{r\eq l}{}
dM_{ij}&\O^\dag dx\O\\& dx_i\d_{ij} + \[x,d\o\]_{ij} + \order(d\o^2)\\
& dx_i\d_{ij} + (x_i - x_j)d\o_{ij}  + \order(d\o^2).
\eea
We substitute this in \re{1MMmeasure} and obtain, to first order in
$d\o$,
\be\label{eq:1dM}
dM=[d\O]_{\gr{SU}{N}}\prod\limits_{i=1}^{N}dx_i\D^2(x)
\ee
where $[d\O]_{\gr{SU}{N}}$ is the Haar
measure\footnote{\parbox[t]{\textwidth}{The left (right) Haar
measure $[d_{\rm L(R)}x]_{\mathsf{G}}$ is an infinitesimal volume
element of $\mathsf{G}$ under which the integral over the whole
group of some continuous function $f$ is invariant under left
(right) group multiplication,
$$
\int\limits_{\mathsf{G}}
[d_{\rm L}x]_{\mathsf{G}} f(cx) =
\int\limits_{\mathsf{G}}
[d_{\rm L}x]_{\mathsf{G}} f(x), \qquad c\in\mathsf{G}
$$
E.g: for $\mathsf{G}=\gr{U}{1}=\lbrace e^{i\phi}\rbrace$ the Haar
measure is $d\th$ since $\int_0^{2\pi} d\th f(\phi+\th)=
\int_0^{2\pi} d\th f(\th) $. A group $\mathsf{G}$ is called
`unimodular' if the left and right Haar measure coincide, all
compact linear Lie groups are unimodular. It is clear that for
consistency with the definition of integration, we must use the Haar
measure when integrating over the angular variables.}} on
$\gr{SU}{N}$ and
\be
\label{eq:Vandermonde}
\D(x) = \prodl_{i<j}(x_i - x_j)
=\prodl_{i=1}^{N}\(\prodl_{j=i+1}^{N}(x_i - x_j)\)
=\det\limits_{i,j}(x_i^{j-1})
\ee
is the Vandermonde determinant\footnote{E.g. for $N=3$, $\D(x) =
\prod\limits_{i<j}(x_i - x_j)=(x_3 - x_2)(x_3 - x_1)(x_2 - x_1)
=\det
\(\begin{array}{ccc}1&x_1&(x_1)^2\\1&x_2&(x_2)^2\\1&x_3&(x_3)^2\\\end{array}\)$.}.

Substituting this into the partition function \re{1MM}, and using
the fact that the potential \re{potential} is a polynomial in $M$
and is hence $\gr{U}{N}$ invariant, the $\O$ unitary matrices (or
``angular degrees of freedom'') decouple and can be integrated out
\beq
Z
&=&\int\!dM\;\exp\[ -N \tr V(M)\] \nn\\
&=&\int\![d\O]_{\gr{SU}{N}}\;
   \int\!\prod\limits_{i=1}^{N}dx_i\;
   \D^2(x)\exp\left( \sum\limits_{k>0}\frac{g_k}{k}\tr(x^k)
   \right)\nn\\
&=&\vol(\gr{SU}{N})
   \int\!\prod\limits_{i=1}^{N}dx_i\;
   \D^2(x)\exp\left( \sum\limits_{i=1}^N V(x_i)\right)\\
&=&\vol(\gr{SU}{N})
   \int\!\prod\limits_{i=1}^{N}dx_i\;    e^{-N \CE},
\eeq
where
\be\label{eq:CE}
\CE= \sum\limits_{i=1}^N V(x_i) - \frac{2}{N}
   \sum_{i<j}\log |x_i - x_j|.
\ee
The group volume $\vol(\gr{SU}{N})$ is a function of $N$ only
\cite{Tilma02} and is irrelevant insofar as the statistics of
eigenvalues are concerned (it does affect the dependance of the free
energy on $N$, but as we will always work in the spherical limit
this is not an issue). Thus, the problem of solving a system of
$N^2$ degrees of freedom is reduced to a problem of solving for $N$
eigenvalues. Models allowing such a {\em reduction to eigenvalues}
are integrable.

\subsection{Saddle Point Method}
In the limit $N\rightarrow\infty$, we can use the saddle point
method (also known as the ``stationary phase approximation'') to
evaluate the partition function, where we include only the dominant
contribution arising from the eikonal trajectory which satisfies the
classical equation of motion $\der{\CE}{x_i}=0$. This yields the $N$
algebraic equations governing the distribution of eigenvalues in the
spherical limit
\be\label{eq:alV}
\der{V(x_i)}{x_i}=\frac{2}{N}\sum_{i\neq j}\frac{1}{x_i-x_j}
\ee
Omitting the log term in \re{CE}, we would have obtained instead
$\der{V(x_i)}{x_i} = 0$, indicating that all the eigenvalues are
concentrated at the local minima of the potential. However, we see
that the log term is responsible for a Coulomb repulsion between the
eigenvalues, spreading then around the minima as illustrated in fig.
\ref{fig:saddle}.
\begin{figure}[!hc]
\figlabel{fig:saddle}
\begin{center}
\includegraphics[scale=0.5]{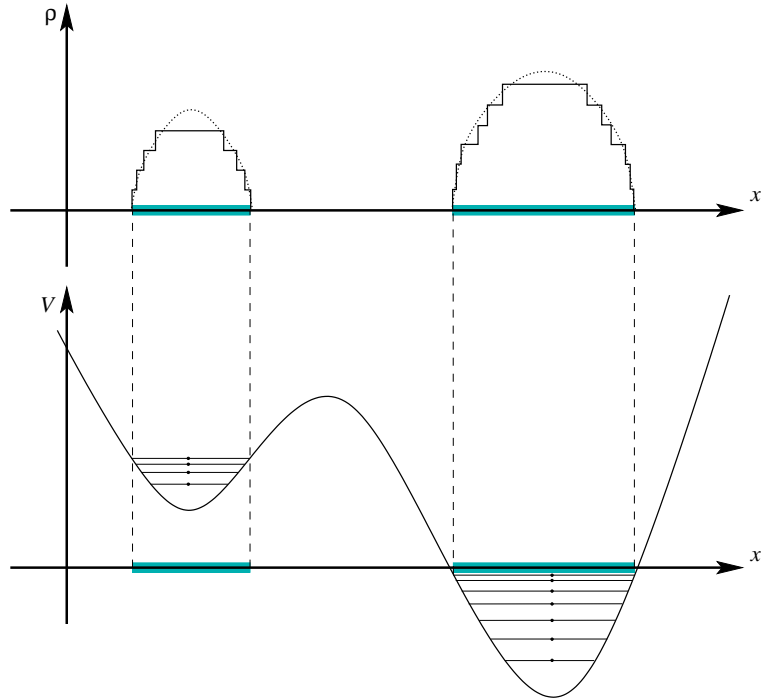}
\bf \caption{ \rm The potential on the lower plot, showing the
eigenvalues distributed around the minima. The eigenvalue density
$\r(x)$ is shown above it, where the dotted line represents the
limit $N\rightarrow\infty$. The shaded regions on the $x$ axes are
the support of $\r$.}
\end{center}
\end{figure}


We can define an eigenvalue density function in the large $N$ limit
\be
\r(x)=\frac{1}{N}\la\tr\d(x-M)\ra,
\ee
where the factor $1/N$ was introduced so that
\be\label{eq:norc}
\int\!dx\;\r(x)=1.
\ee
The disjoint intervals over which it is nonzero form its support,
denoted $\dom(\r)$. Solving for the eigenvalue density is a central
problem as we will then be able to obtain the free energy in the
spherical approximation
\beq
F_0&=&\lim\limits_{N\rightarrow\infty}\frac{1}{N^2}\log Z\nn\\
&=&-\int\!dx\r(x)V(x)+\iint\!dx\,dy\;\r(x)\r(y)\log|x-y|.\label{eq:fr0}
\eeq

Now we can rewrite the $N$ equations in \re{alV} as an principle
value integral equation in the spherical limit
\be\label{eq:pvV}
\der{V(x)}{x}= 2\pv\int\!dy\;\frac{\r(y)}{x-y}.
\ee
where
\be
\pv\int\!dx\,f(x)\equiv\lim\limits_{\eps\rightarrow0^+}
\(\int\limits_a^{x_0-\eps} +  \int\limits^b_{x_0+\eps}\)\!dx\,f(x),
\ee
is the Cauchy principle value for a function $f(x)$ unbounded at
$x=x_0\in(a,b)$. The ``resolvent function'' of the matrix $M$ is
\be\label{eq:resolvant}
\o(z)=\frac{1}{N}\lla\tr\frac{1}{z-M}\rra=\int\!dx\;\frac{\r(x)}{z-x},\qquad
z\in\bbC,
\ee
which is analytic except when $z$ belongs to $\dom(\r)$ where the
eigenvalues reside. From distribution theory,
\be
\lim\limits_{\eps\rightarrow0}\frac{1}{x+i\eps}=\pv\(\frac{1}{x}\)-\pi
i\d(x)
\ee
which shows that the discontinuity is
\be\label{eq:discon}
\o(x+i0)-\o(x-i0)=-2\pi i\r(x),\qquad x\in\dom(\r).
\ee
On the other hand,
\be\label{eq:osol}
\o(x+i0)+\o(x-i0)=\der{V(x)}{x},\qquad x\in\dom(\r).
\ee

When there is only one cut in which the eigenvalues
reside\footnote{The generalisation to several cuts, as in fig.
\ref{fig:saddle}, is relatively straightforward but will have no
bearing on the conclusions of this section.}, say $\r(x)\neq0\
\forall\ x\in(a,b)$ as illustrated in fig \ref{fig:onecut}, one can
establish the general form of the resolvent from the inhomogeneous
linear equation \re{osol} as the sum of a particular solution
$\o_{\rm p}(z)= V'(z)/2$ and the homogeneous solution $\o_0(z)$
which solves $\o_0(z+i0) + \o_0(z-i0)=0$. We write it as
\be
\o_0(z+i0) = -\hf P(z)\sqrt{(z-a)(z-b)}
\ee
where $P(z)$ is an entire function, i.e. a polynomial in $z$. This
simplifies the problem to one of determining coefficients of $P(z)$
in the general solution
\be
\o(z) = \hf\[\der{V(z)}{z} - P(z)\sqrt{(z-a)(z-b)}\].
\ee
It must obey the asymptotics of $\o(z)$ in the large $z$ limit as
derived from \re{resolvant}
\be
\o(z)\sim\frac{1}{z},\qquad z\rightarrow\infty,
\ee
and the normalisation condition, \re{norc}, so
$P(z)\sqrt{(z-a)(z-b)}$ should cancel all non--negative powers of
$\der{V(z)}{z}$. If $V(z)$ is a polynomial of degree $n$ in $z$,
say, then $V'(z)$ is of degree $n-1$ and $P(z)$ should thus be a
polynomial of degree $n-2$.
\begin{figure}[ht]
\figlabel{fig:onecut}
\begin{center}
\includegraphics[scale=0.55]{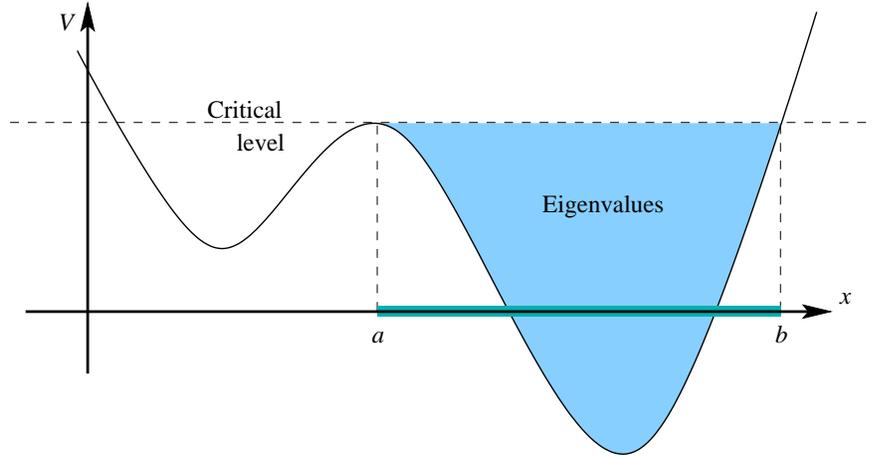}
\bf \caption{ \rm One cut approximation: All the eigenvalues reside
in one of the potential's minima, in the interval bounded by $a$ and
$b$. A critical point occurs when the highest filled level reaches
an extremum in the potential.}
\end{center}
\end{figure}

Then, from \re{discon} we get the density of eigenvalues
\be\label{eq:rhox}
\r(x)=\frac{1}{2\pi}P(x)\sqrt{(x-a)(b-x)}.
\ee
Note that this makes sense only if $P(x)$ is positive which follows
from the normalisation condition \re{norc}.

Through \re{fr0}, we can get the free energy
\beq
F_0&\stackrel{{\rm\footnotesize\re{pvV}}}{=}&-\int\!dx\;\r(x)V(x) + \hf\int\!dx\;\der{V(x)}{x} \nn\\
&=&-\hf\int\!dx\;\r(x)V(x)\nn\\
&=&-\frac{1}{4\pi}\int_a^b\!dx\;P(x)V(x)\sqrt{(x-a)(b-x)}.\label{eq:FE1MM}
\eeq

\subsection{Universal Behaviour}
\label{sec:u}
Universal behaviour can be seen from \re{rhox}, where the eigenvalue
density has square root singularities at the ends of the support
independently of the form of the potential.

There is an exception to this when the polynomial $P(z)$ vanishes at
the edges. In that case, $\r(x)\sim(x-a)^{m+\nf{1}{2}}$ where
$P(a)=0$ and $m$ is the order of the root. This can occur when the
highest filled eigenvalue reaches the top of an extremum of the
potential as illustrated in fig. \ref{fig:onecut}. Any higher than
this, and the one cut assumption would no longer be valid.

The critical configurations correspond to different universality
classes (\S\ref{sec:DSL}) which, in turn correspond to the continuum
theories associated to different minimal conformal field theories
coupled two--dimensional gravity. The latter are characterised by
two relatively prime numbers $p$ and $q$ which related to the
central charge and string susceptibility \cite{Kazakov89} by
\be
c=1-6\frac{(p-q)^2}{pq},\qquad\g_{\rm str}=-\frac{2}{p+q-1}.
\ee

In our case, critical points occur when the couplings $g_k$ of the
potential approach their critical values (see \S\ref{sec:contL}). In
this situation, the expressions for the free energy \re{FE1MM} and
\re{Fg} can be shown to be equivalent \cite{Belavin84}, with
\be\label{eq:asss}
\chi=2, \qquad \g_{\rm str}=\frac{1}{m+1}.
\ee
whence we get $q=2$ and $p=2m+1$ leading to
\be
c=1-3\frac{(2m-1)^2}{2m+1}.
\ee
In the case $m=1$ the central charge vanishes and we get pure
two--dimensional quantum gravity.

\subsection{Other Methods}
There are several other methods for finding the solutions of matrix
models including the method of loop equations, the supersymmetric
Efetov's method applicable only for Gaussian potentials, the
character expansion method and the renormalization method.

Another more common method worth mentioning here is the {\em
orthogonal polynomials method}, where the partition function
(\ref{eq:1MM}) is rewritten in terms of polynomials orthogonal with
respect to its measure. The problem of calculating the partition
function thus reduces to finding the exact form of the orthogonal
polynomials. The solution often involves a large system of finite
difference equations which may be reduced to a recursion relation of
differential equations, called a {\em hierarchy} and the eigenvalues
often behave as a system of free fermions. In fact, the existence of
a representation in terms of free fermions is a sign of
integrability. In fact, MQM turns out to be similar to a particular
two matrix model which is solvable by the method of orthogonal
polynomials. Its action is
\be
\label{eq:2MM}
Z = \int d\MA d\MB\,e^{-N\tr[\MA \MB -V(\MA)-\tlV(\MB)]},
\ee
$V(\MA)$ and $\tlV(\MB)$ being potentials of the same form as
(\ref{eq:potential}), and the measures $d\MA$, $d\MB$ being as in
(\ref{eq:measure}). As we deal with GUEs, both matrices are
diagonalisable, with unitary angular variables $\O_{\rm A}$ and
$\O_{\rm B}$,
\be
\MA = \O^\dag_{\rm A} x \O^{\phantom{\dag}}_{\rm A},\qquad \MB
=\O^\dag_{\rm B} y \O^{\phantom{\dag}}_{\rm B},\qquad \O^\dag_{\rm
A} \O^{\phantom{\dag}}_{\rm A}= \O^\dag_{\rm B}
\O^{\phantom{\dag}}_{\rm B} = \bbI
\ee
with $x$ and $y$ being diagonal $N\times N$ matrices with
eigenvalues of $\MA$ and $\MB$ along the diagonal, respectively.
Given that the action in (\ref{eq:2MM}) is invariant under the
common unitary transformation
\be
\MA \mapsto \O^\dag \MA \O,\qquad \MB \mapsto \O^\dag \MB \O,\qquad
\O^\dag\O = \bbI,
\ee
we may integrate out one of $\O_{\rm A}$ or $\O_{\rm B}$ and we are
left with a partition function involving one of the unitary angular
matrices. It is contained in an Itzykson--Zuber integral \cite{IZ},
the solution of which is known. The orthogonal polynomials method
can then be used to calculate free energy, see
\cite{SergePhD,GinMoore,Gerasimov}. The solution involves solving a
hierarchy of differential equations, the Toda hierarchy in this
case.

With this in mind, we now briefly review some of the results of the
Toda lattice hierarchy.

\newpage
\section{Toda Lattice Hierarchy}
\label{sec:Toda}
In 1967, M. Toda considered a one--dimensional crystal with
nonlinear couplings between nearest neighbour atoms. It was
described by a system of coupled equations of motion
\be\label{eq:todacrystal}
\frac{m}{a}\frac{d^2r_n}{dt^2}= 2e^{-r_n} - e^{-r_{n-1}} -
e^{-r_{n+1}}
\ee
where, writing $s_n(t)$ as the displacement of the $n^{\rm th}$ atom
of mass $m$ from its equilibrium position, $r_n=s_{n+1}-s_{n}$ and
$a$ is some constant. As this is a nonlinear differential equation,
the principle of superposition (ie: that new solutions can be formed
as a linear combination of known solutions of the equation) does not
apply. On the other hand, (\ref{eq:todacrystal}) is ``completely
integrable''.

The relevance of the Toda lattice hierarchy for us is that we will
consider compactified 2D string theory chapter \ref{ch:TAC} on a
where the momenta take on only discrete values as in
\S\ref{sec:pertCFT}, in other words, we have a system on momentum
lattice instead of a crystal lattice. Firstly, we shall need some
general properties of integrable systems.

\subsection{Integrable Systems and Hierarchies}
\label{sec:integr}
A mechanical system with $D$ degrees of freedom,
\be
\frac{dq_i}{dt} = \frac{\p H}{\p p_i},\quad
\frac{dp_i}{dt} = -\frac{\p H}{\p q_i},\quad
\forall i=1,\ldots,D
\ee
is {\em completely integrable} if there exist $D$ conserved,
independent first integrals such that the Poisson bracket of any two
of them vanishes,
\beq\label{eq:firstint}
F_i(q,p) &=& C_i,\quad i = 1,\ldots,D \\
\lbrace F_i,F_j\rbrace_{\rm P.B.}&=&0,\quad i\neq j. \nn
\eeq
We could set $C_1 = H(q,p)$, for example. In fact, the first
integrals $F_i$ may be regarded as a set of commuting Hamiltonians.
Since they commute, to each there is an associated conserved
quantity $C_i$. The two matrix model mentioned above is an example
of an integrable system; explicit calculation by method of
orthogonal polynomials demonstrates that there exists a Toda lattice
description \cite{SergePhD}.

The equations of motion of an integrable system usually form a {\em
hierarchy}; they are usually solved recursively, the solution of one
being substituted into the next, as in finite--difference equations like
(\ref{eq:todacrystal}) describing a system on a lattice.

Integrable systems are classified in terms of the hierarchy they
generate. The hierarchy of interest here with sufficient generality
is the Toda hierarchy. In limiting cases, this hierarchy reduces to
the KdV and KP hierarchies. One way to formulate the Toda hierarchy
that will prove particularly useful is through the Lax formalism.

\subsection{Lax Formalism}
\label{sec:Lax}
Consider an infinite one--dimensional lattice whose nodes are
labeled by a discrete variable $s$ as in fig.\ref{fig:lax}. The
shift operator in $s$ is $\ho = e^{\hbar\nf{\p}{\p s}}$ where
$\hbar$ plays the role of the lattice spacing.
\begin{figure}[!h] \figlabel{fig:lax}
\centering\includegraphics[scale=0.6]{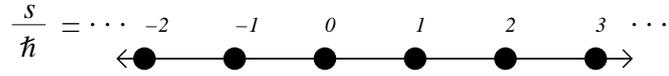} \bf \caption{ \rm
Infinite one--dimensional lattice.}
\end{figure}

The commutation relation between $\ho$ and $s$ is
obviously
\be\label{eq:comm}
[\ho,s]=\hbar\ho.
\ee

The {\em Lax operators} are operators existing on the lattice given
by two semi--infinite series
\bea{r@{\,=\,}c@{\,+\,}l}{eq:Lax}
L  & r(s,t)\ho     &\sum\limits_{k=0}^{\infty}   u_{k}(s,t)\ho^{-k}, \\
\bL& r(s,t)\ho^{-1}&\sum\limits_{k=0}^{\infty} \bu_{k}(s,t)\ho^{k},
\eea
where the coefficients $r(s,t)$, $u_{k}(s,t)$ and $\bu_{k}(s,t)$ are
functions of the lattice node parameter $s$ and an infinite set of
``times'' $t=\lbrace t_{\pm n} \rbrace_{n=1}^{\infty}$. The system
represents a Toda hierarchy if each time evolution is generated by
the Hamiltonians $H_{\pm n}$

\be\label{eq:LaxHams}
H\phm =(L^n)_> +\hf(L^n)_0,\qquad H_{-n}=(\bL^n)_<
+\hf(\bL^n)_0,\quad n>0,
\ee
where ``$(\ )_{\substack{> \\[-3pt] <}}$'' indicates the positive
(negative) part of the series in the shift operator\footnote{In
other words, $(L^n)_{\substack{> \\ <}}$ is a projection of the
Laurent series $L^n$ onto a polynomial in $\ho$ ignoring the
negative (positive) terms.} $\ho$ and ``$(\ )_0$'' denotes the
constant part. The evolution equations are the {\em Lax--Sato
equations}
\bea{r@{\ =\ }l@{\qquad}r@{\ =\ }l}{eq:LaxSato}
\hbar \fracL{\p L  }{\p t\phm }&[H\phm ,L],&
\hbar \fracL{\p L  }{\p t_{-n}}&[H_{-n},L], \\[12pt]
\hbar \fracL{\p \bL}{\p t\phm }&[H\phm ,\bL],&
\hbar \fracL{\p \bL}{\p t_{-n}}&[H_{-n},\bL].
\eea
This system represents the Toda hierarchy since (\ref{eq:Lax},
\ref{eq:LaxHams} and \ref{eq:LaxSato}) together form a collection of
nonlinear finite--difference equations in $s$ for the coefficients
$r(s,t)$, $u_{k}(s,t)$ and $\bu_{k}(s,t)$, the differential being
with respect to the Toda times $t_{\pm n}$.

The integrability of the system becomes apparent when we obtain the
``zero curvature condition'' for the Hamiltonians. Introducing a
covariant derivative
\be
\dd_{n} =\hbar\frac{\p}{\p t_{n}} - H_{n}
\ee
the Lax--Sato equations
\be
[\dd_{n},L]=0,\qquad
[\dd_{n},\bL]=0
\ee
are equivalent to the zero curvature condition
\be
[\dd_{m},\dd_{n}]=0\qquad\mathrm{or}\qquad
\hbar \frac{\p H_n}{\p t_m}-\hbar \frac{\p H_m}{\p t_n}+[H_n,H_m]=0.
\ee
Thus the Toda hierarchy possesses an infinite set of commutative
flows $\dd_{n}$ and is therefore integrable by virtue of
\S\ref{sec:integr}.

As with all hierarchies, we solve the Lax--Sato equations with $n=1$
for the first coefficient $r(s,t)$ to substitute this solution into
subsequent equations. The equations (\ref{eq:LaxSato}) with $n=1$
yield,
\bea{r@{\;=\;}l@{\qquad}r@{\;=\;}l}{eq:shmeah}
\hbar \fracL{\p \log r^{2}(s)}{\p t_1}
&
u_0(s+\hbar)-u_0(s), &
\hbar \fracL{\p \log r^{2}(s)}{\p t_{-1}}
&
\bu_0(s)-\bu_0(s+\hbar),\\[12pt]
\hbar \fracL{\p \bu_0(s)}{\p t_{1}}
&
r^2(s)-r^2(s-\hbar), &
\hbar \fracL{\p u_0(s)}{\p t_{-1}}
&
r^2(s-\hbar)-r^2(s).
\eea
which combine to give the Toda lattice equation,
\be\label{eq:LaxToda}
\hbar^2 \frac{\p^2 \log r^{2}(s)}{\p t_1 \p_{-1}}=
2r^{2}(s)-r^{2}(s+\hbar)-r^{2}(s-\hbar).
\ee
Note the similarity between (\ref{eq:LaxToda}) and (\ref{eq:todacrystal}).

The connection with matrix models can be more explicitly seen if the
problem is reformulated as an eigenvalue problem in terms of
``Baker--Akhiezer wave functions'', $\Psi(x;s)$, which are seen to
contain all the information about the system \cite{Takasaki},
\be\label{eq:LaxEV}
L\Psi(x;s)=x\Psi(x;s)
\ee
with the Schr\"{o}dinger equations
\be\label{eq:LaxSE}
\hbar \frac{\p \Psi }{ \p t_n}=H_n\Psi,\qquad \hbar \frac{\p \Psi }{
\p t_{-n}}=H_{-n}\Psi.
\ee
Indeed, differentiating (\ref{eq:LaxEV}) and using (\ref{eq:LaxSE})
we get back the Lax--Sato equations (\ref{eq:LaxSato}). The Toda
hierarchy can now be represented in terms of infinite matrices. The
function $\Psi$ is treated as a vector whose entries are labeled by
the discrete variable $s$ causing the positive, constant and
negative part of the series in $\ho$ to be mapped to the upper,
diagonal, and lower triangular parts of matrices respectively.

\subsection{The $\t$--Function}
\label{sec:tau}
The Lax operators (\ref{eq:Lax}) can be written as dressed shift
operators $\ho$
\be\label{eq:dressedshift}
L=W \ho W^{-1},\qquad \bL=\bW \ho^{-1} \bW^{-1}.
\ee
where
\be\label{eq:dressing}
W=\sum\limits_{n\geq 0}w_n\ho^{-n},\qquad
\bW=\sum\limits_{n\geq 0}\bw_n\ho^{n}
\ee
Now, the dressing operators (\ref{eq:dressing}) are determined by
(\ref{eq:dressedshift}) up to a factor which commutes with the shift
operator $\ho$. In particular, they may be replaced by ``wave
operators'' \cite{Kostov02} which satisfy \re{dressedshift} with
$W\rightarrow\CW$, $\bW\rightarrow\bCW$
\bea{r@{\;=\;}c@{\;=\;}l}{eq:wave}
\CW  & W \exp \( \sum\limits_{n \geq 1} t_n \ho^n\)
& e^{\frac{1}{2\hbar}\phi} \(1 + \sum\limits_{n\geq 1}w_k\ho^{-n}\)
 \exp\(\frac{1}{\hbar} \sum\limits_{n\geq 1}t_n\ho^n\)\\[12pt]
\bCW & \bW \exp \( \sum\limits_{n \geq 1} t_{-n} \ho^{-n}\)
& e^{-\frac{1}{2\hbar}\phi}\(1 + \sum\limits_{n\geq 1}\bw_n\ho^{n}\)
 \exp\(\frac{1}{\hbar} \sum\limits_{n\geq 1}t_{-n}\ho^{-n}\)
\eea
where the zero mode was absorbed in the prefactor
$e^{-\frac{1}{2\hbar}\phi}$ causing a redefinition of the expansion
coefficients $w_n$, $\bw_n$ from (\ref{eq:dressing}) and $\phi(s,t)$
is related to $r(s,t)$ by
\be\label{eq:rphi}
r^2(s)=e^{\[\phi(s)-\phi(s+\hbar)\]/\hbar}.
\ee
Wave operators will be used in chapter \ref{ch:TAC} to introduce
tachyon perturbations by acting on unperturbed wave functions.

The Lax system may be extended by adding Orlov--Shulman operators
$M$, $\bM$ such that
\bea{r@{\;=\;}l@{,\qquad}r@{\;=\;}l}{eq:dressedl}
L&\CW \ho\CW^{-1}& \bL & \bCW \ho\bCW^{-1},\\
M&\CW s\CW^{-1}& \bM & \bCW s\bCW^{-1},
\eea
the commutator with $L$ and $\bL$ being a version of \refeq{eq:comm}
dressed as in (\ref{eq:dressedl});
\be\label{eq:commLM}
[L,M]=\hbar L,\qquad [\bL,\bM]=-\hbar \bL.
\ee
The commutation relations (\ref{eq:commLM}) serve to determine the
coefficients $v_{\pm k}$ in the explicit expansion for the Orlov--Shulman
operators \cite{OrlavShulman}:
\bea{r@{\;=\;}l}{eq:OS}
M  & \;\;\;\sumL\limits_{n\ge 1}nt_n L^n\;+\;s+\sumL\limits_{n\ge
1}v_n L^{-n},
\\[15pt]
\bM&-\sumL\limits_{n\ge 1}nt_{-n}\bL^n+s-\sumL\limits_{n\ge
1}v_{-n}\bL^{-n}.
\eea
which satisfy the Lax equations \cite{Takasaki}
\bea{r@{\ =\ }l@{\qquad}r@{\ =\ }l}{}
\hbar\derL{M}{t_n} & \[H_n,M\] & \hbar\derL{M}{t_n} & \[H_{-n},M\] \\[\arskip]
\hbar\derL{\bM}{t_n} & \[H_n,\bM\] & \hbar\derL{\bM}{t_n} & \[H_{-n},\bM\]
\eea
Differentiating (\ref{eq:dressedl}) for $L$ and using
(\ref{eq:LaxSato});
\be
\frac{\p L}{\p t_n} = \frac{\p \CW}{\p t_n}\ho\CW^{-1} +
\CW\ho(-\CW^{-2})\der{\CW}{t_n} =
\frac{1}{\hbar}[H_n,L]=\frac{1}{\hbar}H_n\CW\ho\CW^{-1} -
\frac{1}{\hbar}\CW\ho\CW^{-1}H_n,
\nn\ee
it becomes apparent that the Hamiltonians of the
system can be written in terms of the wave operators
\bea{r@{\ =\ }l@{\quad}r@{\ =\ }l}{eq:Hdressed}
H_n & \hbar(\p_{t_n}\CW)\CW^{-1} & H_{-n} & \hbar(\p_{t_{-n}}\CW)\CW^{-1} \\[\arskip]
\bH_n & \hbar(\p_{t_n}\bCW)\bCW^{-1} & \bH_{-n} & \hbar(\p_{t_{-n}}\bCW)\bCW^{-1}
\eea
where the Lax--Sato equations \re{LaxSato} imply that
\be\label{eq:hamcon}
H_{\pm n}=\bH_{\pm n}.
\ee

The condition \re{hamcon} is equivalent the condition that
$\CW^{-1}\bCW$ be independent of the Toda times $t_{\pm n}$;
\be\label{eq:condress}
\p_{\t_{\pm n}}(\CW^{-1}\bCW)=0.
\ee
Indeed, the wave matrices satisfy the following bilinear relations
\cite{Takebe}
\beq
\bCW(t_{\pm n})\bCW(t'_{\pm n})^{-1} &=& \CW(t_{\pm n})\CW(t'_{\pm n})^{-1}\nn\\[\arskip]
&\Rightarrow& \CW^{-1}\bCW = A
\eeq
where A is a constant matrix, independent on the times $t_{\pm n}$.
This gives the Riemann--Hilbert decomposition of $GL(\infty)$
\cite{Takasaki,SYang}.

A remarkable fact of the Toda hierarchy is that all its ingredients
may be reproduced by a single function $\t$ called the {\em
tau--function}. By Proposition 2.4.3 of \cite{Takasaki}, we
have\footnote{These relations can be obtained by considering
identities such as
\be\der{}{t_n}\[H_m,M\]=\frac{\p^2M}{\p t_n\p t_m}
=\frac{\p^2M}{\p t_n\p t_m}=\der{}{t_m}\[H_n,M\].
\nn\ee}
\be
\der{v_n}{t_m}=\der{v_m}{t_n},\qquad
\der{v_{-n}}{t_{-m}}=\der{v_{-m}}{t_{-n}},\qquad
\der{v_{-n}}{t_{m}}=\der{v_{m}}{t_{-n}}
\ee
which implies that there exists a function $\log \t_s(\t_{\pm 1},
\t_{\pm 2},\ldots)=\log \t_s(t)$ of which the coefficients $v_{\pm
n}$ of the Orlov--Shulman operators (\ref{eq:OS}) are first
derivatives. Omitting the argument $t$ from the variables,
\be
v_n(s) = \hbar^2\der{\log \t_s(t)}{t_n}.
\ee
From the $\t$--function, one is thus also able to reproduce the zero
mode
\be
e^{\phi(s)/\hbar} = \frac{\t_s}{\t_{(s+\hbar)}}
\ee
and, through \refeq{eq:rphi}, the first mode of the Lax operators
\be
r^2(s-\hbar)=\frac{\t_{(s+\hbar)}\t_{(s-\hbar)}}{\t_s^2}.
\ee
which seeds for the entire hierarchy.

An even more pertinent fact is that the $\t$--function was shown to
be equal to the partition function of the associated matrix model.
This thereby provides a short cut to the partition function in
matrix models of string theory, as we will see in the next chapter.

\EndOfChapter

\chapter{Matrix Quantum Mechanics}
\label{ch:MQM}
In chapter \ref{ch:MAT} random matrix theory was introduced their
fat Feynman graphs were shown to be dual to discretised Riemann
surfaces. In particular, we saw that computing the free energy of
the matrix model amounts to summing over Riemann surfaces of
different genera, just like the partition of 2D string theory.

String theory appeared in the double scaling limit where the
spherical limit $N\rightarrow\infty$ and the continuum limit
$g_k\rightarrow g_{\rm c}$ are taken in a coordinated manner.

Here, random matrix techniques of chapter \ref{ch:MAT} which will be
pertinent to string theory are generalised, our ultimate aim being
to formulate a matrix model in the next chapter of the perturbed CFT
of chapter \ref{ch:STR}.

\section{The MQM Model}
\label{sec:MQM}
As mentioned earlier, MQM refers to a generalisation of the linear
matrix chain (\ref{eq:linmch}) where the number of matrices $M_i$,
$i=1,\ldots,\na$ becomes infinite, $\na\rightarrow\infty$, and the
index $i$ labeling the matrices is replaced by a continuous argument
$t\in\bbR$, interpreted as Euclidean time. The $N\times N$ matrices
$M(t)$ form a Gaussian unitary ensemble and they are therefore
hermitian and invariant under global unitary transformations; $M(t)
= \O^\dag M(t) \O$, $\O^\dag\O = \bbI$.

\subsection{Action}
\label{sec:MQMaction}
The action of interest is then a functional of one matrix, where the
matrix is now a function of time. The path integral for the action
of this model is thus
\be\label{eq:MQMZ}
Z_N=\int \CD M(t) \exp \[ -N\tr\int dt\, \( \frac{1}{2} ({\p_t
M(t)})^2+V\[M(t)\]\)\]
\ee
where the potential has the same form as in (\ref{eq:potential})
\be\label{eq:MQMpot}
V\[M(t)\]=\sum\limits_{k>0}\frac{g_k}{k}M(t)^k.
\ee
As we deal only with hermitian matrices, the resulting physical
system described by the path integral (\ref{eq:MQMZ}) thus has $N^2$
degrees of freedom.

Following the same line of reasoning as in
\S\ref{sec:ffd}, we construct a propagator for this
theory which is now time--dependent and includes the Green's
function $G(t-t') = e^{-|t-t'|}$. A scalar field (the time variable)
$t$ resides at each end,
\beq\label{fig:propagt}
\raisebox{-20pt}{
\includegraphics[scale=0.4]{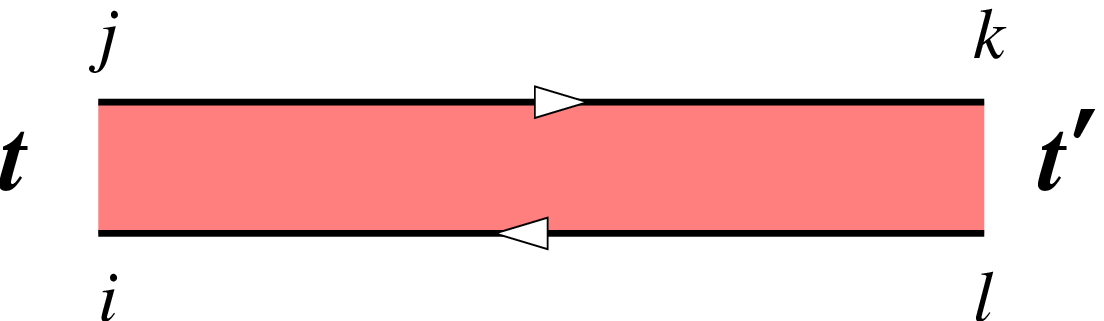}}
&=& \frac{1}{N g_2}\d_{il}\d_{jk} G(t-t').
\eeq
The topological expansion for the free energy (\ref{eq:freeEg}) must
be modified,
\be\label{eq:freeEgt}
F_g = \sum\limits_{\substack{{\rm Discretisations }\\{\rm of\ genus
}\; g}} \frac{1}{s}g_2^{-P}\prod\limits_{k>2} (-g_k)^{n_k}
\int\limits_{-\infty}^\infty dt_i \prod\limits_{\la ij \ra}
G(t_i - t_j)
\ee
where $\prod\limits_{\la ij \ra}$ denotes the product over all edges
connecting the \sth{$i$} and \sth{$j$} adjacent vertices and $P$ is
the number of edges (dual propagators) in the discretisation. The
matrix label $t$ should be viewed as a scalar field $t(\s,\t)$ on
the surface which is discretised along with the surface; $t_i =
t(\s_i,\t_i)$, where $(\s,\t)$ are some choice of coordinates on the
surface which, upon discretisation, turn to labeling the centre of
polygons on the discretised surface. The infinite range of
integration in time variable indicates that the time dimension is
not compact. (One may alternatively consider a semi--infinite time
domain, or even construct an MQM theory on a circle with a
compactified time dimension. We will consider this later.)

By inspecting (\ref{eq:freeEgt}) which is in effect a partition
function of the same standard (Euclidean) form as
\be
Z=\int\CD t(i)\,e^{-S[t(i)]}, \nn
\ee
we can fish out the action for this scalar field. Its discretised
form is
\be
S[t_i] = - \sum\limits_{\la ij\ra} \log G(t_i - t_j).
\ee
Taking the double scaling limit, from the propagator
(\ref{fig:propagt}) we can retrieve the continuum version of the
action
\be\label{eq:contaction}
S[t(\s,\t)] = \int d\s d\t \,
\sqrt{h}\left|h^{ab}\p_at\p_bt\right|^{\nf{1}{2}}
\ee
where $h^{ab}$ is the metric on the surface.

\subsection{MQM as a Description of 2D String Theory}
We already saw that the Feynman diagrams of hermitian matrix models
correspond to a discretisation of two dimensional Riemann manifolds
and the free energy is equivalent to a sum over those discretised
surfaces. In the double scaling limit, the discretisation becomes
continuous and the sum for the free energy becomes a sum over
continuous surfaces. The one--matrix models described discretised
surfaces without any additional structure and multimatrix models
represented surfaces possessing some additional structure. In
\S\ref{sec:DSL} it was argued that the topological expansion of the
one matrix model in the double scaling limit reproduces the
partition function of 2D gravity. In the multimatrix model, we might
expect to get instead 2D gravity coupled to a matter field $t$.

However, the action (\ref{eq:contaction}) is nonlinear --- the usual
kinetic term $h^{ab}\p_at\p_bt$ appears under an exponent of
$\nf{1}{2}$. But this does not undermine the interpretation of $t$
as a matter field. The usual kinetic term is obtained from a scalar
field Green's function of the form $G_{\rm mat}(t)=e^{-t^2}$ which,
in momentum space, has the form
\be
G_{\rm mat}(p)=\int\limits_{-\infty}^\infty\!dt\;G_{\rm
mat}(t)e^{-ipt}=\pi e^{p^2/4} =\pi - \frac{\pi}{4}p^2 + \order(p^4),
\ee
whereas the Green's function of MQM in momentum space is of the form
\beq
G(p)&=&\int\limits_{-\infty}^\infty\!dt\;e^{-|t|}e^{-ipt}=
\int\limits_{0}^\infty\!dt\[e^{-t(1+ip)} + e^{-t(1-ip)}\]\nn\\
&=&\frac{2}{1+p^2}=2-2p^2 + \order(p^4)\\
&\sim&G_{\rm mat}(p)\qquad\text{for small momenta $p$}.\nn
\eeq
So, for small momenta, the nonstandard action (\ref{eq:contaction})
can nevertheless be used to represent a scalar matter field. This
would not be valid at high momenta, but then we would be within the
domain of short--distance physics which is in any case
non--universal and should not affect the critical properties of the
matrix model that survive in the continuum limit. This conclusion
was confirmed through numerical computations and comparisons with
results obtained by CFT and other methods \cite{KazakovMigdal}. We
can therefore interpret the MQM model (\ref{eq:MQMZ}) as two
dimensional gravity coupled to $c=1$ matter which, in turn, can be
interpreted as the sum of surfaces or non--critical strings embedded
in one dimension. And, as we saw in \S\ref{sec:LiouvilleTh}, this is
the same as 2D critical string theory in a linear dilaton
background.

Hence, MQM provides a description of 2D string theory differing from
the common CFT formulation in that integrability is manifest and the
random matrix framework donates many tools enabling one to tackle
otherwise intractable problems.

\newpage
\section{Solving the MQM model}
\label{sec:SolveMQM}
To analyse the dynamics of MQM, we begin by simplifying the
classical problem by decomposing the degrees of freedom exploiting
the symmetry of $M$ and then proceed to quantise the theory. We will
see that the system described by (\ref{eq:MQMZ}) simplifies
considerably and can be described by a system of noninteracting
fermions moving in a potential.
\subsection{Reduction to Eigenvalues}
\label{sec:evals}
We start by decomposing the $N\times N$ hermitian matrices into
their real eigenvalues $x_i$ and angular matrix variables $\O$
diagonalising $M$ as in \S\ref{sec:1MM}, all of which are now
time--dependent\footnote{Explicitly, $M_{ij}(t) =
\sum\limits_{lm}\O^\dag_{il}(t) x_l(t) \d_{lm} \O_{mj}(t)$  where
$M_{ij}(t)$ are the matrix elements.}
\beq\label{eq:MQMdiag}
&M(t)=\O^\dag(t)x(t)\O(t)&\\
&x(t)=\diag\lb x_1(t),\ldots,x_N(t)\rb,\qquad,\O^\dag(t)\O(t)=\bbI.&\nn
\eeq
This transformation induces a Jacobian in the path integral measure.
It can be found by considering a conventional matrix integral
measure $dM$ as was done previously for the one--matrix model
\be\label{eq:dM}
dM = \prod\limits_i dM_{ii} \prod\limits_{i<j}
\Re(dM_{ij})\Im(dM_{ij})=[d\O]_{\gr{SU}{N}}\prod\limits_{i=1}^{N}dx_i\D^2(x)
\ee
where $[d\O]_{\gr{SU}{N}}$ is the Haar measure on $\gr{SU}{N}$ and
$\D(x)$ is the Vandermonde determinant \re{Vandermonde}.

Now, the angular variables from the diagonalisation
(\ref{eq:MQMdiag}) are canceled in the $\gr{U}{N}$--invariant
potential (\ref{eq:MQMpot}) but not in the kinetic term,
\be
\tr (\dM)^2 = \tr (\p_t \dx)^2 + \tr ([x,\dO\O^\dag])^2,
\ee
where a dot over a variable indicates differentiation with respect
to $t$.

Since $\dO\O^\dag$ is anti--Hermitian and traceless, $\det
e^{\dO\O^\dag}=\bbI$ and, by taking Hermitian adjoints term by term
we get $(e^{\dO\O^\dag})^\dag=(e^{\dO\O^\dag})^{-1}$. Therefore,
$e^{\dO\O^\dag}\in\gr{SU}{N}$ and
$\dO\O^\dag\in\mathfrak{su}(N)$, the Lie algebra of
$\gr{SU}{N}$, so it can be decomposed in terms of $\gr{SU}{N}$
generators,
\be
\dO\O^\dag= \sum\limits_{i=1}^{N-1} \dal_{i}H_i+ \frac{i}{\sqrt{2}}\sum\limits_{i<j}
\( \dbe_{ij}T_{ij}+\dtbe_{ij}\tlT_{ij}\),
\ee
where $H_i$ are the diagonal generators of the Cartan
subalgebra\footnote{The Cartan subalgebra of $\gr{SU}{N}$ is the
maximally Abelian subalgebra of $\gr{SU}{N}$; the largest subalgebra
which generates an Abelian subgroup of $\gr{SU}{N}$.} and the other
generators can be represented as
\bea{r@{\;=\;}l}{}
(T_{ij})_{kl}&\delta_{ik}\delta_{jl}+\delta_{il}\delta_{jk}\\
(\tlT_{ij})_{kl}&-i(\delta_{ik}\delta_{jl}-\delta_{il}\delta_{jk}).
\eea
Now, with this decomposition, we rewrite the action of MQM in
(\ref{eq:MQMZ}) as,
\be\label{eq:MQMaction}
S_{\rm MQM}=\int dt\, \[ \sum\limits_{i=1}^N \( \hf{\dot x}_i^2
-V(x_i) \)+
\hf\sum\limits_{i<j}(x_i-x_j)^2({\dbe}_{ij}^2+{\dtbe}_{ij}^2)\].
\ee
where the $N$ appearing in front of this action in the partition
function is now seen to play the same role as $\hbar^{-1}$ in
quantum field theory. Thus, the MQM degrees of freedom previously
contained in $M(t)$ are now embodied in $x_i(t)$, $\a_i(t)$,
$\b_{ij}(t)$ and $\tbe_{ij}(t)$.

\subsection{Quantisation}
\label{sec:quantisation}
To quantise the theory, we start by formulating the classical
conjugate momenta from the Lagrangian of the theory, $\CL$,
\be\label{eq:Lagrangian}
\CL= \sum\limits_{i=1}^N \( \hf{\dot x}_i^2 -V(x_i) \)+
\hf\sum\limits_{i<j}(x_i-x_j)^2({\dbe}_{ij}^2+{\dtbe}_{ij}^2)
\ee
\be\label{eq:conjmomenta}
p_i=\der{\CL}{\dot x_i},\qquad \Pi_{ij}=\der{\CL}{\dot
\b_{ij}},\qquad \tPi_{ij}=\der{\CL}{\dot \tbe_{ij}}
\ee
where $\Pi_{ij}$ and $\Pi_{ij}$ are generators of left rotations on
$\O$, that is, $\O\rightarrow A\O$, $A\in\gr{SU}{N}$. Since $\CL$ is
independent of ${\dot \a_{i}}$, we have the first--class constraint
\be
\Pi_{i}=\der{\CL}{\dot \a_{i}}=0.
\ee
Performing the usual Legendre transform
$H(x_i,p_i,t)=\suml_{k}p_k\dot x_k - \CL(x_i,\dot x_i, t)$ we get
the classical Hamiltonian
\be
\HMQM=\suml_{i=1}^N \( \hf p_i^2 +V(x_i) \)+
\hf\suml_{i<j}\frac{{\Pi}_{ij}^2+{\tPi}_{ij}^2 }{ (x_i-x_j)^2},
\ee

In deriving the quantum Hamiltonian, the measure of integration
expressed in terms of $\O$ and $x$ should assume a similar form as
(\ref{eq:dM})
\be
\CD M = \CD\O\prod\limits_{i=1}^{N}d x_i \D^2(x).
\ee
Since the integration measure of the scalar product in the
coordinate representation coincides with the path integral measure,
they contain the same Jacobian. Under the decomposition
(\ref{eq:MQMdiag}), the scalar product becomes
\be
\label{eq:scalprod}
\langle \Phi|\Phi'\rangle =\int d M\,\Phi^\dag(M)\Phi'(M) =\int
d\O\int \prod\limits_{i=1}^N dx_i\, \Delta^2(x)
\Phi^\dag(x,\O)\Phi'(x,\O).
\ee
Therefore, the transformation to momentum space wave functions is
\beq
\Phi(p,\O)&=&\int \prod\limits_{i=1}^N \(dx_i\,
\frac{e^{-\frac{i}{\hbar}p_ix_i}}{\sqrt{2\pi\hbar}}\) \D(x)
\Phi(x,\O).\\
\Rightarrow p_l\Phi(p,\O)&=&\int \prod\limits_{i=1}^N \(dx_i\,p_l
\frac{e^{-\frac{i}{\hbar}p_ix_i}}{\sqrt{2\pi\hbar}}\) \D(x)
\Phi(x,\O).\nn\\
&=&\int \prod\limits_{i=1}^N \(dx_i\,i\hbar\der{}{x_l}
\frac{e^{-\frac{i}{\hbar}p_ix_i}}{\sqrt{2\pi\hbar}}\) \D(x)
\Phi(x,\O).\nn\\
&=&-\int \prod\limits_{i=1}^N \(dx_i\,
\frac{e^{-\frac{i}{\hbar}p_ix_i}}{\sqrt{2\pi\hbar}}\) \D(x)\(\frac{i\hbar}{\D(x)}\der{}{x_l}\D(x)\)
\Phi(x,\O).\nn\\
\eeq
from which we see that the quantum momentum operator generating
spacial transformations, in the coordinate representation, is
\be
\hp_i=\frac{-i \hbar}{\D(x)} \der{}{x_i}\D(x).
\ee
Hence, the Hamiltonian of the quantised system is
\be\label{eq:HMQM}
\hHMQM=\sum\limits_{i=1}^N
\(\frac{-\hbar^2}{2\D(x)} \frac{\p^2}{\p x^2_i}\D(x)+V(x_i)\)+
\hf\sum\limits_{i<j}
\frac{\hPi_{ij}^2 + \widehat{\tPi^{\lefteqn{_{2}}}_{\lefteqn{_{ij}}}}
}{ (x_i-x_j)^2}.
\ee
and the wave functions are governed by the Schr\"{o}dinger wave equation
with a linear constraint
\be
i\hbar\frac{\p \Phi(x,\O)}{\p t}=\hHMQM \Phi(x,\O),
\qquad \hPi_i\Phi(x,\O)=0.
\ee
\subsection{Equivalence to Noninteracting Fermions}
\label{sec:nonintferm}
The partition function (\ref{eq:MQMZ}) can now be rewritten as
\be
Z_N=\Tr e^{-\frac{1}{\hbar}\CT\hHMQM},\qquad\CT\rightarrow\infty
\ee
where $\CT$ is the time interval over which the partition function
is defined. If we consider the partition function to represent a sum
over surfaces imbedded in one dimension, this time interval should
be infinite as indicated. Clearly, only the ground state of the
Hamiltonian contributes to the free energy in this limit
\beq
F&=&\lim\limits_{\CT\rightarrow\infty}\frac{\log Z_N}{\CT}\nn\\
&=&\lim\limits_{\CT\rightarrow\infty} \frac{1}{\CT}\log
\[e^{-\frac{\CT}{\hbar} E_0}\(1+\sum_{i>0}e^{-\frac{\CT}{\hbar}(E_i -E_0)}\)\] \nn\\
&=&-\frac{E_0}{\hbar}.
\eeq
So all we need is the ground state wave function with the energy
eigenvalue $E_0$. Since the angular terms in (\ref{eq:HMQM}) are
positive definite, this ground state must lie amongst the wave
functions that are annihilated by $\Pi_{ij}$ and $\tPi_{ij}$, that
is, amongst the wave functions $\Phis(x)$ belonging to the singlet
representation of $\gr{SU}{N}$ which do not depend on the angular
degrees of freedom\footnote{As will be more clearly explained in
\S\ref{sec:sunrep}, the wave function can be decomposed
into irreducible representations of $\gr{SU}{N}$
$$
\Phi(x,\O)=\sum_r\sum^{\dim(r)}_{I,J=1} D_{IJ}^{(r)}(\O)
\Phi^{(r)}_{I,J} (x)
$$
where $r$ labels the representation and $D_{IJ}^{(r)}(\O)$ is an
element indexed by $I,J$ of the representation matrix isomorphic to
$\O \in \gr{SU}{N}$ in the representation $r$. Given that the wave
function should be invariant under the generators of left rotations,
$\Pi_{ij}\Phi(x,\O)=\tPi_{ij}\Phi(x,\O)=0$, the only term remaining
in this sum belongs to the trivial or {\em singlet} representation
where $D^{(\rm sing)}(\O)=\bbI$.} \cite{Klebanov91,Brezin78}.
Writing
\be
\Psis(x)=\D(x)\Phis(x),\qquad
{\hPi}_{ij}\Phis=\htlPi_{ij}\Phis=0,
\ee
the Hamiltonian (\ref{eq:HMQM}) of the eigenvalue problem
$\hHMQM\Psis = E\Psis$ in the singlet sector assumes a much simpler
form
\be\label{eq:Hff}
\hHMQM\stackrel{\substack{\rm singlet\\\rm sector}}{\longrightarrow}
\hHMQM^{\rm (sing)}=\sum\limits_{i=1}^N\Hss_i,\qquad
\Hss_i=-\frac{\hbar^2}{2}\frac{\p^2}{\p x_i^2}+V(x_i),
\ee
where $\Hss_i$ are single--particle Hamiltonians. The wave functions
$\Phis(x)$ should not change under permutation of the eigenvalues
which constitutes a unitary transformation of the same form as
(\ref{eq:MQMdiag}) that leaves the system invariant. This makes
$\Phis(x)$ symmetric and therefore $\Psis(x)$ completely
antisymmetric. Remarkably, in this way, the original problem of
solving for $N^2$ bosonic degrees of freedom in (\ref{eq:MQMZ}) has
been reduced in (\ref{eq:Hff}) to a system of $N$ noninteracting
fermions moving non--relativistically in a potential $V(x)$. This
striking observation was first made in \cite{Brezin78}.

The one--particle wave function for the \sth{$i$} fermion with
energy $\e_i$ of the $N$ fermion system is
\be
\Hss_i\psi_i(x)=\e_i\psi_i(x).
\ee
When the system is in its ground state, the fermions occupy all the
lowest energy levels up to the Fermi energy and the collective wave
function is a fully antisymmetrised product of single particle
states; a Slater determinant:
\be\label{eq:Slater}
\Psi_{\rm
GS}(x)=\frac{1}{\sqrt{N!}}\det\limits_{i,j}\psi_i(x_j),\qquad
E_0=\suml_{i=1}^{N}\e_i,
\ee
where $x_j$ is the position of the $j^{\rm th}$ particle and the
index $i$ labels the energy eigenstates.

\subsection{The Fermi Sea}
\label{sec:fermisea}
Since $N$ was associated with $\nf{1}{\hbar}$, the spherical limit
$N\rightarrow\infty$ of \S\ref{sec:topexpan} corresponds to the
classical limit $\hbar\rightarrow0$. In this limit, the fermions
become classical objects in the sense that they reside in a two
dimensional phase space spanned by the usual $(x,p)$ coordinates.
However, they still obey the Pauli exclusion principle so the $N$
fermion system may be thought of classically as an incompressible 2D
liquid called the {\em Fermi sea}. Given that, by the Pauli
exclusion principle, each fermion occupies an area of $2\pi\hbar$ in
the phase space, the total volume of this liquid is $2\pi\hbar\times
N = 2\pi$.

The energy of the excitation level on the boundary of the Fermi sea
in the phase space should be equal to the Fermi energy. This gives
us an equation for the Fermi ``shore'', as illustrated in fig.
\ref{fig:fs_ex}, which we can derive from the classical Hamiltonian
(the potentials considered here are not explicitly dependent on time
so the Hamiltonian is equal to the total energy)
\be\label{eq:boundary}
h(x,p)=\hf p^2 + V(x)=\fermiE = \e_i |_{_{i=N}}.
\ee
\begin{figure}[h]
\figlabel{fig:fs_ex}
\begin{center}
\includegraphics[scale=0.65]{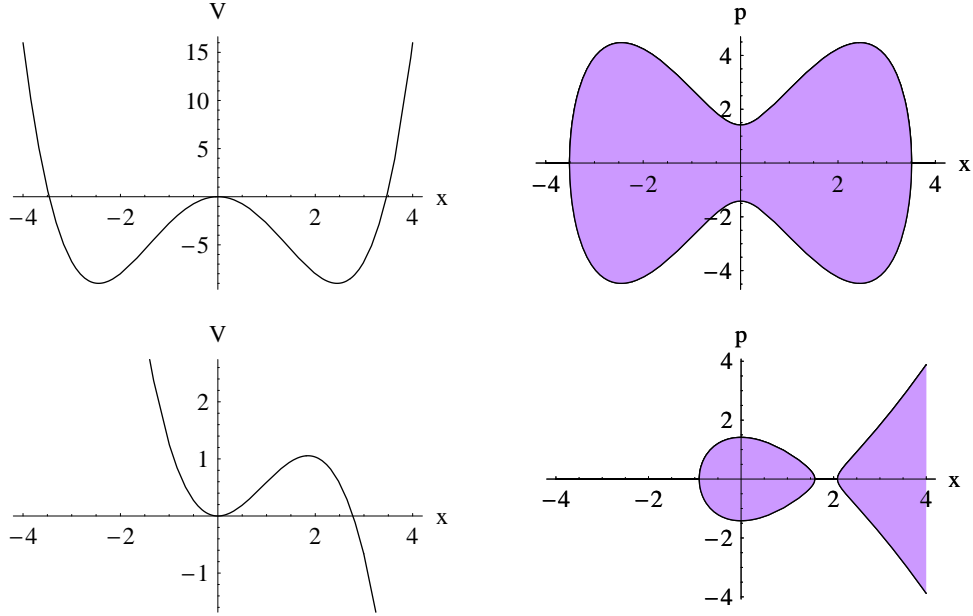}
\bf \caption{ \rm Some potentials on the left and their
corresponding Fermi sea with $\fermiE=1$ on the right. (Above:
$V(x)=-3x^2+\frac{1}{4}x^4$. Below:
$V(x)=\frac{37}{40}x^2-\frac{1}{3}x^3$)}
\end{center}
\end{figure}

The number of fermions $N$ and the energy of the ground state Slater
determinant $E_0$ are given by
\beq
N&=&\iint \frac{dxdp}{2\pi\hbar}\th(\fermiE-h(x,p)),\label{eq:N}
\\
E_0&=&\iint
\frac{dxdp}{2\pi\hbar}h(x,p)\th(\fermiE-h(x,p)).\label{eq:E0}
\eeq
where
\be
\th(\xi)=\left\lb\begin{array}{c}1,\quad\xi>0\\0,\quad\xi<0\end{array}\right.
\ee
is the Heaviside step function.

We can find the energy in an inexplicit way by differentiating
\re{N} and \re{E0} with respect to the Fermi level,
\beq
\hspace{-6ex}\hbar\der{N}{\fermiE}&=&
\iint\frac{dxdp}{2\pi}\d\big(\fermiE-h(x,p)\big)
=\frac{1}{\pi}\int\limits_{x_1}^{x_2}\frac{dx}{\sqrt{2(\fermiE-V(x))}},
\equiv\hbar\rho(\fermiE)\label{eq:dN}
\\
\hspace{-6ex}\hbar\der{E_0}{\fermiE} &=&
\iint\frac{dxdp}{2\pi}h(x,p)\d\big(\fermiE-h(x,p)\big)
=\hbar\fermiE\,\r(\fermiE),\label{eq:dE}
\eeq
where $x_1$ and $x_2$ are the turning points of the classical
trajectory, as illustrated in fig. \ref{fig:fermipot}.
\begin{figure}[!ht] \figlabel{fig:fermipot}
\begin{center}
\includegraphics[scale=0.5]{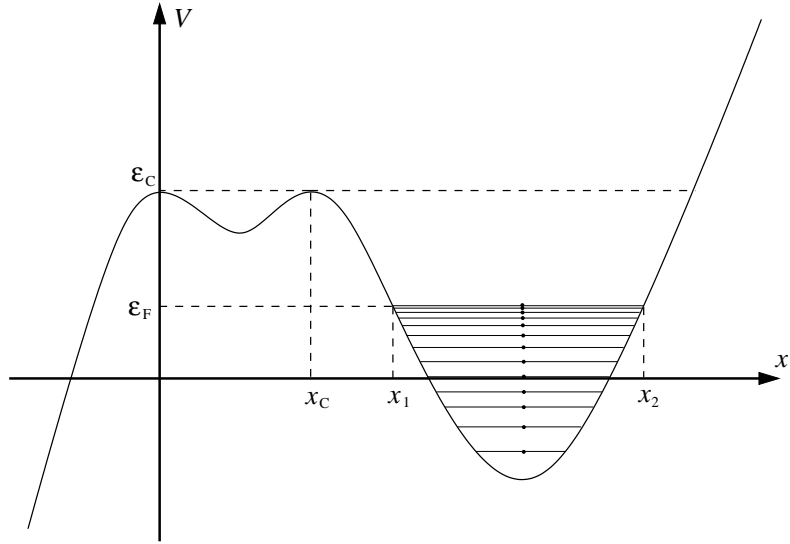}
\end{center}
\bf \caption{\rm The ground state Fermions are trapped a potential.}
\end{figure}

Note that since the potential depicted in fig. \ref{fig:fermipot} is
not bounded from below, there are, strictly speaking, no lowest
energy states in the region $x<0$ for this particular potential.
This is of no real concern, however, since we are only interested in
perturbative effects from which we model our physics --- the
amplitudes of tunneling from an unstable state are suppressed by
$\sim e^{\nf{-1}{\hbar}}$ and are thus not seen in any perturbative
theory where $\nf{1}{\hbar}=N\rightarrow\infty$. We can thus
consider such unbounded potentials by ignoring the unbounded
regions, and concentrating on the perturbative expansion around the
local extrema, neglecting nonperturbative effects such as tunneling.

Now, we would like to find the energy eigenvalue density and the
free energy by integrating \re{dE} but first the unknown constant
$\fermiE$ should be eliminated. This can be done by applying the
normalisation condition \re{N}. As stressed previously, in the
continuum limit, the sums over surfaces in string theory should be
independent on the discretisation used, in other words, independent
on the form of the potential, and the universal information is
contained in the singular or universal part of the free energy
\re{Fg}.
This singularity occurs when the Fermi level reaches a critical
level $\e_{\rm c}$ at the top of the potential.\footnote{In this
limit, the free energy for a triangulated surface looks like $F_g
\sim F_g^{\rm(reg)} + (g_c - g_3)^{\frac{\chi}{2}(2-\g_{\rm str})}$
with $\chi=2$ and $\g_{\rm str}=\nf{-1}{m+1}$ where $m$ is related
to the behaviour of the density near this critical point: $\r(x)
\sim (x-x_{\rm c})^{m + \nf{1}{2}}$ as explained in \S\ref{sec:u}.}

Thus, the singular contribution to the integral \re{dN} comes from
the local maxima in the potential and they can be well approximated
as quadratic. Writing $\e = \e_{\rm c} - \fermiE$, where $\e_{\rm
c}$ is the critical level where the density diverges
\cite{KazakovMigdal}, we get
\be
\r(\e)=\frac{1}{\pi}\int\limits_{x_1}^{x_2}\frac{dx}{\sqrt{2(\fermiE-V(x))}}
\sim-\frac{1}{2\pi}\log \e
\ee
\be
F_0=-\frac{E_0}{\hbar}
=-\frac{1}{\hbar}\int\!d\e\;\e\frac{\r(\e)}{\hbar}
\sim\frac{1}{2\pi\hbar^2}\e^2\log\e.
\ee

\subsection{Double Scaling Limit --- The Inverse Oscillator Potential}
\label{sec:MQMDSL}
Above, we saw that the spherical limit ($N\rightarrow\infty$)
corresponds to the classical limit of a system of noninteracting
fermions. However, we are interested in the continuum limit of
discretisations obtained when the coupling constants of the
potential approach their critical values, as explained in
\S\ref{sec:contL}. It was subsequently explained that
those two limits must be taken in a correlated manner in order to
include contributions to the string perturbation expansion from all
genera.

As explained in \S\ref{sec:MM}, the continuum limit is achieved when
the coupling constants approach their critical values. In that case,
the density behaves as $\r(x)\sim(x-x_{\rm c})^{m+\nf{1}{2}}$. The
correct variable to use here it therefore $x-x_{\rm c}$ where we
expand the potential about its local extrema located at $x_{\rm c}$.
Writing $y=\frac{1}{\sqrt{\hbar}}(x-x_{\rm c})$,
\be
V(y)=\e_{\rm c}-\frac{\hbar}{2}y^2+\frac{\hbar^{\nf{3}{2}}\l}{3}y^3+\cdots,
\ee
where $\e_{\rm c} = V(x_{\rm c})$. With this, the Schr\"{o}dinger
equation for the eigenfunction at the Fermi level is
\be\label{eq:FLSE}
\(-\hf\frac{\p^2}{\p y^2}-\hf y^2 +\frac{\hbar^{\nf{1}{2}}\l}{3} y^3
+\cdots\) \psi_N(y)=-\frac{1}{\hbar}\e\psi_{N}(y).
\ee
whence we define a rescaled energy eigenvalue
\be
\mu = \frac{1}{\hbar}(\e_{\rm c} - \fermiE)
\ee
which should remain fixed as we take the spherical limit
$N=\frac{1}{\hbar}\rightarrow\infty$ and the continuum limit
$\e_{\rm c} \rightarrow \fermiE$. This defines the double scaling
limit of MQM.

From \refeq{eq:FLSE}, we see that the limit $\hbar\rightarrow 0$
annuls all the terms in the potential containing $\hbar$. This is
another remarkable property of MQM --- in the double scaling limit,
MQM reduces to a problem of fermions in an inverse oscillator
potential. From \re{FLSE}, it is described by the following equation
\be\label{eq:seqp}
-\hf\(\frac{\p^2}{\p x^2} + x^2\)\psi_{\e}(x)=\e\psi_{\e}(x).
\ee
This relates to the independence of matrix models on the form of the
potential and is a manifestation of their universality.

The two independent solutions of (\ref{eq:seqp}) are  \cite{GinMoore}
\be\label{eq:buddah}
\psi_\e^+(x)=D_1 (\e,x),\quad \psi_\e^-(x)=D_{-1}(\e,x)
\ee
where
\be\label{eq:PCF}
D_\nu(\e,x)=\fracL  {2^{\frac{2-\nu}{4}}
x^{\frac{1-\nu}{2}} e^{\frac{-ix^2}{2}}
}  {\sqrt{4\pi(1+e^{2\pi\e})^{\nf{1}{2}}}}
\left| \fracL{
\G\(\fracL{2-\nu}{4}+\fracL{i\e}{2}\)}{
\G\(\fracL{2+\nu}{4}+\fracL{i\e}{2}\)}\right|^{\nf{1}{2}}
\!\!{_1}F_1(\frac{1}{4}-i\e;\frac{2-\nu}{2};ix^2)
\ee
are parabolic cylinder functions and
\be
{_1}F_1(a;b;x)=\suml_{k=0}^{\infty}\frac{(a)_k}{(b)_k}\frac{x^k}{k!}
\ee
are confluent hypergeometric functions of the first kind with
$(a)_k$ and $(b)_k$ being the Pochhammer symbols\footnote{
$(a)_k=\frac{\G(a+k)}{\G(a)}$.}.

Thus, we have at hand a microscopic description of MQM. From
asymptotics of \re{PCF}, the free energy of the matrix model can be
found \cite{SergePhD,KazakovMigdal}
\be\label{eq:MQMfe}
F(\D)=\frac{1}{4\pi}\( \frac{\D^2}{\log\D}
-\frac{1}{12}\log\D
+\sum\limits_{n=1}^{\infty}\frac{(2^{2n+1}-1) |\mathscr{B}_{2n+2}|}{
4n(n+1)(2n+1)}
\(\frac{2\D}{\log\D}\)^{-2n}\),
\ee
where $\D$ is related to the string coupling $\k$ and to the
rescaled energy $\mu$ by
\be
\k=\D^{-1},\qquad
\der{\D}{\mu}=2\pi\r(\mu)
\ee
and $\mathscr{B}_{2n+2}$ are the Bernoulli
numbers\footnote{\parbox[t]{\textwidth}{Bernoulli numbers can be
defined by the contour integral or sum
\be
\mathscr{B}_n=\frac{n!}{2\pi
i}\oint\limits_{\rm L}\frac{dz}{z^{n+1}}\frac{z}{e^z
-1}=\suml_{k=0}^{n}\frac{1}{k+1}\suml_{r=0}^k(-1)^r\,C_r^k\,
r^n\nn
\ee where the contour L is a circle of radius $<2\pi i$ enclosing
the origin oriented counterclockwise, and $C_r^k$ are the
binomial coefficients.}}.

As previously emphasised, this free energy corresponds to the
partition function of 2D quantum gravity. It is not evident at this
stage how the tachyons emerge from MQM, however. In the next
section, we will formulate an effective field theory and see how the
tachyon vertex operators can be realised in terms of matrix
operators.

\newpage
\section{Das--Jevicki String Field Theory}
\label{sec:DJ}
In the previous section, a microscopic description of MQM was
obtained in the double scaling limit leading to a free energy which
is equivalent to the partition function of 2D string theory.
However, the tachyon born out of the world--sheet dynamics is still
not apparent. Being a target space field, it can be viewed as an
effective degree of freedom described by an effective field theory.
In MQM, the effective degrees of freedom are the collective
excitations of the singlet sector MQM fermions and are described by
a {\em collective theory} formulated by Das and Jevicki
\cite{DasJevicki}. As this theory encompasses all interactions of 2D
string theory, it forms a string field theory in the target space.
The tachyon will be seen to emerge from this collective field
theory.

\subsection{Effective Action and Collective Field}
The Hamiltonian of MQM in the singlet representation is given by the
energy of the Fermi sea \re{E0}, to which we add a chemical
potential term which allows the number of fermions to vary. The
double scaled Hamiltonian is thus
\be\label{eq:hcol1}
\Hcol=\iint\limits_{\substack{\rm Fermi \\\rm
Sea}}\!\!\frac{dxdp}{2\pi}\;(\frac{p^2}{2} + V(x) + \mu)
\ee
where the integration is over the whole Fermi sea (explaining the
absence of the step function) and $$V(x)=-\frac{x^2}{2}$$ is the
inverse oscillator potential

We describe the Fermi sea by two functions; $\ppl(x,t)$ which traces
out the shore of the sea that is above the turning point, and
$\pmi(x,t)$ describing the shore below this point as in fig.
\ref{fig:Jsea}(a). Profiles such as in fig. \ref{fig:Jsea}(b) which
can not be described in such a way shall be not be considered as
they give rise to complications and will not be relevant in any case
\cite{SergePhD}. Thus we arrive at the boundary condition
\be\label{eq:bc}
\pmi(x_*,t)= \ppl(x_*,t)
\ee
where $x_*$ is the leftmost\footnote{In chapter \ref{ch:TAC} we
shall see it suffices to consider only the Fermi sea to the right of
the inverse oscillator potential, so the Fermi sea on left side of
the potential has been omitted from Fig.\ref{fig:Jsea}.} point of
the Fermi sea as indicated in Fig.\ref{fig:Jsea}(a).
\begin{figure}[!ht] \figlabel{fig:Jsea}
\begin{center}
\begin{tabular}{cc}
\includegraphics[scale=0.4]{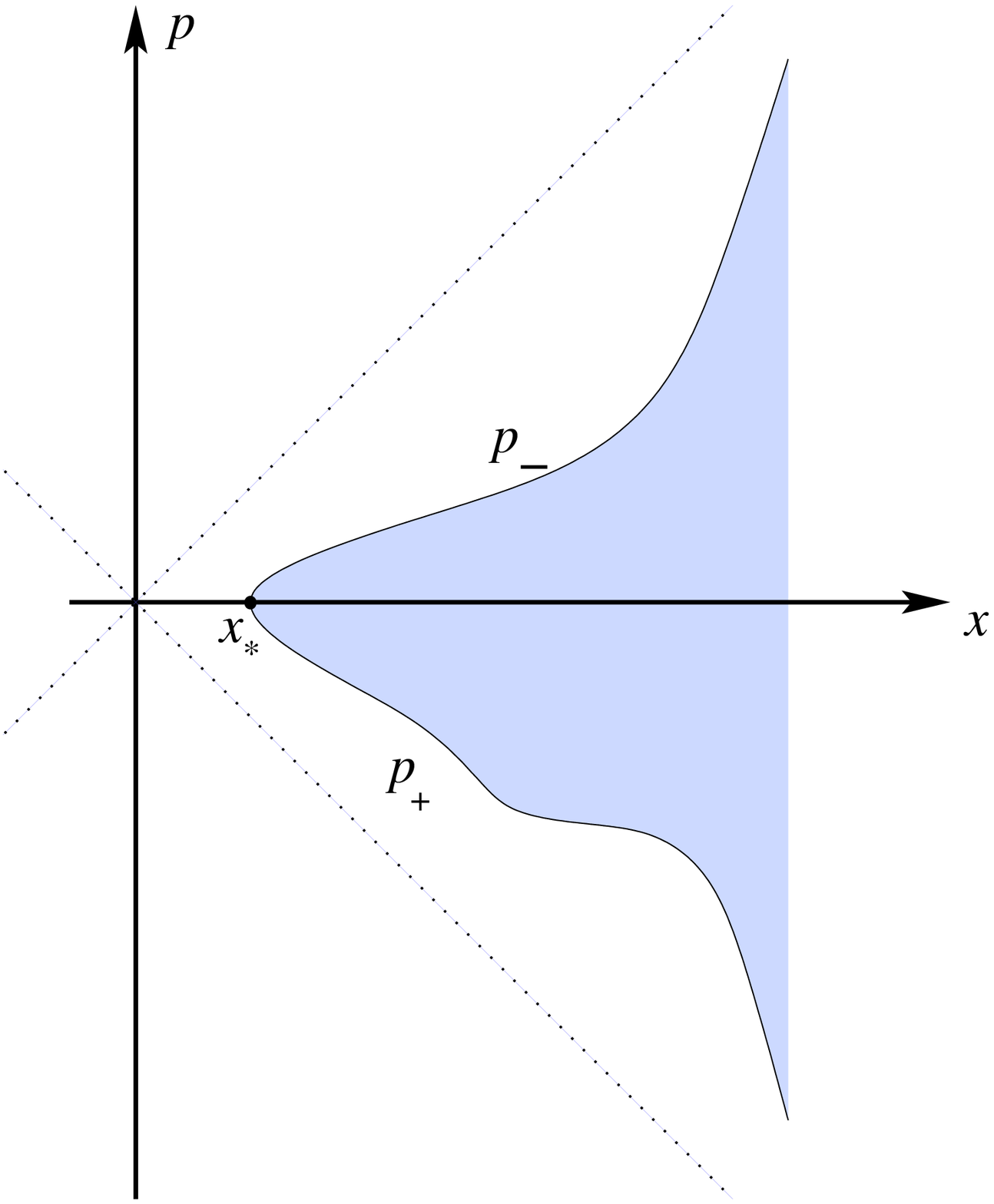}&
\includegraphics[scale=0.4]{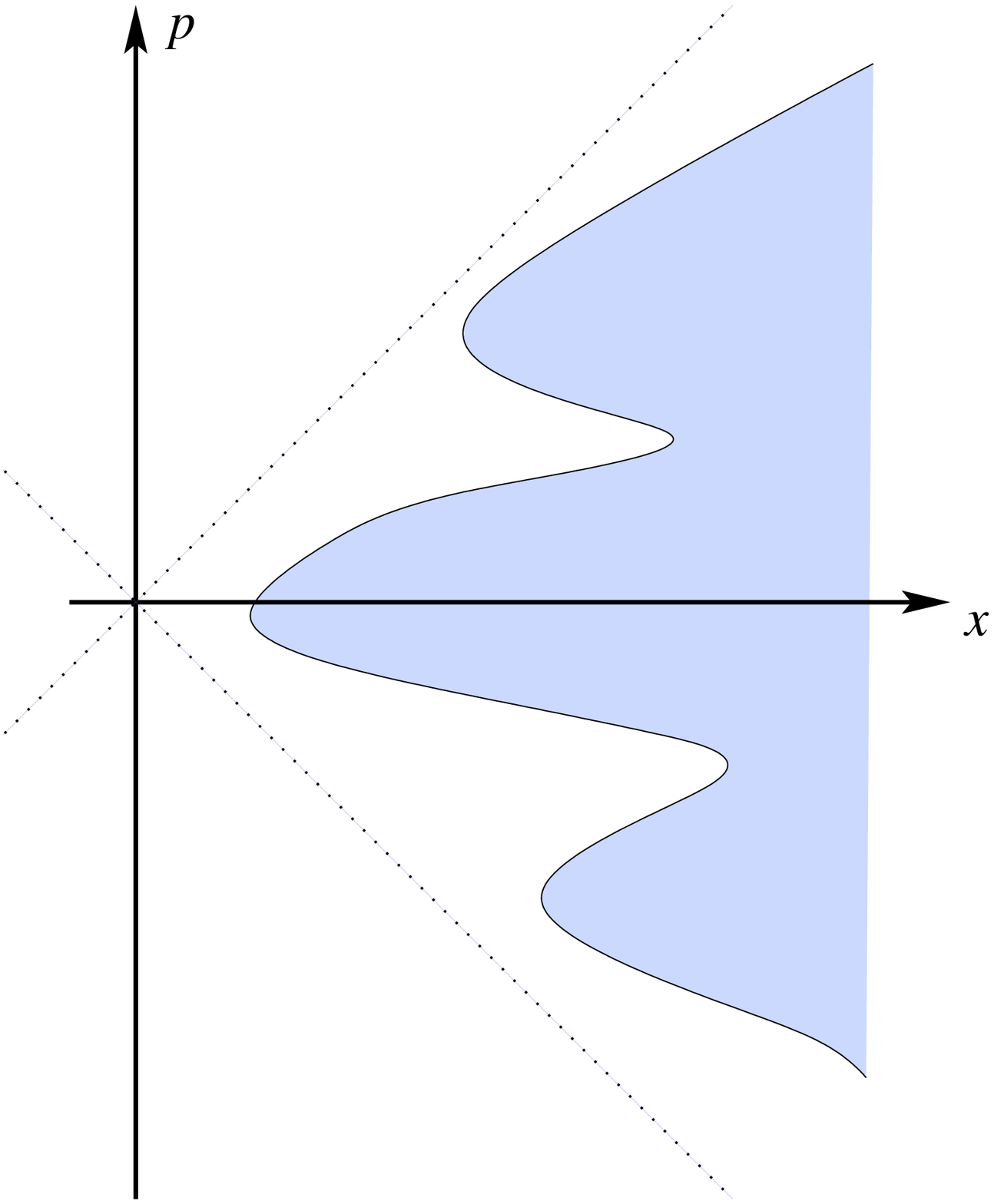}\\
{\bf (a)} & {\bf (b)}
\end{tabular}
\bf \caption{ \rm {\bf (a)} The shore of the Fermi sea is described
by $\ppl(x,t)$ and  $\pmi(x,t)$, and the leftmost point is $x_*$.
Eventual more general profiles shown in {\bf (b)} will not be
considered.}
\end{center}
\end{figure}
With the boundary condition \re{bc} in hand, we can perform the
integral over $p$ in \re{hcol1}
\be\label{eq:Hcol2}
\Hcol=\int_{x_*}^{\infty} \!\!\frac{dx}{2\pi}\;
\(\frac{1}{6}(\ppl(x,t)^3 -
\pmi(x,t)^3)+(V(x)+\mu)(\ppl(x,t)-\pmi(x,t)\).
\ee

The appropriate collective field is the density of
eigenvalues
\be\label{eq:denseval}
\colfield(x,t)=\tr \d (x\bbI_{N\times N} - M(t)),\qquad M(t)=\diag
(x_1(t),\ldots,x_N(t))
\ee
where the matrix variable has been diagonalised as in \re{MQMdiag}.
It becomes a continuous field in the double scaling limit
$N\rightarrow\infty$. The introduction of a conjugate field
$\Pi(x,t)$ with the Poisson brackets
\be
\lb \colfield(x,t),\Pi(x',t) \rb = \d(x-x')
\ee
gives a canonical phase space.

Since the density of the Fermi sea is proportional to its width, we
write
\be
\colfield(x,t)=\frac{1}{\pi}(\ppl(x,t)-\pmi(x,t)),
\ee
and the conjugate variable is defined as
\be\label{eq:conjvar}
\ppl(x,t)+\pmi(x,t)=\der{\Pi}{x}
\ee
so that
\be\label{eq:Hcol3}
\Hcol=\int\limits_{x_*}^{\infty} \!\! dx\;
\[\hf\colfield\(\der{\Pi}{x}\)^2 +
\frac{\pi^2}{6}\colfield^3+(V(x)+\mu)\colfield\].
\ee
Solving the canonical equation of motion
\be
\frac{\d \Hcol}{\d (\p_x\Pi)}=\der{}{t}\int\!dx\;\colfield
\Rightarrow -\p_x\Pi=\frac{1}{\colfield}\int\!dx\;\p_t\colfield
\ee
and substituting this into the Hamiltonian \re{Hcol3} we get the
effective action
\be\label{eq:DJSeff}
\Seff=\int\!dt\;\int\!dx\;\[\frac{1}{2\colfield}\(\int\!dx\;\p_t\colfield\)^2-\frac{\pi^2}{6}\colfield^3-(V(x)+\mu)\]
\ee

This nonlinear effective field theory results from the matrix model
with an inverse oscillator potential
\be
S=\int\!dt\;\hf\tr(\dot{M}^2+M^2)
\ee
and linear equations of motion $\ddot{M}(t) - M(t)$. Through the
nonlinear transformation \re{denseval}, MQM provides an exact
solution for the nonlinear theory \re{DJSeff}.

The effective action \re{DJSeff} is a background independent
formulation of 2D string field theory \cite{Jevicki93}; the cubic
interaction term describes the splitting and joining of strings, and
the linear tadpole term describes string creation and annihilation
into the vacuum.

The effective field theory is indeed integrable. The two equations
of motion from \re{Hcol2}, namely,
\be\label{eq:ppmconstraints}
\p_t p_\pm + p_\pm\p_x p_\pm +\p_x V(x)=\p_t p_\pm + p_\pm\p_x p_\pm
-x=0,
\ee
are large--wavelength KdV type equations which are known to be
integrable. Also, we get our infinite set of conserved, commuting
flows (as required for integrability in \S\ref{sec:integr}) from
\be
H_n(\ppl,\pmi,t))=\frac{1}{2\pi}\int dx
\int\limits_{\pmi(x,t)}^{\ppl(x,t)}\!dp\;(p^2 - x^2)^n
\ee
because, up to surface terms,
\be
\lb H_n,H_m\rb=0\quad{\rm and}\quad\frac{dH_n}{dt}=\int\!
dx\;\p_x(p^2-x^2)(p^2-x^2)^n=0.
\ee

\subsection{MQM and 2D String Theory}
\label{sec:tinmqm}
To introduce a background, we should introduce perturbations about
some classical solution $\colfield_0$ of \re{DJSeff}.  These
perturbations are embodied in $\eta(x,t)$, where
\be\label{eq:pertubaba}
\colfield(x,t)=\colfield_0(x)+\frac{1}{\sqrt{\pi}}\p_x\eta(x,t)
\ee
For this purpose the classical solution was taken to be the static
trajectory of the single fermion on the boundary of the Fermi sea,
from \re{boundary}, the momentum of which is $p=\ppm$ in the
notation of this section, and we thus take
\be\label{eq:DJbg}
\colfield_0(x)=\frac{1}{\pi}\sqrt{\ppm^2} =
\frac{1}{\pi}\sqrt{x^2-2\mu},
\ee
as our background.

Substituting \re{pertubaba} into \re{DJSeff} to get the action for
the perturbations,
\beq\label{eq:seff1}
\Seff[\eta]&=&\hf\int\!dt\int\!dx\,\(
\frac{(\p_t\eta)^2}{(\pi\colfield_0+{\sqrt{\pi}}\p_x\eta)}
-\pi\colfield_0(\p_x\eta)^2-\frac{\sqrt{\pi}}{3}(\p_x\eta)^3 \)\nn\\
&=&\hf\int\!dt\int\!dx\,\( \frac{(\p_t\eta)^2}{\pi\colfield_0}
\[\sum_{r=0}^{\infty}\frac{-\p_x\eta}{\sqrt{\pi}\colfield_0}\]
-\pi\colfield_0(\p_x\eta)^2 -\frac{\sqrt{\pi}}{3}(\p_x\eta)^3 \).\nn\\
\eeq
The quadratic part of \re{seff1} can be rewritten
\be
\hf\int\!dt\int\!dx\,\(\frac{(\p_t\eta)^2}{\pi\colfield_0}-\pi\colfield_0(\p_x\eta)^2\)=
\hf\int\!dt\int\!dx\,(\p_\mu\eta) g_{\mu\nu}^{(0)}(\p_\nu\eta)
\ee
where the massless field $\eta(x,t)$ is viewed as propagating in the
background metric $g_{\mu\nu}^{(0)}$,
\be
(\p_\mu\eta) =
\(\begin{array}{c}\p_x\eta\\\p_t\eta\end{array}\),\qquad
g_{\mu\nu}^{(0)}=\(\begin{array}{cc}-\pi\colfield_0&0\\0&(\pi\colfield_0)^{-1}\end{array}\).
\ee
This metric can be removed by introducing a ``time--of--flight''
coordinate \cite{Jevicki93} $q(x)$ defined as
\be\label{eq:tofmqm}
\frac{dx(q)}{dq}=\pi\colfield_0,\quad{\rm or}\quad
q(x)=\int\limits_0^x\!\frac{dx}{\pi\colfield_0},
\ee
whence, using \re{DJbg}, the coordinates become
\be\label{eq:tof}
x(q)=\sqrt{2\mu}\cosh q,\qquad p(q)=\sqrt{2\mu}\sinh q.
\ee
And \re{seff1} becomes
\be\label{eq:seff2}
\Seff= \hf\int\!dt\int\!dq\,\( (\p_t\eta)^2 -(\p_{q}\eta)^2
-\frac{\sqrt{\pi}}{3}\gstr(q)\[(\p_q\eta)^3+3(\p_q\eta)(\p_t\eta)^2
+ \cdots\]\)
\ee
which is the action for a massless scalar field $\eta(x,t)$ in
Minkowski space--time with a spatially dependant coupling constant
\beq
\gstr(q)&=&\frac{1}{(\pi\colfield_0(q))^2} \nn\\
&=& \frac{1}{2\mu\sinh^2 q}\sim\frac{1}{\mu}e^{-2q}\quad{\rm
as}\quad q\rightarrow\infty
\eeq

This is asymptotically the same coupling as the string coupling
constant \re{strdilatoncoupling}, with
\be
t\mapsto -iX,\qquad q\mapsto \phi
\ee
where $X$ and $\phi$ are the matter and Liouville fields
respectively on the world sheet. By comparing \re{seff2} to
\re{redefTs}, it is seen that the massless field $\eta$ corresponds
to the redefined tachyon of \refeq{eq:redefT}.
\be\label{eq:Romania}
\eta(X,\phi)=e^{Q\phi}T
\ee

We now see how the target space of 2D string theory can emerge from
MQM. The MQM time and time--of--flight coordinates, the latter of
which is related to the matrix eigenvalues through \re{tofmqm}, form
the flat target space of 2D string theory in the linear dilaton
background \re{LinDilBG}.

\subsection{Tachyon Vertex Operators}
\label{sec:mqmtvo}
The tachyon vertex operators were found in the Minkowskian form
\re{vertexops}. We should find the corresponding matrix vertex
operators describing left-- and right--moving tachyons.

Consider
\be\label{eq:pookaka}
T^{\pm}_n=e^{\pm nt}\tr(M\mp P)^n.
\ee

In the collective field theory formulation,
\beq\label{eq:colvop}
\tr(M\mp P)^n &=& \int\! \frac{dx}{2\pi}
\;\int\limits_{\pmi(x,t)}^{\ppl(x,t)}\!dp\;(x\mp
p)^n \nn\\
&=&\frac{1}{n+1}\int\!\frac{dx}{2\pi}\;(x\mp
p)^{n+1}\Big]_{\pmi(x,t)}^{\ppl(x,t)} \nn\\
&=&\frac{1}{2\pi(n+1)}\int\!dx\;\[\(x\mp\(\pi\colfield_0+\frac{\alm}{\pi\colfield_0}\)\)^{n+1}-
\right. \nn\\
&&\qquad\qquad\qquad\left.
\(x\mp\(-\pi\colfield_0+\frac{\alp}{\pi\colfield_0}\)\)^{n+1}\],
\eeq
where we rewrote the second quantised fields in terms of their
classical solution $\colfield_0(x)$ and quantum fluctuations $\alpm
(x,t)$
\be
\ppm(x,t)=\pm\pi \colfield_0(x) + \frac{\almp(x,t)}{\pi
\colfield_0(x)}.
\ee
From \re{ppmconstraints} we obtain
\be\label{eq:afluct}
(\p_t\mp\p_q)\alpm = 0
\ee
indicating that the fluctuations $\alpm$ are chiral;
$\alp=\alp(t+q)$ are left-- and $\alm=\alm(t-q)$ right--moving
solutions.

Now consider the asymptotics
\be
q\rightarrow\infty\Rightarrow \pi
\colfield_0\stackrel{\rm\footnotesize \re{DJbg}}{\longrightarrow}x
\stackrel{\rm\footnotesize\re{tof}}{\longrightarrow}\frac{1}{2}e^q.
\ee
Substituting for $\colfield_0$ and $x$ in \re{colvop}, we see that
the operators \re{pookaka} in this limit become
\be\label{eq:ta}
T^{\pm}_n \sim \int\!\frac{dq}{2\pi}e^{n(q\pm t)}\alpm.
\ee
The integral here projects out modes from the operator $\alpm$ which
create chiral field components that behave as $e^{n(q\pm t)}$. The
comparison of the vertex operators in \re{ta} with ones in
\re{vertexops} show that they obey the same general properties,
although there are a few differences.

Firstly, there is apparently a factor of $e^{-2\phi}$ lacking from
the former. This may be attributed to the use of the redefined
tachyon in drawing any equivalence between MQM in the collective
field theory and 2D string theory, as noted in the previous
subsection \re{Romania}. Secondly, the integral in the collective
field theory vertex operators is over the target space field $q$
defined on the world--sheet, whereas in the string theory
formulation, the integral is over the world--sheet coordinates. To
confirm agreement beyond any doubt means to compare correlation
functions of MQM with those of 2D quantum gravity \cite{Moore91}.

Thus the matrix operators \re{pookaka} correspond to tachyon vertex
operators in Minkowski space--time with the imaginary momenta on a
lattice $k=in$. Other momenta can be realised by analytically
continuing to the entire complex plane
\be\label{eq:MQMtachV}
V^{\pm}_k \sim T^{\pm}_{-ik}=e^{\mp ikt}\tr(M\mp P)^{-ik}.
\ee
A Euclidean version is obtain by migrating to the Euclidean time
coordinate of 2D string theory, $t\rightarrow -iX$. In that case, we
get vertex operators with Euclidean real momenta $p=\pm n$ as in
\S\ref{sec:pertCFT}. We will see in the next chapter
that considering the momentum to be on a lattice will lead to an
interpretation in terms of the Toda lattice hierarchy.

It is also possible to write vortex operators that to show that
discrete states also appear in MQM \cite{SergePhD}, although, for
our purposes, the vertex operators are all that we need.

\EndOfChapter

In conclusion, we see that the singlet sector of MQM affords us
sufficient insight to describe tachyons. Higher representations
contain vortex operators \cite{SergePhD} but would add an
unnecessary complication to our formalism. We now turn our attention
to the tachyon perturbations themselves. We shall see how to
simplify MQM by using a certain set of canonical coordinates known
as light--cone coordinates, and study how the tachyon perturbations
can be described using the Toda lattice hierarchy.

\chapter{Tachyon Perturbations}
\label{ch:TAC}
The necessary background knowledge in string theory was provided in
the review of chapter \ref{ch:STR}, in chapter \ref{ch:MAT} the
mathematical machinery of matrix models was introduced, and its
relation to string theory elucidated in chapter \ref{ch:MQM}. In the
present chapter, tachyon perturbations will be introduced in order
to produce a non--trivial background in 2D string theory, as
discussed in \S\ref{sec:pertCFT}. It will then be shown that they
are described by the Toda hierarchy.

\section{Chiral Representation of MQM}
Before treating tachyon perturbations, it will be convenient to
introduce a representation of MQM which shall greatly facilitate
certain calculations --- MQM in chiral phase space coordinates
\cite{AKK02}. Computation of the $S$--matrix of 2D string theory
will then prove trivial and we shall use this formalism to perturb
the ground state of the theory, introducing a time--dependant
tachyon condensate.

\subsection{Left and Right Coordinates}
\label{sec:MQMlight}
We wish to work in a representation where the creation and
annihilation operators for the tachyons are diagonal. So, we begin
by defining the following matrix quantum operators associated to
left-- and right--moving tachyons
\be\label{eq:Xpm}
\Xpm=\frac{M\pm P}{\sqrt{2}},\qquad P\equiv-i\der{}{M},
\ee
obeying
\be\label{eq:lightcom}
[(\Xp)^i_j,(\Xm)^k_l]=-i  \;\d^{i}_{l}\d^k_j.
\ee
We define ``right'' and ``left'' Hilbert spaces $\mathcal{H}_+$ and
$\mathcal{H}_-$ whose respective elements $\Phi_+(X_+)$ and
$\Phi_-(X_-)$ are functions of $X_+$ or $X_-$ only. Then
\be
\Xpm\Phipm(\Xpm)=\xpm\Phipm(\Xpm).
\ee
and the inner product is defined as
\be\label{eq:IP}
\la \Phipm|\Phipm' \ra = \int \!d\Xpm
\;\Phipm^\dag(\Xpm)\Phipm'(\Xpm),\qquad
\Phipm(\Xpm)\in\mathcal{H}_\pm.
\ee

The eigenvalues $\xpm$ of the matrices $\Xpm$ are chiral coordinates
not in the target space but in the {\em phase space} spanned by the
eigenvalues of $X$ and $P$, namely, by $x$ and $p$. From
\re{lightcom} we see that the momentum operator of the left Hilbert
space is the position operator of the right one and the wave
functions of the two spaces are thus related by Fourier
transformation.

The inverse oscillator Hamiltonian of the double scaling limit in
these chiral variables is written
\be
\label{eq:LCH}
H=\hf\tr (P^2 - M^2) = -\hf\tr (\Xp\Xm+\Xm\Xp)
\ee
and the Schr\"{o}dinger equation pertaining to it is of first
order in those variables
\be
\p_t\Phipm(\Xpm,t)=\mp\tr\(\Xpm\der{}{\Xpm}+\frac{N}{2}\)\Phipm(\Xpm,t)
\ee
which we solve to get the time dependence of the wave functions
\be
\Phipm(\Xpm,t)=e^{\mp\hf N^2t}\Phipm(e^{\mp t}\Xpm).
\ee

\subsection[Decomposition into Irreducible $\gr{SU}{N}$ Representations]{
Decomposition of Wave Function into Irreducible $\gr{SU}{N}$ Representations}
\label{sec:sunrep}
The matrix argument of the wave functions can be diagonalised into
eigenvalues $\xpm$ and angular variables $\O_\pm$ as in
\S\ref{sec:evals}. Since we have axiomatically $\gr{U}{N}$ invariant
matrices, the angular argument of the wave functions belongs to
$\gr{SU}{N}$. Hence, the wave functions are functions on the group
and may be decomposed into a sum of irreducible $\gr{SU}{N}$
representations\footnote{The existence of such a decomposition is
justified by the Peter--Weyl theorem.}
\be
\label{eq:sudecomp}
\Phipm(x,\O)=\sum_r\sum^{\dim(r)}_{I,J=1} D_{IJ}^{(r)}(\O)
\Phi_{\!_\pm\, I,J}^{(r)} (x)
\ee
where the representations are classified by their highest weight
$r$, $D_{IJ}^{(r)}(\O)$ is an element indexed by $I,J$ of the
representation matrix corresponding to $\O \in \gr{SU}{N}$ in the
representation $r$, and $\dim (r)$ denotes the dimension of the
representation $r$. The coefficients $\Phi^{(r)}_{\!_\pm\,I,J} (x)$
are functions only of the $N$ eigenvalues
$x_{\!_\pm\,1},\ldots,x_{\!_\pm\,N}$.

The Hilbert space structure is induced by the inner product
(\ref{eq:IP}) which is written under the decomposition
(\ref{eq:sudecomp}) as
\be\label{eq:suIP}
\la \Phipm | \Phipm' \ra = \sum_r \frac{1}{\dim(r)}
\int\prod_{k=1}^{N}\!d x_{\pm,k} \; \D^2(\xpm)\sum_{I,J=1}^{\dim(r)}
\Phi_{\!_\pm\, I,J}^{(r)\dag} (\xpm)
\Phi_{\!_\pm\, I,J}^{\prime (r)}(\xpm),
\ee
where the angular part of the wave functions was integrated out
leaving the square of the Vandermonde determinant $\D(\xpm)$. This
motivates the usual redefinition, as in \S\ref{sec:nonintferm},
\be
\label{eq:redef}
\Psi_{\!_\pm\, I,J}^{(r)}(\xpm)=\D(\xpm)\Phi_{\!_\pm\, I,J}^{(r)}(\xpm).
\ee
which absorbs the Vandermonde from the scalar product.

The Hilbert spaces $\mathcal{H}_\pm$ now decompose into a direct sum
of invariant proper subspaces which carry the irreducible
representations of $\gr{SU}{N}$, namely,
\be
\mathcal{H}=\bigoplus_r \mathcal{H}_{\!_\pm}^{(r)}
\ee
where the inner product of subspace $r$ is given by the
corresponding term in the outer sum of (\ref{eq:suIP}). These
subspaces are eigenspaces of the Hamiltonian \re{LCH} and are
therefore invariant under its action, so the decomposition is
preserved by the dynamical evolution.

Now, the action of the Hamiltonian on the wave functions
$\Psi^{(r)}_{\!_\pm\,I,J}$ is given by a matrix differential
operator $H^{(r)}_{\pm,I,J}$ and the Schr\"{o}dinger equation
becomes
\be\label{eq:shishkabab}
i\hbar\der{\Psi^{(r)}_{\!_\pm\,I,J}}{t}=\sum_{K=1}^{\dim (r)}
H^{(r)}_{\!_\pm\,I,K}\Psi^{(r)}_{\!_\pm\,K,J}.
\ee
The outer index $J$ is completely free and can be omitted --- we
shall now speak of functions $\vec{\Psi}^{(r)}_{\pm}=
\lb\Psi^{(r)}_{\pm,J}\rb^{\dim(r)}_{J=1}$ belonging to the
irreducible representation $r$ which transform as\footnote{Omitting
one of the indices will cause the inner product \re{suIP} to be
multiplied by a factor of $\dim(r)$ from the implied sum over the
omitted index.}
\be
\vec{\Psi}^{(r)}_{\!_\pm\,I}(\O^\dag X_\pm \O) = \sum\limits_J
D_{IJ}^{(r)}(\O) \vec{\Psi}^{(r)}_{\!_\pm\,J} (X_\pm).
\ee

An advantage of the chiral formulation is that for any given
representation $r$, the Hamiltonian \re{LCH} reduces on the
functions $\vec{\Phi}^{(r)}_{\!_\pm}=
\lb\Phi^{(r)}_{\!_\pm\,J}\rb^{\dim(r)}_{J=1}$ to its radial part
only and does not contain any angular interaction like the usual
Hamiltonian \re{HMQM}
\be
H^{\pm}_0 = \mp i \sum\limits_k \(x_{\!_\pm\,k}\frac{\p}{\p
x_{\!_\pm\,k}}+ \frac{N}{2}\).
\ee
With the redefinition (\ref{eq:redef}) and the explicit form of the
Vandermonde (\ref{eq:Vandermonde}), the constant term changes and
the Hamiltonian on the functions $\vec{\Psi}^{(r)}_{\!_\pm}$ is
\be
\label{eq:hambam}
\tlH^{\pm}_0=\mp i\sum_k (x_{\!_\pm\,k}\frac{\p}{\p x_{\!_\pm\,k}} +
\hf).
\ee
This can be proved as follows: Omitting the $\pm$ and the $(r)$
representation indices for the sake of clarity, from the eigenvalue
equation
\beq\label{eq:eve}
&&H_0 \vec{\Phi} = E \vec{\Phi}\nn\\&\Leftrightarrow& \tlH_0
\vec{\Psi} = E \vec{\Psi} = \D E\Phi = \D H_0 \vec{\Phi}
\eeq
we have\footnote{Note that $$\sum_{k=1}^N x_k\p_{x_k}\prod_{i<j}^N(x_j-x_i) =
\sum_{k=1}^{N-1} k \prod_{i<j}^N(x_j-x_i) = \frac{N(N-1)}{2}\prod_{i<j}^N(x_j-x_i)$$}
\beq
\tlH_0 \D \vec{\Phi}
&=& \mp i\sum_{k=1}^N (x_{k}\frac{\p}{\p x_{k}}+\hf) \D \vec{\Phi}\nn\\
&=& \sum_{k=1}^N \left\lb(x_k\p_{x_k}\D)\vec{\Phi} + \D(
x_k\p_{x_k}\vec{\Phi})\right\rb
+ \frac{N}{2}\D \vec{\Phi}\nn\\
&=& \frac{N(N-1)}{2}\D \vec{\Phi}
+ \D\(\sum_{k=1}^N x_k\p_{x_k}\) \vec{\Phi}
+ \frac{N}{2}\D \vec{\Phi}\nn\\
&=& \D\sum_{k=1}^N \(x_k\p_{x_k} + \frac{N}{2}\)\vec{\Phi}\nn\\
&=&\D H_0\Phi\nn
\eeq
as expected from \re{eve} proving that the Hamiltonian \re{hambam}
is correct for the wave functions $\vec{\Psi}$.

As usual, in the singlet sector, the wave functions $\Psi_\pm^{\rm
(singlet)}$ are completely antisymmetric (\S\ref{sec:nonintferm}).
And in \S\ref{sec:DJ}, we saw that the singlet sector of MQM
contains tachyon excitations we are interested in, so we restrict
our subsequent analysis to this sector. Let us first consider the
ground state of the system and examine then relationship between
$\psip(\xp)$ and $\psim(\xm)$.

\subsection{$S$--Matrix}
\label{sec:smatrix}
The scattering matrix of 2D string theory can be calculated from the
parabolic cylinder functions \re{buddah}. But its computation in the
chiral formulation of MQM is much more straightforward and arrives
at the same result.

The ground state of the MQM is obtained by filling all energy levels
up to some fixed energy; the Fermi energy $\fermiE$, which is
written in terms of the chemical potential $\mu$ as $\fermiE=-\mu$.
It represents the unperturbed 2D string theory background and
consists of a system of noninteracting fermions described by a
Slater determinant as in \re{Slater},
\be\Psipm^{\rm GS}(\xpm)=\frac{1}{\sqrt{N!}}\left|
\begin{array}{ccc}
\psi_{\!_\pm\,E_1}(x_{\!_\pm\,1}) & \cdots & \psi_{\!_\pm\,E_N}(x_{\!_\pm\,1}) \\
\vdots && \vdots\\
\psi_{\!_\pm\,E_N}(x_{\!_\pm\,1}) &\cdots& \psi_{\!_\pm\,E_N}(x_{\!_\pm\,N}) \\
\end{array}
\right|,\qquad E_N=\fermiE
\ee

The one particle wave functions
$\psi_{\!_\pm\,E_k}(x_{\!_\pm\,k})\equiv\psipm^E(x_{\!_\pm\,})$
which constitute the Slater determinant are solutions of the
Schr\"{o}dinger equation with the Hamiltonian $H_0$ from \re{hambam}
\beq
H_0 \psipm^E(\xpm)&=& E \psipm^E(\xpm) \nn\\
\Rightarrow \p_{\xpm}\psipm^E(\xpm)&=& \frac{1}{\xpm}(\pm iE - \hf)
\psipm^E(\xpm) \nn\\
\Rightarrow \psipm^E(\xpm) &=& \frac{1}{\sqrt{2\pi}}\xpm^{\pm iE -
\hf}\label{eq:lrwf} \nn
\eeq
where the normalisation is chosen such that we get the proper orthonormality relation,
\be
\int\!d\xpm\psipm^E(\xpm)\psipm^{E'}(\xpm)=\d(E-E').
\ee
Since $\xpm$ are conjugate canonical variables, the left and right
chiral representations are related by a unitary Fourier
transformation,
\be
\psim(\xm)=[\hat{S}\psip](\xm)=
\int\limits_{-\infty}^{\infty}\!d\xp\,K(\xm,\xp)\psip(\xp).
\ee
The kernel is
\be
K(\xm,\xp)=\frac{1}{\sqrt{2\pi}}e^{i\xp\xm},\qquad \xpm\in\bbR
\ee
where the unrestricted domain for $\xpm$ indicates the Fermi sea
straddles the inverse oscillator potential, the two sides of the
potential being connected only through nonperturbative tunneling
processes. As mentioned in \S\ref{sec:fermisea} we are only
interested in perturbative effects, so it would suffice to consider
fermions from just one side of the inverse oscillator potential,
giving us the liberty of restricting the domain for $\xpm$ to
positive semidefinite values. In this case, the kernel is given by
one of two equally valid possibilities, depending on whether the
wave function is symmetric or antisymmetric, which in turn depends
on the two independent boundary conditions which would be imposed on
the original second order Hamiltonian\footnote{\label{fn:n}In other
words, $
\int\limits_{-\infty}^{\infty}\!d\xp\,e^{i\xp\xm}\psip(\xp)=
\int\limits_{0}^{\infty}\!d\xp\,\(e^{i\xp\xm}\psip(\xp)\pm
e^{-i\xp\xm}\psip(\xp)\) $ where we take the plus (minus) if
$\psip(\xp)$ is even (odd).}
\beq\label{eq:kern1}
K(\xm,\xp)&=&
\sqrt{\fracL{2}{\pi}}\cos(\xm\xp)  ,\qquad\quad \xpm\geq 0,\\
&{\rm or}& \nn\\
K(\xm,\xp)&=&i\sqrt{\fracL{2}{\pi}}\sin(\xm\xp)
 ,\qquad\quad \xpm\geq 0.\label{eq:kern2}
\eeq
Choosing for the sake of simplicity to work with the first (cosine)
kernel, the explicit relation between the left and right wave
functions in \re{lrwf} is
\beq
[\hat{S}^{\pm1}\psipm^E](\xmp)
&=& \frac{1}{\sqrt{2\pi}}\sqrt{\frac{2}{\pi}}
\intl_0^\infty\!d\xpm\;\cos(\xmp\xpm) \xmp^{\pm iE -\hf} \nn\\
&=& \frac{1}{\pi}\intl_0^\infty\!d\xmp\;\hf (e^{i\xp\xm}+e^{-i\xp\xm})
\xpm^{\pm iE -\hf} \nn\\
&=&\frac{1}{2\pi}\intl_0^\infty\!d\xmp
\[\(\frac{i}{\xmp}\)^{\pm iE +\hf} \!\!\!+ \(\frac{-i}{\xmp}\)^{\pm iE +\hf}\] e^{-\xp}
\xpm^{(\pm iE+\hf)-1} \nn\\
&=& \frac{1}{2\pi}\G(\pm iE+\hf) \[(e^{i\nf{\pi}{2}})^{\pm iE +\hf} + (e^{i\nf{\pi}{2}})^{-(\pm iE +\hf)}\]
\xmp^{(\mp iE+\hf)-1}\nn\\
&=& \sqrt{\frac{2}{\pi}}\G(\pm iE +\hf)
\cosh\[\frac{\pi}{2}\(\frac{i}{2}\mp E\)\]\psimp^E(\xmp).\nn
\eeq
where
$$
\G(z)=\int_0^\infty\!dx\;e^{-x}x^{z-1}
$$
is the Gamma function.

In other words, we have the fermionic $S$--matrix of MQM in the
double scaling limit in the energy
representation
\be
\psimp(\xmp)=[\hat{S}^{\pm1}\psipm^E](\xmp)= \CR(\pm E) \psimp^E(\xmp),
\ee
where the reflection coefficient $\CR(E)$ is given by the pure phase ($\CR^\dag(E)\CR(E)=\CR(-E)\CR(E)=1$),
\be\label{eq:CR}
\CR(E)=\sqrt{\frac{2}{\pi}}\G(iE +\hf)
\cosh\[\frac{\pi}{2}\(\frac{i}{2}- E\)\].
\ee
making the $S$--matrix unitary.
If the sine kernel \re{kern2} were chosen instead, then we would have
\be
\CR'(E)=\sqrt{\frac{2}{\pi}}\G(iE +\hf)
\sinh\[\frac{\pi}{2}\(\frac{i}{2}- E\)\].
\ee

The scattering amplitude between any
two in and out states is
\be
\la\psim | \hat{S}\psip\ra = \la \hat{S}^{-1}\psim|\psip\ra = \la
\psim|K|\psip\ra \equiv \intl_0^\infty\!d\xp d\xm\;\psim^\dag(\xm)
K(\xm,\xp)\psip(\xp)
\ee
where $K$ is the Fourier kernel \re{kern1}. The orthogonality
relation between the wave functions is thus
\be
\la\psim^E|\psim^{E'}\ra= \la\psim^E|K|\psip^{E'}\ra=\CR(E)\d(E-E').
\ee

Equipped with the $S$--matrix, we are now in a position to introduce
a nonzero tachyon background to the system.

\newpage
\section{Tachyon Perturbations}

At first sight, it might be tempting to introduce perturbing terms
directly into to the MQM action as matrix realisations of the
tachyon vertex operators (see \S\ref{sec:pertCFT}). Indeed, they are
known \re{pookaka} albeit only asymptotically, however, they disappear in
the double scaling limit where only the quadratic part of the
potential contributes. Instead, we change not the potential of the
system but its state directly; it is the state which describes the
background.

\subsection{The Perturbed Wave Functions}
Write the perturbed single particle wave functions as \cite{AKK02}
\be\label{eq:pertwf}
\tpsipm^E(\xpm)=e^{\mp\vphipm(\xpm;E)}\psipm^E(\xpm).
\ee
If we wish to consider tachyons with imaginary momenta on an equally
spaced lattice as in the compactified Euclidean theory we are
interested in, the phase is
\be\label{eq:phase1}
\vppm(\xpm;E) = \hf \phi(E) + V_{\!_\pm\;}(\xpm) +
v_{\!_\pm\;}(\xpm;E),
\ee
where $V_{\!_\pm\;}$ are polynomials in $\xpm^{\nf{1}{R}}$ which vanish at
$\xpm=0$
\be
V_{\!_\pm\;}(\xpm)=\sum_{k=1}^\infty\tpmn{k}\xpm^{\nf{k}{R}},
\ee
and the function
\be
v_{\!_\pm\;}(\xpm;E) = -\sum_{k=1}^\infty\frac{1}{k}v_{\pm k}(E)\xpm^{\nf{-k}{R}}
\ee
becomes zero as $\xpm\rightarrow\infty$.

The motivation for introducing such a perturbation is as follows: To
produce a tachyon background, we wish to create a coherent state of
tachyons which makes nonzero the expectation value of all
orders\footnote{Note the subtlety here that one point correlation
functions are always zero in CFT.} of the tachyon vertex
operators\footnote{Recall that we chose $k\geq0$, defining left and
right vertex operators instead of a single vertex operator for all
momenta --- \S\ref{sec:vervor}.}; $\la
(V_k^{\pm})^n\ra\neq0\;\forall \;n>0$ for some $k$. Initially we
might try to perturb the state thus:
$$
\tpsipm^E(\xpm)=e^{\mp\vphipm}\psipm^E(\xpm),\qquad
\vphipm=\sum\limits_k \tpmn{k}V^{\pm}_k.
$$
This can be viewed as a mapping from one Hilbert space containing
$\psipm^E(\xpm)$ to another which contains $\tpsipm^E(\xpm)$. In the
asymptotic limit it has a realisation in terms of matrix operators,
$$
\xpm\rightarrow\infty\quad\Rightarrow\quad\vphipm \stackrel{{\rm\footnotesize
\re{MQMtachV}}}{\longrightarrow} \sum\limits_k \tpmn{k} \Tr(M\pm
P)^{\nf{k}{R}} \stackrel{{\rm\footnotesize \re{Xpm}}}{\longrightarrow}
\sum\limits_k \tpmn{k} \xpm^{\nf{k}{R}}
$$
which explains the form of $V_{\!_\pm\;}(\xpm)$ in \re{phase1}.
However, this form of the vertex operators is valid only
asymptotically and the tail $v_{\!_\pm\;}(\xpm;E)$ is therefore
included to take into account possible corrections for the exact
form of $V^{\pm}_k$ in the non--asymptotic region.

As we shall soon see, this case \re{phase1} has a description in
terms of the Toda lattice hierarchy, where the couplings $\tpmn{k}$
of the tachyons will be related to the Toda times. The parameter $R$
measures the spacing of the momentum lattice and plays the role of
the compactification radius of the corresponding Euclidean theory.

The zero mode $\phi(E)$ and tail $v_{\!_\pm\;}(\xpm;E)$ are subject
to the requirement that the $S$--matrix remains diagonal on the
perturbed wave functions,
\be\label{eq:kakapoo}
\hat{S}\tpsip^E(\xp)\equiv\tpsim^E(\xm).
\ee

The condition \re{kakapoo} can be seen as a statement that the left
and right modes are alternative representations of the same physical
state.

Evidentally, the perturbed wave functions are no longer
eigenfunctions of the one particle Hamiltonian \re{hambam}, but are
instead eigenfunctions of the operators
\be
H_{\!_\pm}(E)=H^{\pm}_0+\xpm\p\vppm(\xpm;E).
\ee
The $H_{\!_\pm}(E)$ have an explicit energy dependence through
$v_{\!_\pm\;}(\xpm;E)$ and can not therefore be considered as
Hamiltonians, but were we to define new Hamiltonians $H_{\!_\pm}$
acting on $\tpsipm^E(\xpm)$ which solve
\be\label{eq:hamstring}
H_{\!_\pm}=H^{\pm}_0+\xpm\p\vppm(\xpm;H_{\!_\pm}),
\ee
then $\tpsipm^E(\xpm)$ would be eigenstates of these Hamiltonians.
The new Hamiltonians do not describe the time evolution of the
system, indeed, the Fermi sea is stationary with respect to the time
defined by these Hamiltonians. However, the wave functions
$e^{\mp\nf{t}{2}}\tpsipm^E(e^{\mp t}\xpm)$ do solve a one particle
Schr\"{o}dinger equation which is of the same form as
\re{shishkabab} and it is from this which we compute the time
evolution of the Fermi sea.

The condition \re{kakapoo} implies that the chiral Hamiltonians
$H_{\!_+}$ and $H_{\!_-}$ are physically indistinct --- the phase
space trajectories they generate coincide and they both define the
same action of a self--adjoint operator $H$ in the
$\pm$--representations.

In the quasiclassical limit $\fermiE\rightarrow-\infty$, each energy
level in the Fermi sea corresponds to a certain one--particle
trajectory in the phase space of $\xm,\xp$ variables. The Fermi sea
in the ground state can be viewed as a stack of all classical
trajectories with $E < \fermiE=-\mu$, it acts as an incompressible
fluid. If the filled region of phase space is simply connected, it
is completely characterised by the boundary of the Fermi sea; the
curve representing the trajectory of the fermion with highest energy
$\fermiE$.

To elucidate the meaning of the condition \re{kakapoo}, we examine
the orthonormality relation $\la\tpsim^{\Em}|\tpsim^{\Ep}\ra$ under
this trajectory in the quasiclassical limit: (For the purposes of
computing the saddle point equation which determines the zero mode
$\phi(E)$ and tail $v_{\!_\pm\;}(\xpm;E)$, we need only consider one
of the complex exponentials in the cosine below.)
\beq
\la\tpsim^{\Em}|\tpsim^{\Ep}\ra&=&\la\tpsim^{\Em}|K|\tpsip^{\Ep}\ra\nn\\
&=&\!\!\frac{1}{\pi\sqrt{2\pi}}\!\iint\limits_0^\infty\!\!d\xp d\xm
\cos(\xp\xm)\xm^{i\Em-\nf{1}{2}}\xp^{i\Ep-\nf{1}{2}}e^{-i\vpm(\xm;\Em)-i\vpp(\xp;\Ep)}\nn\\
&\approx&\!\!\frac{1}{2\pi\sqrt{2\pi}}\!\iint\limits_0^\infty\!\frac{d\xp
d\xm}{\sqrt{\xp\xm}}
\exp\Big[i\xp\xm +i\Ep\ln\xp + i\Em\ln\xm \nn\\
&&\qqquad\qqquad\qqquad-i\vpm(\xm;\Em)-i\vpp(\xp;\Ep)\Big].\nn
\eeq
Equating the partial derivatives of the exponent with respect to
$\xp$ and $\xm$ to zero, we find the two saddle point equations
\be\label{eq:saddle}
\xpm\xmp=-\Epm+\xpm\der{}{\xpm}\vppm.
\ee
These two equations should be consistent and describe the same
trajectory in the phase space. Thus, since we are considering the
trajectory for the fermion of highest energy in the quasiclassical
limit, we have $\Ep=\Em=\fermiE=-\mu\rightarrow-\infty$ and
\re{saddle} effectively becomes
\be\label{eq:Fred}
\xpm\xmp=\mu+\xpm\p_{\!_\pm}\vppm(\xpm;-\mu).
\ee
That $\Ep$ and $\Em$ should coincide is nothing but a restatement of
the fact that the $H_{\!_+}$ and $H_{\!_+}$ Hamiltonians are
physically indistinct,
\be
H_{\!_+}=E \Leftrightarrow H_{\!_-}=E.
\ee

We now move on to show how the perturbed system has a description in
terms of the Toda lattice hierarchy \cite{AKK02}.

\subsection{Toda Lattice Description}
Remaining in the energy representation where $\hat{S}$ is diagonal,
the energy $E$ plays the role of the coordinate along the lattice of
tachyon energies; $E_n=ip_n$.

Define
\be
\hxpm\equiv\xpm, \qquad \delpm \equiv \der{}{\xpm}.
\ee
It follows immediately, through their action on the non--perturbed
wave functions \re{lrwf}, that they are related by\footnote{Strictly
speaking, in order to satisfy \re{Jazz}, it would be necessary to
utilise the cosine kernel \re{kern1} for even $n$ and the sine
kernel \re{kern2} for odd $n$ in the definition of the Fourier
transform $\hS^{\pm 1}$ (see footnote \ref{fn:n} on page
\pageref{fn:n}). This detail will not change the results here, but
it should be understood that any rigourous proof would entail a
partition of the problem into cases of even and odd $n$.}
\be\label{eq:Jazz}
\pm i\delpm=\hS^{\mp1}\hxmp\hS^{\pm1}.
\ee
Now, starting with the obvious Heisenberg commutator relation
$[\delpm,\hxpm]=1$, we get, with $\tlH^{-}_0$ from \re{hambam},
\beq
\hxm\hS\hxp\hS^{-1}-\hS\hxp\hS^{-1}\hxm&=&i\label{eq:IneedCoffee}\\
\Rightarrow \tlH^{-}_0 &=& i (\xm \frac{\p}{\p \xm} + \hf)\nn\\
&=& -\hf\(\hxm\hS\hxp\hS^{-1}+\hS\hxp\hS^{-1}\hxm\)\label{eq:Fiji}\\
&=& \hS\tlH^{+}_0\hS^{-1}
\eeq
And since $\tlH^{\pm}_0\psipm^E(\xpm)=E\psipm^E(\xpm)$ we may simply
write
\be\label{eq:eo}
\tlH^{\pm}_0=\hE
\ee
where $\hE$ is the energy operator.

In the energy representation, acting on the unperturbed wave
functions \re{lrwf}, we see that the position operator can be
written in terms of the energy shift operator,
\be
\hxpm\psipm^E(\xpm)=\xpm\psipm^E(\xpm)
=\psipm^{E\mp i}(\xpm)
=e^{\mp i\p_E}\psipm^E(\xpm)
\equiv\ho^{\pm1}\psipm^E(\xpm)\nn
\ee
\be\label{eq:eeso}
\Rightarrow \hxpm=\ho^{\pm1},\qquad \ho\equiv e^{\mp i\p_E}.
\ee
The energy operator \re{eo} can be identified with the discrete node
variable $\hbar s$ of the standard Toda lattice described in
\S\ref{sec:Lax} and the energy shift operator is conjugate to the
node shift operator,
\beq\label{eq:esopa}
E   \longleftrightarrow  -\hbar s,   &\qquad&
\ho \longleftrightarrow \ho
\eeq

With the choice for the perturbing phase being \re{phase1}, the
perturbed wave functions \re{pertwf} can now be constructed from the
unperturbed ones through the action of a unitary operator, called a
``wave operator''\footnote{In some references, they are called
``dressing operators'' \cite{Takasaki} but we used this term to
describe a slightly different operator in \S\ref{sec:Lax}. The
terminology here is consistent with \cite{Kostov02}.}. The perturbed
state is then called a ``dressed state''
\be
\tpsipm^E(\xpm)=e^{\mp\vphipm(\xpm;E)}\psipm^E(\xpm)=\CWpm\psipm^E(\xpm)
\ee
\be\label{eq:miniskirt}
\CWpm =e^{\mp i\nf{\phi}{2}}\ \(1+\sum\limits_{k\ge 1}\wpm{k}\ho^{\mp \nf{k}{R}}\)
\exp\(\mp i \sum\limits_{k\ge 1} \tpmn{k} \ho ^{\pm \nf{k}{R}}\).
\ee

By comparison with \re{wave}, we see that the coupling constants
$\tpmn{k}$ play the role of Toda times $t=\lbrace t_{\pm n}
\rbrace_{n=1}^{\infty}$ of \S\ref{sec:Lax}. And the perturbed wave
functions $\tpsipm^E(\xpm)$ are none other than the Baker--Akhiezer
wave functions $\Psi$ of \re{LaxEV}. From \re{LaxEV}, we find that
the Lax operators act as the phase space variables on the perturbed
wave function,
\be\label{eq:LtoX}
L_{\pm}\tpsipm^E(\xpm)= \xpm \tpsipm^E(\xpm)
\ee

The canonical coordinates congruent to $\hxpm$ in the perturbed wave
function basis are then precisely the Lax operators as defined in
\re{dressedl},
\be\label{eq:Numea}
\Lpm=\CWpm\ho^{\pm1}\CWpm^{-1} = e^{\mp i\phi/2} \ho^{\pm
1}e^{\pm i\phi/2} \(1 +\sum_{k\ge 1} a(E)_{\pm k}   \ho^{\mp
\nf{k}{R}}\)
\ee

Likewise, the energy operator \re{eo} is written in terms of the
Orlov--Shulman operators \re{OS}, with an extra minus sign owing to
the association between $E$ and $s$ \re{esopa},
\be\label{eq:Tahiti}
M_\pm = -\CWpm \,\hE\, \CW^{-1}_\pm = \frac{1}{R} \sum _{k\ge 1} k
t_{\pm k}   \Lpm^{ k/R}- \hE + \frac{1}{R}\sum_{k\ge 1} v_{\pm k}
\Lpm^{-k/R}.
\ee

The single particle Hamiltonian can be written in terms of the Lax
operators also, as in \re{LaxHams},
\be
H_{\pm n} =\pm(L^{\nf{n}{R}}_\pm)_{\substack{> \\[-3pt] <}} +
\hf(L^{\nf{n}{R}}_{\pm})_0,\qquad n>0
\ee
These Hamiltonians constitute one of the ingredients of the Toda
hierarchy as mentioned in \S\ref{sec:integr}, namely, a set of
commuting flows.

Strictly speaking, since the expansions if $L_\pm$ and $M_\pm$
contain powers of $\ho^{\nf{1}{R}}$ rather than powers of $\ho$, the
correct identification of the Toda lattice operators used here with
the original operators used in chapter \ref{ch:MAT} should read
\beq
L   \longleftrightarrow  L_{+}^{\nf{1}{R}},  &\qquad&
\bL \longleftrightarrow  L_{-}^{\nf{1}{R}}\\
M   \longleftrightarrow  R M_+,   &\qquad& \bM \longleftrightarrow R
M_-.
\eeq
Together with \re{esopa} we now have all the elements of the Toda
hierarchy, as summarised in table \ref{tab:toda} and all subsequent
relations follow in the same manner as in chapter \ref{ch:MAT}.

\begin{table}[!ht]\tablabel{tab:toda}
\begin{center}
\begin{tabular}{|ccc||ccc|}
\hline
\bf Ch. \ref{ch:MAT} &\vline& \bf Ch. \ref{ch:TAC} & \bf Ch. \ref{ch:MAT} &\vline& \bf Ch. \ref{ch:TAC}\\
\hline\hline
$-\hbar s$   &$\leftrightarrow$& $E$          & $\ho$  &$\leftrightarrow$& $\ho$              \\
\hline
$L$   &$\leftrightarrow$& $L_{+}^{\nf{1}{R}}$ & $\bL$  &$\leftrightarrow$& $L_{-}^{\nf{1}{R}}$\\
\hline
$M$   &$\leftrightarrow$& $RM_{+}$            & $\bM$  &$\leftrightarrow$& $RM_{-}$           \\
\hline
$\CW$ &$\leftrightarrow$& $\CW_{+}$           & $\bCW$ &$\leftrightarrow$& $\CW_{-}$          \\
\hline
$t_{\pm n}$ &$\leftrightarrow$& $\pm t_{\pm n}$& $\hbar$&$\leftrightarrow$& $i$          \\
\hline
\end{tabular}
\bf\caption{\rm Relationship between the MQM Toda hierarchy
presented here and the standard Toda hierarchy which is described in
chapter \ref{ch:MAT}.}
\end{center}
\end{table}

\subsection{String Equation}
The Toda hierarchy exhibits an infinite set of partial differential
equations for the coefficients $w_{\pm k}$ of the wave operators
\re{miniskirt}. The first of those, in particular, regards the zero
mode,
\be
i\der{}{t_{1\phantom{-}}}\der{}{t_{-1}}\phi= A[\phi]-A^*[\phi],\qquad
A[\phi]=e^{i\phi(E) - i\phi(E+\nf{i}{R})}.
\ee

Ordinarily, one should have at hand some kind of initial condition
which selects a unique solution for the equations of the hierarchy.
This role can be played by the partition function of the system in
the absence of any tachyon background \cite{SergePhD}. Or, the
initial condition can come under the vise of some constraints on the
Lax and Orlov--Shulman operators which preserve the hierarchial
structure of the system, called the {\em string equations}. String
equations can provide a shortcut to the $\tau$ function, yielding a
set of algebraic or first--order partial differential equations in
place of second--order partial differential equations.

The condition \re{condress} on the wave operators that
$\CW^{-1}_{+}\CW_{-}$ be independent from the couplings $t_{\pm n}$
is equivalent to the condition \re{kakapoo}: Define $\hat{\i}$ such
that $\hat \i \psipm^E=\psimp^E$, then $\hat{S}=\hCR\hat{i}$, where
$\hCR = \CR(\hE)$ with $\CR(E)$ from \re{CR}. From these
definitions, \re{kakapoo} can be recast as
\beq
\hat{S}\tpsip^E(\xp) &=&\tpsim^E(\xm) \nn\\
\Rightarrow \hCR\hat{\i}\CW_{+}\psip^E(\xp) &=& \CW_{-}\psim^E(\xm) \nn\\
\Rightarrow \CW^{-1}_{+}\CW_{-} &=& \hCR\label{eq:Tonga}
\eeq
which is independent of the couplings since $\hCR$ does not depend
on the couplings. It also yields the missing relationship between
the left and right wave operators.

It is precisely from this which we obtain the string equations:
\beq
[\delpm,\hxpm]&=&1\nn\\
\Rightarrow\  i &\stackrel{{\rm \footnotesize\re{IneedCoffee}}}{=}&
\hxm\hS\hxp\hS^{-1}-\hS\hxp\hS^{-1}\hxm \nn\\
\Rightarrow\ \hE &\stackrel{{\rm\footnotesize \re{Fiji}}}{=}&
\llb
\begin{array}{l}
  \fracL{1}{2}(i + 2\hS\hxp\hS^{-1}\hxm) \\[12pt]
  \fracL{1}{2}(2\hxm\hS\hxp\hS^{-1} - i)
\end{array}\right.
\nn.
\eeq
Now, noting that $\hxm\hS\hxp\hS^{-1}$ has the same effect on
$\psim$ as $\hS^{-1}\hxp\hS\hxm$ has on $\psip$ and using the
relationship \re{Tonga}, we get
\beq
-\hE - \frac{i}{2} \ =\  \hxm\hS^{-1}\hxp\hS \ =\  \ho^{-1}\hCR^{-1}\ho\hCR
\ =\ \ho^{-1}\CW_-^{-1}\CW_+\ho\CW_+^{-1}\CW_-, \nn\\
-\hE + \frac{i}{2} \ =\  \hS^{-1}\hxp\hS\hxm \ =\  \hCR^{-1}\ho\hCR\ho^{-1}
\ =\ \CW_-^{-1}\CW_+\ho\CW_+^{-1}\CW_-\ho^{-1}.\hspace*{2pt} \nn
\eeq
Multiplying to the right and left by $\CW_{\pm}^{\pm1}$ we retrieve
the string equations written in terms of Lax and Orlov--Shulman
operators as defined in \re{Numea} and \re{Tahiti} \cite{Kostov02}
\be\label{eq:string1}
\Rightarrow L_+L_- = M_+ + \frac{i}{2},\qqquad L_-L_+ = M_- -
\frac{i}{2}.
\ee
And since $[\CR(\hE),\hE]=0$,
\be\label{eq:string2}
M_+=M_-
\ee
In the notation of the standard Toda hierarchy of chapter
\ref{ch:MAT}, the string equations are $R$--dependant;
\be
L^R\bL^R=M+\frac{i}{2},\qquad\bL^RL^R=\bM-\frac{i}{2},\qquad\bM=M.
\ee

\subsection{Toda Description of the Fermi Sea}
In the quasiclassical limit $E\rightarrow-\infty$, the Fermi sea is
completely characterised by the trajectory of the fermion with the
highest energy. The Toda hierarchy in this limit becomes the {\em
dispersionless Toda hierarchy} \cite{Takasaki} where $\hE$ and $\ho$
can be considered as classical canonical variables with the Poisson
bracket being the symplectic form derived from the commutators
\re{comm},
\be\label{eq:fish}
\lb\o,E\rb_{\rm PB}=\o.
\ee
All other commutators between operators now become Poisson brackets
via \re{fish}. By virtue of \re{LtoX}, the Lax operators can be
identified with $\xpm$ in the classical limit, and from the string
equations \re{string1} and \re{string2} we have
\be
\lb \xm, \xp\rb_{\rm PB} = 1,
\ee\be
\xp\xm=M_{\pm}(\xpm)=\frac{1}{R}\sum\limits_{k\ge 1}  k t_{\pm
k} \ \xpm^{ k/R}  +\mu  + \frac{1}{R}\sum\limits_{k\ge 1} v_{\pm k}\
\xpm^{-k/R}.\label{eq:string3}
\ee
where the dispersionless string equation \re{string3} coincides with
\re{Fred} which explicitly contains the perturbing phase. Therefore,
that is the equation which describes the dynamics of the Fermi sea.

It can be shown\footnote{It suffices to show that the right hand
side of \re{Tao} is a closed form, see \S\ref{sec:tau} in this
thesis, or \cite{Takasaki}.} that there exists a free energy,
namely, the $\t$--function, defined in the dispersionless limit as
\be\label{eq:Tao}
d\log\t=
\sum\limits_{k=1}^\infty v_{ k} dt_k -
\sum\limits_{k=1}^\infty v_{- k}dt_{-k}
+ \phi(E) dE
\ee
from which we get the coefficients
\be
v_n=\der{\log\t}{t_n},\qquad \phi(E)=\der{\log\t}{E}
\ee
and, by \re{LtoX}, the functions $\xpm(\o,E)$ are given by the Lax
operators from \re{Numea},
\beq
\xpm &=& e^{\mp i\phi/2} e^{\mp i\p_E}   e^{\pm i\phi/2}
\(1 +\sum_{k\ge 1} a_{\pm k}   \ho^{\mp \nf{k}{R}}\) \nn\\
&=& e^{\mp i\phi/2}  e^{\pm i\phi/2}  e^{[\p_E,\phi]} e^{\mp i\p_E}
\(1 +\sum_{k\ge 1} a_{\pm k}   \o^{\mp \nf{k}{R}}\) \nn\\
&\stackrel{(|E|\rightarrow\infty)}{\longrightarrow}& e^{-\frac{1}{2R}\chi}\o^{\pm1}
\(1 +\sum_{k\ge 1} a_{\pm k}   \o^{\mp \nf{k}{R}}\)\label{eq:Insomnia}
\eeq
where the {\em string susceptibility} is
\be
\chi = -R\p_E\phi = \p_E^2\log\t.
\ee

\newpage
\section{Exact Solution For Two Nonzero Couplings}
The aim is to find the coefficients $a_{\pm k}(E)$ in \re{Insomnia}.
This will be done for the case where the first two couplings are
nonzero; $t_{\pm1},t_{\pm2} \neq 0$. In this case, \re{Insomnia}
becomes
\be \label{eq:Anzaas}
\xpm = e^{-\frac{\chi}{2R}}\omega^{\pm 1} \left( 1 + a_{\pm1}
\omega^{\mp\frac{1}{R}} + a_{\pm2} \omega^{\mp\frac{2}{R}} \right)
\ee
The string equations \re{string3} provide a constraint on $x_\pm$
which determines the constants $a_{\pm 1}$, $a_{\pm 2}$:

\be \label{eq:streq}
\xp \xm =   \left\lbrace
\begin{array}{lr}
\frac{t_{+1}}{R} \xp^{\frac{1}{R}} + \frac{t_{+2}}{R}
\xp^{\frac{2}{R}} + \mu &  \makebox[15mm]{\dotfill}
\makebox[16mm][c]{Large $\omega$} \sep \frac{t_{-1}}{R}
\xm^{\frac{1}{R}} + \frac{t_{-2}}{R} \xm^{\frac{2}{R}} + \mu
&  \makebox[15mm]{\dotfill} \makebox[16mm][c]{Small $\omega$} \\
\end{array} \right.
\ee

\subsection{Profiles of the Fermi Sea}
Defining, for convenience
\be
Q_1 = \frac{1}{R} e^{\chi \frac{R - \nicefrac{1}{2}}{R^2}}, \qquad
Q_2 = \frac{1}{R} e^{\chi \frac{R - 1}{R^2}},
\ee
we insert $\xpm$ into \re{streq} on both sides and expand the powers
of $\o^{\nf{1}{R}}$ (to $2^{\text{nd}}$ order):
\beq
\lefteqn{1+ a_{+1} a_{-1} + a_{+2} a_{-2} +
(a_{-1} + a_{+1} a_{-2})\o^\frac{1}{R}} \nn\\
&+& (a_{+1} + a_{-1} a_{+2})\o^\frac{-1}{R} + a_{-2}
\o^{\frac{2}{R}} + a_{+2} \o^{\frac{-2}{R}}  \nn\sep &\hspace*{-12pt}=&\hspace*{-12pt}
\left\lbrace
\begin{array}{l}
\!\!\!
t_{+1} Q_1 \o^{\nf{1}{R}} \bigg[ 1 + \frac{1}{R}(a_{+1} + a_{+2}\o^{\nf{-1}{R}})\o^{\nf{-1}{R}}
+ \frac{1}{R}\left(\frac{1}{R} - 1\right)\frac{1}{2 !} (a_{+1} + a_{+2}\o^{\nf{-1}{R}})^2 \o^{\nf{-2}{R}} \bigg]  \\
\!\!\!
\phantom{.} + t_{+2} Q_2 \o^{\nf{2}{R}}\bigg[ 1 + \frac{2}{R} (a_{+1} + a_{+2}\o^{\nf{-1}{R}})\o^{\nf{-1}{R}}
+ \frac{2}{R}\left(\frac{2}{R} - 1\right)\frac{1}{2 !} (a_{+1} + a_{+2}\o^{\nf{-1}{R}})^2 \o^{\nf{-2}{R}} \bigg] \\
\!\!\!
\phantom{.}+ \mu e^{\nf{\chi}{R}} + \order (\o^{\nf{-3}{R}}) \xstrut\xstrut\xstrut\sep
\!\!\!
t_{-1} Q_1 \o^{\nf{-1}{R}} \bigg[ 1 +
\frac{1}{R}(a_{-1} + a_{-2}\o^{\nf{1}{R}})\o^{\nf{1}{R}}
+ \frac{1}{R}\left(\frac{1}{R} - 1\right)\frac{1}{2 !} (a_{-1} + a_{-2}\o^{\nf{1}{R}})^2 \o^{\nf{2}{R}} \bigg] \\
\!\!\!
\phantom{.}+ t_{-2} Q_2 \o^{\nf{-2}{R}}\bigg[ 1 + \frac{2}{R} (a_{-1} + a_{-2}\o^{\nf{1}{R}})\o^{\nf{1}{R}}
+ \frac{2}{R}\left(\frac{2}{R} - 1\right)\frac{1}{2 !} (a_{-1} + a_{-2}\o^{\nf{1}{R}})^2 \o^{\nf{2}{R}} \bigg] \\
\!\!\!
\phantom{.}+ \mu e^{\nf{\chi}{R}} + \order (\o^{\nf{3}{R}})
\end{array}
\right.
\label{eq:big}
\eeq
Matching up the coefficients of the positive powers of $\o$ of the
left hand side of \re{big} with those of the upper equality on the
right hand side and the negative powers of $\o$ with those of the
lower equality, we obtain
\be \label{eq:a12}
a_{\pm 2} = t_{\mp2} Q_2,\qqquad a_{\pm 1} = t_{\mp 1} Q_1 +
\frac{2}{R} t_{\mp 2} Q_2 a_{\mp 1} - a_{\mp 1} a_{\pm 2}
\ee\be
\label{eq:chi1}
\frac{1}{R} t_{\pm 1} Q_1 a_{\pm 1} + t_{\pm 2}Q_2 \left[
\frac{2}{R}a_{\pm2} + \frac{1}{R}\left( \frac{2}{R}
-1\right)a_{\pm1}^2\right]
 + \mu e^{\nf{\chi}{R}} - a_{\mp 1} a_{\pm 1} - a_{\mp 2} a_{\pm 2} = 1
\ee

The equations \re{a12} allow us to solve explicitly for the
coefficients $a_{+1}$ and $a_{-1}$:
\be \label{eq:a2}
a_{\pm 2} = t_{\mp2} Q_2  = \frac{t_{\mp2}}{R} e^{\chi \frac{R -
1}{R^2}}
\ee
\beq
a_{\pm 1} &=& \frac{t_{\mp1}Q_1 +
t_{\mp2}t_{\pm1} Q_1 Q_2 \left(\frac{2}{R} - 1\right) }{ 1 -
t_{\pm2}t_{\mp2} Q_2^{\phantom{2}2} \left(\frac{2}{R} - 1\right)^2}\nn\\
&=& \frac{ t_{\mp1} \frac{1}{R} e^{\chi \frac{R -
\nf{1}{2}}{R^2}} + t_{\mp2}t_{\pm1} e^{\chi \frac{2R -
\nf{3}{2}}{R^2}}  \left(\frac{2-R}{R^3}\right) }{ 1 -
t_{\pm2}t_{\mp2} e^{\chi \frac{2R - 2}{R^2}} \left(\frac{2 -
R}{R^3}\right)^2 }\label{eq:a1}
\eeq
And the string susceptibility, $\chi$, is found by solving \re{chi1}
given values for the perturbation coupling constants, $t_{\pm 1,2}$,
the chemical potential, $\mu$, and compactification radius, $R$:
\beq
\label{eq:chi}
1&=&\mu e^{\frac{\chi}{R}} + \frac{ \left(t_{-1} + t_{-2}t_{1}
\frac{2-R}{R^2} e^{\chi \frac{R - 1}{R^2}}\right) \left( t_{1} +
t_{2}t_{-1} \frac{2-R}{R^2} e^{\chi \frac{R - 1}{R^2}}\right)
}{\left[1 - t_{-2}t_{2}\left(\frac{2-R}{R^2}\right)^2  e^{\chi
\frac{2R - 2}{R^2}} \right]^2} \frac{1-R}{R^3} e^{\chi
\frac{2R-1}{R^2}} \nn\\&& \qquad + \,\,t_{2}t_{-2}\frac{2-R}{R^3}
e^{\chi \frac{2R - 2}{R^2}}.
\eeq
With these coefficients, the explicit solution \re{Anzaas} provides
a parameterization for the boundary of the Fermi sea.

By setting $t_{\pm2} = 0$  in equations \re{a2}, \re{a1} and
\re{chi} we check that the correct solutions of Sine-Liouville
theory \cite{AKK02} are reproduced
\beq
a_{\pm 2} &=& 0 \nn\\
a_{\pm 1} &=& \frac{t_{\mp1}}{R} e^{\chi \frac{R - \nicefrac{1}{2}}{R^2}} \label{eq:SineGordon}\\
1 &=& \mu e^{\frac{\chi}{R}} + \frac{1}{R^2}\left(\frac{1}{R} -
1\right)t_{+1}t_{-1} e^{\chi \frac{2R - 1}{R^2}}\nn
\eeq

The numerical solutions for $\chi$ are displayed in table
\ref{tab:lin}(a), and we verify in Fig. \ref{fig:profilesa}(a) that
the profiles for $t_{\pm2}=0$ conform with those previously found in
\cite{AKK02}.
\begin{table}[t]\tablabel{tab:lin}
\begin{center}
\fontsize{9pt}{13pt} \fontseries{tt} \selectfont
\begin{tabular}{cc}
    \begin{tabular}{|c|c|r@{.}l|r@{.}l |}
    \hline
    \multicolumn{6}{c}{$R=\nf{2}{3}, t_{\pm1}=2, t_{\pm2}=0$}\\
    \hline No.&  $\mu$&
    \multicolumn{4}{c}{$\chi$ Solutions $\in\bbR$}\vline \\
    \hline\hline
    1 &-1  &-1&9341  &1&9341 \\
    2 &39  &-2&91293 &\omit&---\\
    3 &79  &-3&24699 &\omit&---\\
    4 &119 &-3&45918 &\omit&---\\
    5 &159 &-3&61594 &\omit&---\\
    6 &199 &-3&74064 &\omit&---\\
    7 &239 &-3&84435 &\omit&---\\
    8 &279 &-3&93321 &\omit&---\\
    9 &319 &-4&01099 &\omit&---\\
    10&359 &-4&08018 &\omit&---\\
    \hline
    \end{tabular}

&
    \begin{tabular}{|c|c|r@{.}l|r@{.}l|}
    \hline
    \multicolumn{6}{c}{$R=\nf{4}{3}, t_{\pm1}=2, t_{\pm2}=0$}\\
    \hline No.&  $\mu$&
    \multicolumn{4}{c}{$\chi$ Solutions $\in\bbR$}\vline \\
    \hline\hline
    1 &3   &-1&25078 &8&92567 \\
    2 &43  &-5&0081  &23&1283 \\
    3 &83  &-5&88879 &26&6358 \\
    4 &123 &-6&41441 &28&7336 \\
    5 &163 &-6&79038 &30&2353 \\
    6 &203 &-7&0833  &31&4057 \\
    7 &243 &-7&3233  &32&3649 \\
    8 &283 &-7&52662 &33&1777 \\
    9 &323 &-7&70299 &33&8828 \\
    10&363 &-7&85873 &34&5054 \\
    \hline
    \end{tabular}

\end{tabular}\bf \caption{\rm Numerical solutions for $\chi$
with first couplings only.}
\end{center}
\end{table}
\begin{figure}[!hb]
\figlabel{fig:profilesa}
\begin{center}
\begin{tabular}{cc}
 \includegraphics[scale=0.8]{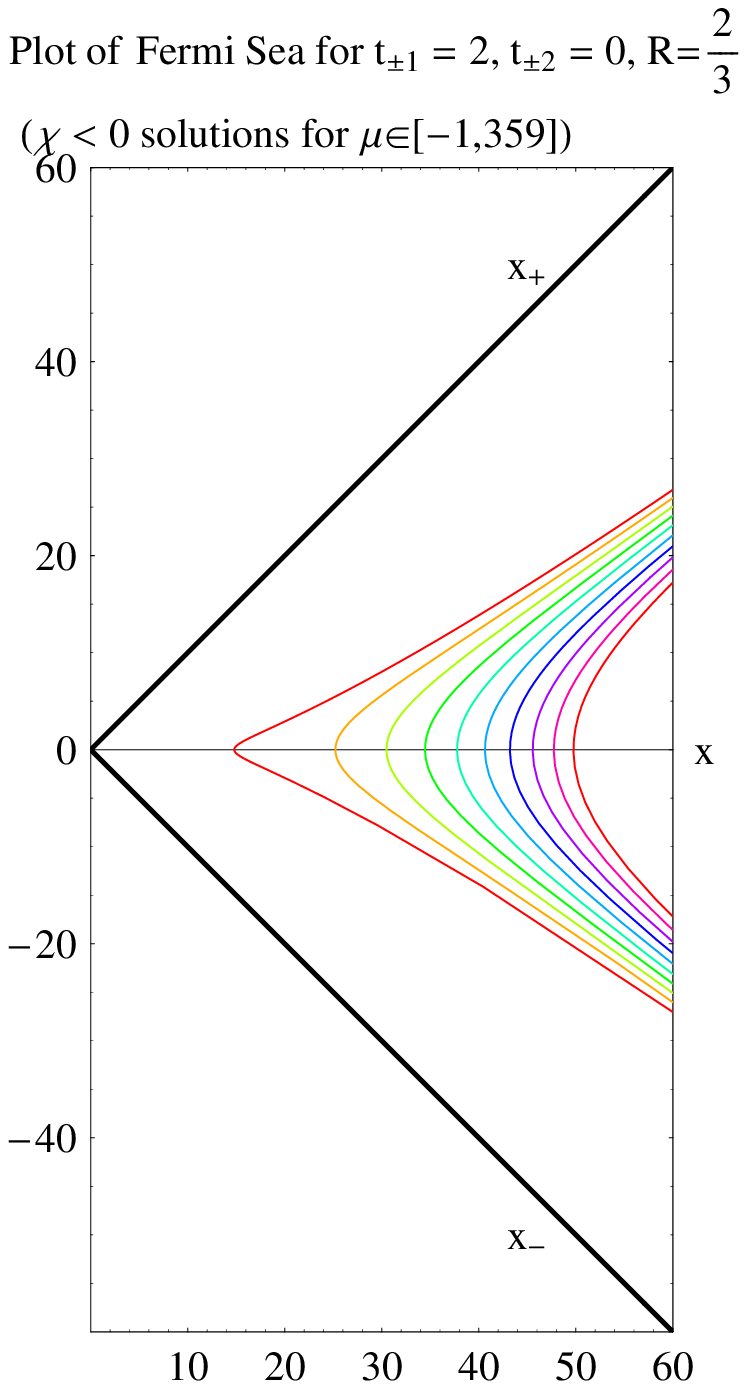}&
\includegraphics[scale=0.8]{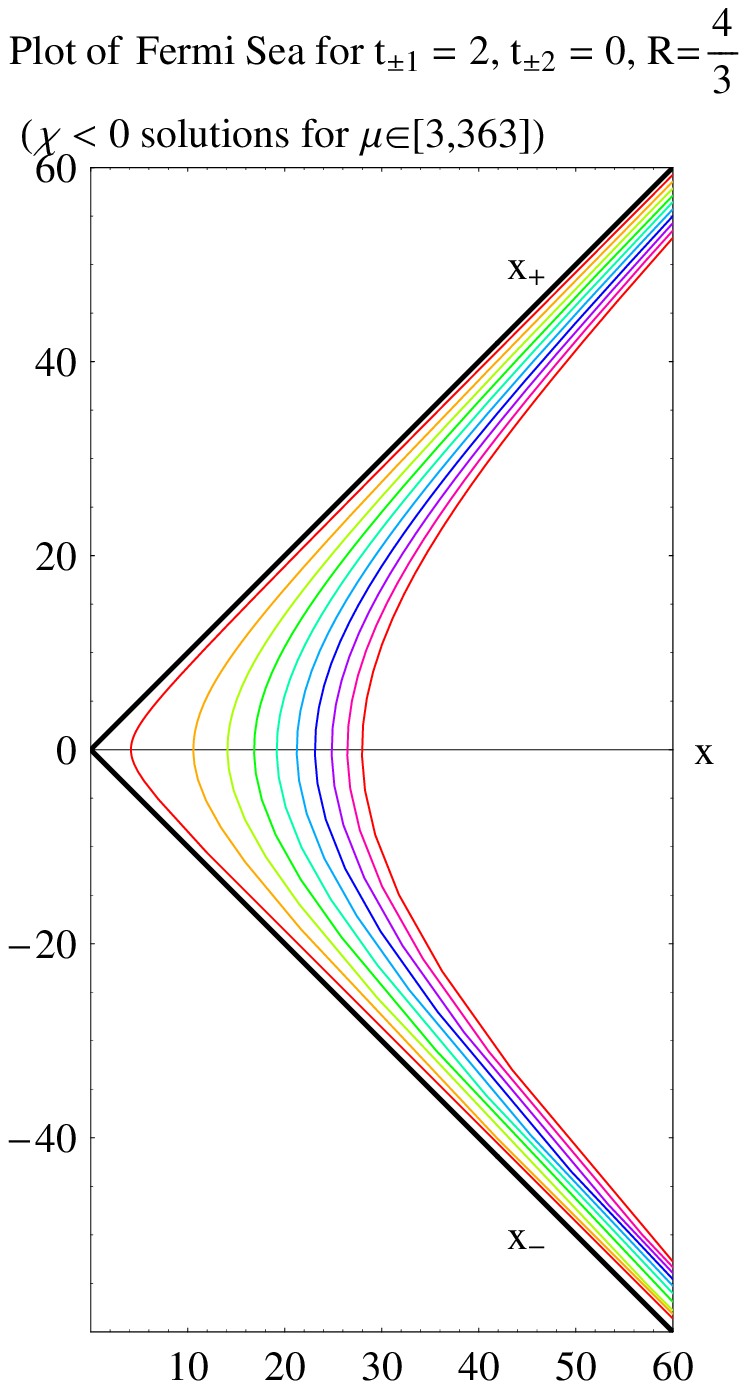}\\
 {\bf (a)} & {\bf (b)}
\end{tabular}
\bf \caption{\rm Profiles of the Fermi sea for linear order
perturbations: {\bf (a)} With $R=\nf{2}{3}$ reproduces previous
results \cite{AKK02}, {\bf (b)} with $R=\nf{4}{3}$. }
\end{center}
\end{figure}

There may be multiple negative numerical solutions for $\chi$ for a
given value of the chemical potential, $\mu$. We choose the $\chi$
solutions which reduce to the correct unperturbed behaviour as the
couplings vanish and ignore the additional ones. The solutions from
the correct branch are listed the first column of solutions in the
tables.

Considerable simplification of \re{chi} is achieved if we set $t_1 =
t_{-1}$ and $t_2 = t_{-2}$, in which case we are considering
profiles that are symmetric about the x--axis.
\beq
\label{eq:chisimp}
1 &=& \mu e^{\frac{\chi}{R}} + t_{2}^2 \frac{2-R}{R^3} e^{\frac{2R - 2}{R^2}\chi } +
\frac{
t_1^2 \frac{1-R}{R^3} e^{\frac{2R - 1}{R^2} \chi}
}{
\left(1 - t_2 \frac{2-R}{R^2} e^{\frac{R - 1}{R^2}\chi} \right)^2
}
\eeq
Numerical solutions for $\chi$ of this situation with nonzero second
couplings are displayed in table \ref{tab:quad} and example profiles
are plotted in fig.\ref{fig:profilesb}.
\begin{table}[!hb]\tablabel{tab:quad}
\fontsize{9pt}{13pt} \fontseries{tt} \selectfont
\begin{center}
\begin{tabular}{cc}
\begin{tabular}{|c|c|r@{.}l|}
\hline
\multicolumn{4}{c}{$R=\nf{4}{3}, t_{\pm1}=0, t_{\pm2}=2$}\\
\hline No.&  $\mu$&
\multicolumn{2}{c}{$\chi$ Solutions $\in\bbR$}\vline \\
\hline\hline
    1 &2   &-1&95871 \\
    2 &42  &-5&21472 \\
    3 &82  &-6&04117 \\
    4 &122 &-6&54111 \\
    5 &162 &-6&90127 \\
    6 &202 &-7&1832 \\
    7 &242 &-7&41499 \\
    8 &282 &-7&61185 \\
    9 &322 &-7&78298 \\
    10&362 &-7&93435 \\
\hline
\end{tabular}

&
\begin{tabular}{|c|c|r@{.}l|r@{.}l|r@{.}l|}
\hline
\multicolumn{8}{c}{$R=\nf{4}{3}, t_{\pm1}=2, t_{\pm2}=2$}\\
\hline No.& $\mu$&
\multicolumn{6}{c}{$\chi$ Solutions $\in\bbR$}\vline\\
\hline\hline
    1 &3   &-1&80551 &-0&308277&6&70678 \\
    2 &43  &-5&23261 &2&13603  &4&02611 \\
    3 &83  &-6&05183 &2&39344  &3&7565 \\
    4 &123 &-6&54877 &2&51264  &3&63311 \\
    5 &163 &-6&90728 &2&58499  &3&55865 \\
    6 &203 &-7&18815 &2&63485  &3&50752 \\
    7 &243 &-7&4192  &2&67189  &3&46962 \\
    8 &283 &-7&61552 &2&70081  &3&44009 \\
    9 &323 &-7&78624 &2&7242   &3&41625 \\
    10&363 &-7&93728 &2&74362  &3&39647 \\
\hline
\end{tabular}
\label{tab:chi22R4}
\end{tabular}
\bf \caption{\rm Numerical solutions for $\chi$ with second
couplings, $R=\nf{4}{3}$.}
\end{center}
\end{table}
\begin{figure}[!ht]
\figlabel{fig:profilesb}
\begin{center}
\begin{tabular}{cc}
 \includegraphics[scale=0.8]{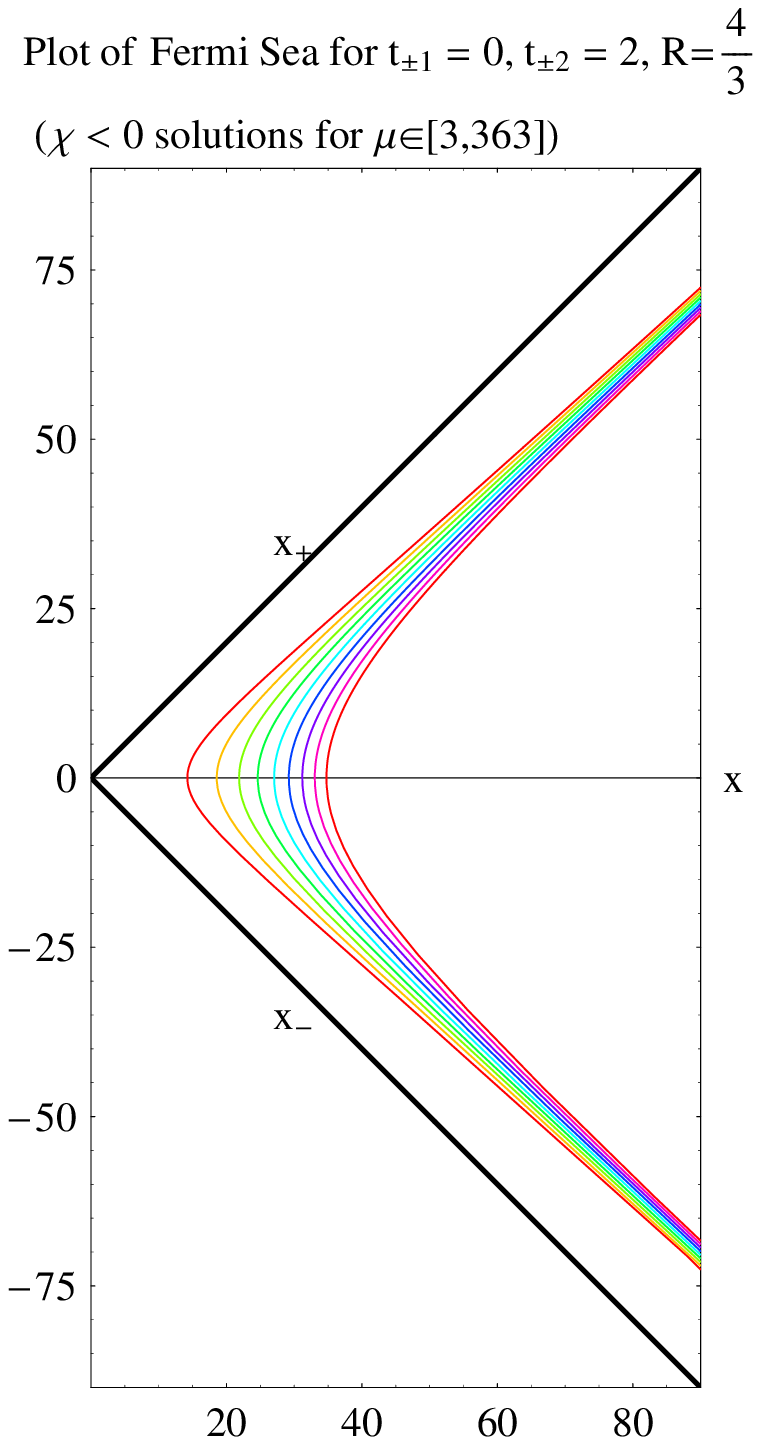}&
\includegraphics[scale=0.8]{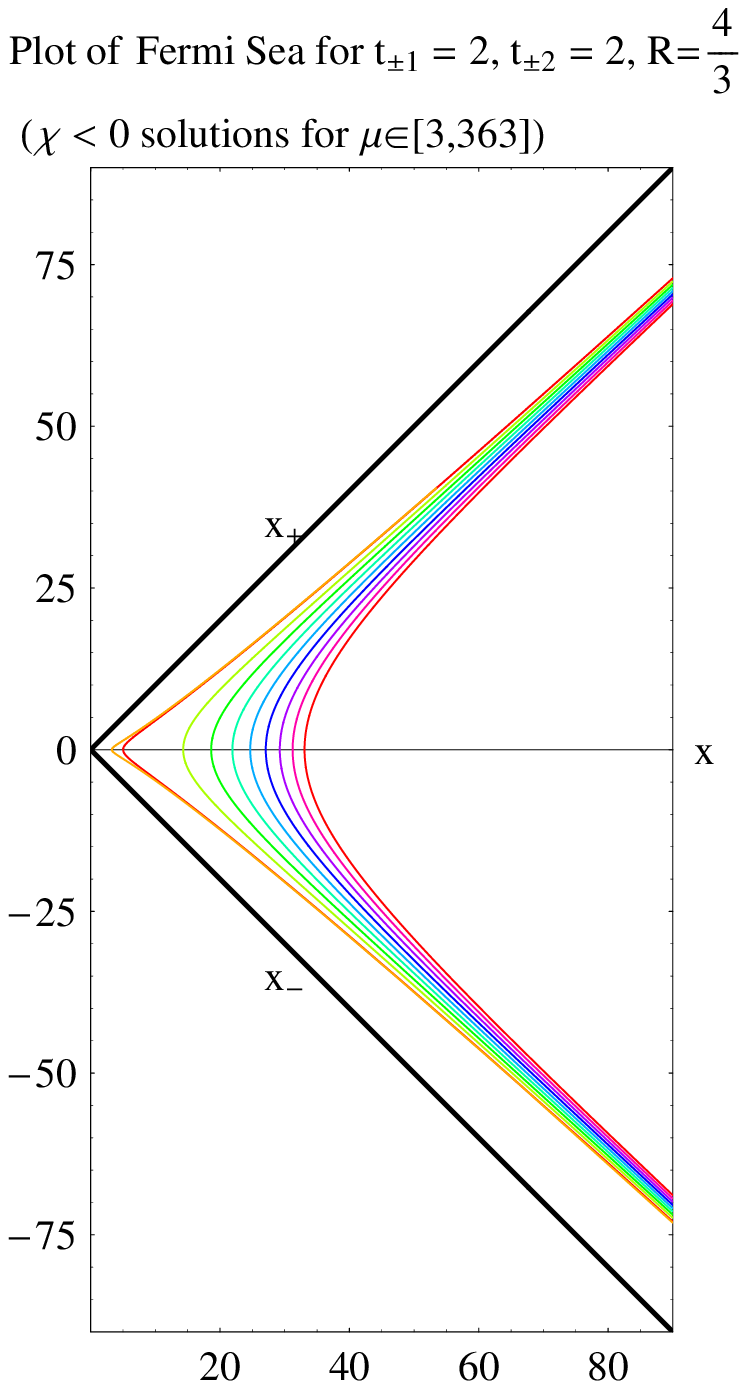}\\
 {\bf (a)} & {\bf (b)}
\end{tabular}
\bf \caption{\rm Profiles of the Fermi sea for perturbations up to
quadratic order, $R=\nf{4}{3}$: {\bf (a)} $t_{\pm1}=0$, $t_{\pm2}=2$
{\bf (b)}
 $t_{\pm1}=t_{\pm2}=2$.}
\end{center}
\end{figure}

\subsection{Free Energy}
Our ultimate aim is to compute the free energy of the matrix model,
and thereby the partition function of 2D string theory. Introducing
the variables
\be
\label{eq:dimlessvars}
y_1 = \l_1 \sqrt{\frac{1-R}{R^3}} \mu^{\frac{1-2R}{2R}}, \qquad
y_2 = \l_2 \frac{2-R}{R^2} \mu^{\frac{1-R}{R}},
\ee
with
\be
\label{eq:lambdas}
\l_1 =\sqrt{t_1 t_{-1}},\qquad
\l_2 =\sqrt{t_2 t_{-2}},
\ee
we see that \re{chisimp} can be rewritten
\beq
\label{eq:ys}
1 &=& e^{\frac{X}{R}} + y_2^2 \frac{R}{2-R} e^{\frac{2R - 2}{R^2}X } +
\frac{
y_1^2 e^{\frac{2R - 1}{R^2} X}
}{
\left(1 - y_2 e^{\frac{R - 1}{R^2} X} \right)^2
} \\
\chi &=& -R \log \mu + X(y_1,y_2)\label{eq:xub}.
\eeq
As a check, we may set $t_{\pm1} = 0$. Then, by transforming $R \mapsto 2R$
and $t_{\pm2} \mapsto t_{\pm1}$, eq. (\ref{eq:ys}) becomes:
\be
1 = e^\frac{X}{R} + t_{1}t_{-1} \frac{1-R}{4R^3} \mu^\frac{1-2R}{R}
e^{\frac{2R-1}{2R^2}X}
\ee
substituting for $X$ \re{xub}, we get back the result for
sine--Gordon gravity \re{SineGordon}.

To calculate the free energy we must integrate the function
$X(y_1,y_2)$;
\be
\label{eq:freeergry}
\CF = \int\!d\mu\!\int\!d\mu\; \chi = -\frac{R}{2}\mu^2(\log \mu -
1) + \int\!d\mu\!\int\!d\mu \; X[y_1(\mu), y_2(\mu)].
\ee
Further simplification may be achieved by introducing a third
variable $y_3$, not dependent on $\mu$, defined in terms of $y_1$
and $y_2$
\be
y_3 = \frac{y_1^{\phantom{1}a}}{y_2^{\phantom{2}b}},\qquad \frac{a}{b} = \frac{1-2R}{2-2R}
\ee
Then, substituting $y_2 = y_1^\frac{a}{b} y_3^\frac{-1}{b}$ into
\re{ys} we get
\be\label{eq:deadend}
\left(1 - y_1^\frac{a}{b} y_3^\frac{-1}{b} e^{\frac{R - 1}{R^2}X
}\right)^2 \left(1 - e^\frac{X}{R} - y_1^\frac{2a}{b}
y_3^\frac{-2}{b} \frac{R}{2-R} e^{\frac{2R - 2}{R^2}X }\right) -
y_1^2 e^{\frac{2R-1}{R^2}X} = 0.
\ee
This is a polynomial in $y_1$ with irrational powers and may not be
solved explicitly for $y_1$. It is therefore not possible to solve,
even implicitly, for the free energy (\ref{eq:freeergry}) by
performing the integral as was done in the previously considered
case of sine-Gordon gravity \cite{KKK,AKK02,AK02}.

\EndOfChapter

We extended the case of sine--Gordon gravity previously considered
in \cite{AKK02} where the CFT action is perturbed only by terms with
first couplings to the case where it is perturbed by terms with
nonzero second couplings as well. Unfortunately, we were not able to
obtain an explicit expression for the free energy as was done in
earlier work. We may, however, hunt for phase transitions occurring
for specific values of the couplings. They are signaled by a
discontinuity in the curve $X(\mu)$, which causes the total
derivative $\frac{d\mu}{dX}$ to vanish. In other words, with the
functional $K[y_1(\l_1,\mu(X)),y_2(\l_2,\mu(X)),X]$ designating the
right hand side of \re{ys} we should solve
\be
\left.\frac{d}{dX} K[y_1(\l_1,\mu(X)),y_2(\l_2,\mu(X)),X]\right|_{
\frac{d\mu}{dX}=0}=0.
\ee

Near these critical values of $t_{\pm 1}$, $t_{\pm 2}$, the CFT
action perturbed by a \snd{2} order potential \re{pertCFT} should
describe pure 2D quantum gravity ($c=0$ theory, see \S\ref{sec:u})
by analogy with the one--matrix model. Similarly, double--critical
points of the one--matrix model corresponding to the Ising model
should appear under additional constraints \cite{Serge}. Such an
Ising model would be equivalent to Majorana fermions interacting
with 2D quantum gravity \cite{KazakovMigdal}. Possible signs of
critical behaviour appear in fig.\ref{fig:profilesb}(b) where the
first two profile curves do not follow the same general trend as the
other curves.


\phantomsection
\addcontentsline{toc}{chapter}{Conclusions and Outlook}
\chapter*{\centerline{Conclusion and Outlook}}
We have seen how to construct a critical string theory in the
presence of background fields, and how the cosmological constant
term leads to a tachyon background which introduces a potential wall
in the target space precluding strings from venturing into the
strong coupling region. Then it was shown how noncritical string
theory could be related to critical string theory in a linear
dilaton background.

From the effective action of 2D string theory in non--trivial
backgrounds, we saw that, in the low energy limit, the target space
could be curved with a Euclidean black hole background. We
constructed a time--dependant tachyon background using tachyon
vertex operators to perturb the CFT describing Liouville gravity
coupled to $c=1$ matter and explained the FZZ conjecture which
relates the T--dual of this perturbed CFT with one non-zero tachyon
coupling, which is a sine--Liouville CFT, to the exact black hole
background valid for all energies that describes the black hole
background.

Following an introduction to general random matrix theory, we showed
how the partition function of 2D quantum gravity was related to the
free energy of the one--matrix model in the double scaling limit. In
chapter \ref{ch:MQM}, we formulated matrix quantum mechanics and saw
that the free energy of 2D string theory could be calculated from
MQM. The dynamics in the target space (namely, the dynamics of the
tachyon) could be described in terms of matrix operators by
considering a collective field theory --- Das--Jevicki string field
theory.

In the last chapter, we recited the formulation of MQM in chiral
coordinates and demonstrated that certain quantities, such as the
$S$--matrix, can be calculated quite easily from this model. We then
described tachyon perturbations in terms of the Toda lattice
hierarchy and computed the phase space profiles of MQM for
perturbations with two nonzero couplings.

It was not possible to solve explicitly for the free energy as was
done in the case of sine--Liouville gravity where only the first
couplings are nonzero. It may be possible, however, to determine the
behaviour of the free energy near critical points and thus determine
which theories arise from MQM in this critical regime. This is an
interesting problem for future research.

Many pertinent topics were not treated in this thesis; winding
modes, discrete states and non--perturbative effects to mention a
few. The application of matrix models to their investigation has
only relatively recently begun. These are exciting, rich areas of
research which show little sign of abating.

\renewcommand{\baselinestretch}{1} \normalsize
\clearemptydoublepage
\phantomsection
\addcontentsline{toc}{chapter}{References}
\newpage\thispagestyle{empty}

\bibliographystyle{osa}
\bibliography{biblio}
\end{document}